\PassOptionsToPackage{unicode}{hyperref}
\PassOptionsToPackage{hyphens}{url}
\PassOptionsToPackage{dvipsnames,svgnames,x11names}{xcolor}
\documentclass[
  8pt,
  pa4paper,
  twocolumn]{extarticle}
\usepackage{amsmath,amssymb}
\usepackage{lmodern}
\usepackage{iftex}
\ifPDFTeX
  \usepackage[T1]{fontenc}
  \usepackage[utf8]{inputenc}
  \usepackage{textcomp} 
\else 
  \usepackage{unicode-math}
  \defaultfontfeatures{Scale=MatchLowercase}
  \defaultfontfeatures[\rmfamily]{Ligatures=TeX,Scale=1}
\fi
\IfFileExists{upquote.sty}{\usepackage{upquote}}{}
\IfFileExists{microtype.sty}{
  \usepackage[]{microtype}
  \UseMicrotypeSet[protrusion]{basicmath} 
}{}
\usepackage{xcolor}
\IfFileExists{xurl.sty}{\usepackage{xurl}}{} 
\IfFileExists{bookmark.sty}{\usepackage{bookmark}}{\usepackage{hyperref}}
\hypersetup{
  pdftitle={MARF: The Medial Atom Ray Field Object Representation},
  pdfauthor={Peder Bergebakken Sundt; Theoharis Theoharis},
  pdflang={en\_US},
  colorlinks=true,
  linkcolor={Maroon},
  filecolor={Maroon},
  citecolor={Blue},
  urlcolor={Blue},
  pdfcreator={LaTeX via pandoc}}
\usepackage[margin=1.3cm,columnsep=0.8cm]{geometry}
\usepackage{longtable,booktabs,array}
\usepackage{multirow}
\usepackage{calc} 
\usepackage{etoolbox}
\makeatletter
\patchcmd\longtable{\par}{\if@noskipsec\mbox{}\fi\par}{}{}
\makeatother
\IfFileExists{footnotehyper.sty}{\usepackage{footnotehyper}}{\usepackage{footnote}}
\makesavenoteenv{longtable}
\usepackage{graphicx}
\makeatletter
\def\maxwidth{\ifdim\Gin@nat@width>\linewidth\linewidth\else\Gin@nat@width\fi}
\def\maxheight{\ifdim\Gin@nat@height>\textheight\textheight\else\Gin@nat@height\fi}
\makeatother
\setkeys{Gin}{width=\maxwidth,height=\maxheight,keepaspectratio}
\makeatletter
\def\fps@figure{htbp}
\makeatother
\usepackage[normalem]{ulem}
\providecommand{\tightlist}{%
  \setlength{\itemsep}{0pt}\setlength{\parskip}{0pt}}
\newlength{\cslhangindent}
\newlength{\csllabelwidth}
\newlength{\cslentryspacingunit} 
\newenvironment{CSLReferences}[2] 
 {
  \setlength{\parindent}{0pt}
  \ifodd #1
  \let\oldpar\par
  \def\par{\hangindent=\cslhangindent\oldpar}
  \fi
  \setlength{\parskip}{#2\cslentryspacingunit}
 }%
 {}
\usepackage{calc}

\newcommand{\CSLLeftMargin}[1]{\parbox[t]{\csllabelwidth}{#1}}
\newcommand{\CSLRightInline}[1]{\parbox[t]{\linewidth - \csllabelwidth}{#1}\break}

\ifLuaTeX
\usepackage[bidi=basic]{babel}
\else
\usepackage[bidi=default]{babel}
\fi
\babelprovide[main,import]{american}

\def\languageshorthands#1{}
\usepackage{libertine}\usepackage{enumitem}\setenumerate{listparindent=\parindent}\setitemize{listparindent=\parindent, parsep=0pt, leftmargin=*}\setlist{listparindent=\parindent, parsep=0pt,}\PassOptionsToPackage{labelfont=sc}{caption}\usepackage{float}\makeatletter\def\fps@figure{t}\def\fps@table{t}\makeatother\usepackage[nodisplayskipstretch]{setspace}\AtBeginDocument{\abovedisplayshortskip=-6pt}\makeatletter
\@ifpackageloaded{subfig}{}{\usepackage{subfig}}
\@ifpackageloaded{caption}{}{\usepackage{caption}}
\AtBeginDocument{%

}
\AtBeginDocument{%

}
\newcounter{pandoccrossref@subfigures@footnote@counter}
\newenvironment{pandoccrossrefsubfigures}{%
\setcounter{pandoccrossref@subfigures@footnote@counter}{0}
\begin{figure}\centering%
\gdef\global@pandoccrossref@subfigures@footnotes{}%
\DeclareRobustCommand{\footnote}[1]{\footnotemark%
\stepcounter{pandoccrossref@subfigures@footnote@counter}%
\ifx\global@pandoccrossref@subfigures@footnotes\empty%
\gdef\global@pandoccrossref@subfigures@footnotes{{##1}}%
\else%
\g@addto@macro\global@pandoccrossref@subfigures@footnotes{, {##1}}%
\fi}}%
{\end{figure}%
\addtocounter{footnote}{-\value{pandoccrossref@subfigures@footnote@counter}}
\@for\f:=\global@pandoccrossref@subfigures@footnotes\do{\stepcounter{footnote}\footnotetext{\f}}%
\gdef\global@pandoccrossref@subfigures@footnotes{}}
\@ifpackageloaded{float}{}{\usepackage{float}}
\floatstyle{ruled}
\@ifundefined{c@chapter}{\newfloat{codelisting}{h}{lop}}{\newfloat{codelisting}{h}{lop}[chapter]}
\floatname{codelisting}{Listing}

\makeatother  \usepackage{xtab}
  \renewenvironment{longtable}{%
    \begin{center}\begin{xtabular}%
    }{%
    \end{xtabular}\end{center}%
    }
  \renewcommand{\caption}[1]{}
  \renewcommand{\endhead}{}
  
\usepackage{soul}
\ifLuaTeX
  \usepackage{selnolig}  
\fi

\title{MARF: The Medial Atom Ray Field Object Representation}
\author{Peder Bergebakken Sundt \and Theoharis Theoharis}
\date{\today}

\begin{document}
\maketitle
\begin{abstract}
We propose Medial Atom Ray Fields (MARFs), a novel neural object representation that enables accurate differentiable surface rendering with a single network evaluation per camera ray. Existing neural ray fields struggle with multi-view consistency and representing surface discontinuities. MARFs address both using a medial shape representation, a dual representation of solid geometry that yields cheap geometrically grounded surface normals, in turn enabling computing analytical curvature despite the network having no second derivative. MARFs map a camera ray to multiple medial intersection candidates, subject to ray-sphere intersection testing. We illustrate how the learned medial shape quantities applies to sub-surface scattering, part segmentation, and aid representing a space of articulated shapes. Able to learn a space of shape priors, MARFs may prove useful for tasks like shape retrieval and shape completion, among others. Code and data can be found at \href{https://github.com/pbsds/MARF}{github.com/pbsds/MARF}.
\end{abstract}

\newenvironment{widefig}{\renewenvironment{figure}{\begin{figure*}[htb!]\centering}{\end{figure*}}}

\hypertarget{introduction}{%
\section{Introduction}\label{introduction}}

Learning efficient and accurate ways to represent 3D geometry is valuable to applications such as 3D shape analysis, computer graphics, computer vision, and robotics. The recent discovery of \emph{neural fields}, also known as coordinate-based networks or implicit neural representations, has brought a renewed interest in visual computing problems. While simple in construction, neural fields exhibit an impressive ability to compactly represent, manipulate and generate continuous signals of arbitrary resolution and dimensionality across a plethora of modalities, in our case 3D geometry. They can also learn the underlying space of the training shapes, useful in applications such as generative shape modeling, shape infilling/completion, and shape retrieval.

\begin{figure}
\hypertarget{fig:sphere-tracing}{%
\centering
\includegraphics{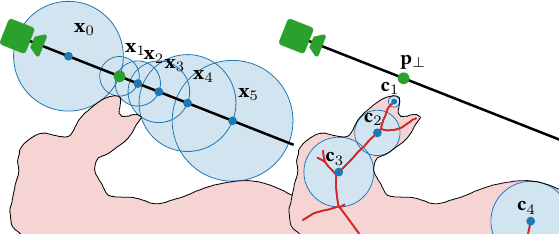}
\caption{A 2D slice of the Stanford bunny. On the left we sphere-trace it. On the right we show its medial axis, four medial atoms, and the projection \(\mathbf p_\perp\) of nearest atom center \(\mathbf c_1\) onto the line. Each tracing step requires an evaluation of the distance field which proves expensive when represented with a neural field. We explore MARFs which map the \emph{line} to \(n\) medial intersection candidates in a single evaluation. The medial representation has many downstream use-cases.}\label{fig:sphere-tracing}
}
\end{figure}

While Cartesian neural fields represent 3D volumes admirably, rendering them requires ray-marching or sphere-tracing, where each sample along the ray in turn requires a full network evaluation which is expensive. In this paper we avoid sphere-tracing entirely, by parameterizing the field in terms of rays instead of points. Visualized in Fig.~\ref{fig:sphere-tracing}, we explore neural fields that map an oriented ray directly to its surface intersection point via an intermediate medial representation. This enables efficient real-time single-evaluation differentiable neural surface rendering and extraction.

\begin{figure}
\hypertarget{fig:nn-architecture}{%
\centering
\includegraphics{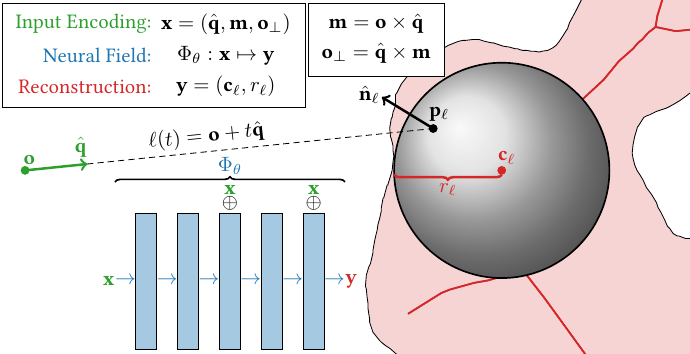}
\caption{A small MARF network \(\Phi_\theta\) illustrated. Given a ray \(\ell\) with origin \(\mathbf o\) and direction \(\hat{\mathbf q}\) it predicts the maximally inscribed medial atom/sphere \((\mathbf c_\ell, r_\ell)\) whose intersection point \(\mathbf p_\ell\) with the ray \(\ell\) corresponds to the intersection between \(\ell\) and the represented object (here shown with its medial axis in red). To uniquely encode rays we normalize \(\hat{\mathbf q}\) and trade \(\mathbf o\) for its moment \(\mathbf m\) and perpendicular foot \(\mathbf o_\perp\). The network is a simple MLP with skip connections, here illustrated with 4 hidden layers, where \(\oplus\) denotes vector concatenation and \(\to\) denotes a linear map.}\label{fig:nn-architecture}
}
\end{figure}

We propose \emph{Medial Atom Ray Fields} (MARFs), visualized in Fig.~\ref{fig:nn-architecture}, which map oriented rays to a set of spherical intersection candidates called \emph{medial atoms}, that are maximally inscribed in the represented shape pinned tangential to the ray-surface intersection point. From a MARF prediction, a simple line-sphere intersection test between the ray and the \(n\) predicted atoms is all one needs to jointly determine \emph{where} and \emph{whether} the ray hits. This medial representation also allows computing the surface normal without analytical network differentiation, which essentially means that we get it for free. This in turn enables computing the surface curvature, a second derivative quantity, despite the second derivative of our piecewise linear network being zero.

We identify two key challenges that hinder the usefulness of ray fields, which we address with the medial shape representation.

The first challenge is that ray fields are not by construction multi-view consistent like their 3D Cartesian counterparts. This is because the four Degrees of Freedom (DoF) of the input rays may cause a predicted 3D point to change appearance across views. Prior works focus on learning a latent manifold of sound ray fields. Our proposed MARF instead phrase the output domain in Cartesian space, which is stable w.r.t. change in incident viewing direction. We further enforce multi-view consistency during training with a novel multi-view loss.

The second challenge is for ray fields to represent discontinuities like sharp edges and overlapping geometry common to depth maps. Neural fields being Lipschitz continuous in their inputs {[}\protect\hyperlink{ref-bartlettSpectrallynormalizedMarginBounds2017}{1},\protect\hyperlink{ref-rahamanSpectralBiasNeural2019}{2}{]} produce interpolation artifacts across such jumps. Prior works either sidestep the issue by relaxing the problem or use a filtering scheme to discard outliers. Our medial representation allows us to regularize multiple predictions to behave well. We alleviate discontinuities by making each candidate specialize on different shape ``limbs'' while adhering to the medial constraints. This is achieved through a principled network initialization scheme and through regularization. Also, by labeling each ray hit by the candidate which produced it, a part segmentation emerges unsupervised.

The learned medial representation is of significant interest in 3D shape analysis, being applicable to classification, semantic manipulation, and segmentation. The represented medial axis, also known as the topological skeleton, produces smooth interpolations in the learned latent space of shapes. The medial radius, also known as the local feature size, is useful in shape analysis and visualization.

In summary, we make the following contributions:

\begin{itemize}
\tightlist
\item
  We propose learning MARFs, which map oriented rays to a set of medial atom intersection candidates that both classify rays as hit/miss and deliver the point of intersection as well as introduce a geometric 3D inductive bias, which oriented ray fields thus far have lacked.
\item
  We demonstrate how MARFs allow computing the analytical surface curvature, despite the network being piecewise linear, using the medial atom normals.
\item
  We introduce a multi-view consistency loss to constrain ray fields to generalize better from sparse set of training views.
\item
  We show that MARFs can learn a space of shape priors.
\item
  We show that MARFs discover a part segmentation unsupervised.
\end{itemize}

\hypertarget{scope.}{%
\paragraph*{Scope.}\label{scope.}}
\addcontentsline{toc}{paragraph}{Scope.}

We target object-centric surface rendering. While one can compose multiple MARFs into a scene, we consider this outside the scope of this paper. We explore a global shape representation, where a single network represents the shapes without spatial partitioning.

\hfill\break
\noindent In Section~\ref{sec:background} we outline prior work and establish key preliminaries, in Section~\ref{sec:method} we discuss our method, in Section~\ref{sec:experiments} we evaluate our method, and in Section~\ref{sec:conclusion} we conclude our work and discuss future directions.

\hypertarget{sec:background}{%
\section{Background and Related Work}\label{sec:background}}

In this section we discuss related works while covering preliminaries about neural (ray) fields and the medial axis.

\hypertarget{representing-3d-geometry-and-scenes-with-neural-fields.}{%
\paragraph*{Representing 3D Geometry and Scenes with Neural Fields.}\label{representing-3d-geometry-and-scenes-with-neural-fields.}}
\addcontentsline{toc}{paragraph}{Representing 3D Geometry and Scenes with Neural Fields.}

Neural fields emerged in 2019 as a compact, continuous and flexible way to represent signals parametrized with spatial and temporal coordinates. The seminal papers {[}\protect\hyperlink{ref-meschederOccupancyNetworksLearning2019}{3}--\protect\hyperlink{ref-parkDeepSDFLearningContinuous2019}{5}{]} use them to represent a set of closed 3D shapes \(\{\mathcal O_i\subset \mathbb R^3\}_{i=1}^n\) of arbitrary topology, either by learning their binary occupancy field or Signed Distance Field (SDF). The SDF \(d_{\partial\mathcal O_i}^\pm:\mathbb R ^3\to\mathbb R\) in particular represents \(\mathcal O_i\) by mapping 3D coordinates to the distance of the nearest surface boundary \(\partial\mathcal O_i\), where interior distances are negative and exterior ones are positive:

\begin{equation}\protect\hypertarget{eq:sdf-def}{}{ d_{\partial\mathcal O_i}^\pm(\mathbf x) =
  \min_{\mathbf x' \in \partial\mathcal O_i} \left\|\mathbf x-\mathbf x'\right\|
  \cdot \begin{cases}
  1  & \text{if }\mathbf x\notin\mathcal O_i \\
  -1 & \text{if }\mathbf x\in\mathcal O_i \\
\end{cases}}\label{eq:sdf-def}\end{equation}

Neural fields may in fact learn any field \(f:\mathcal X\to\mathcal Y\) that maps input coordinates \(\mathbf x\in\mathcal X\) to signal values \(\mathbf y\in\mathcal Y\), given enough \((\mathbf x, \mathbf y)\in\mathcal X\smash\times\mathcal Y\) supervision examples. Examples include: Unsigned distance fields {[}\protect\hyperlink{ref-chibaneNeuralUnsignedDistance2020}{6},\protect\hyperlink{ref-venkateshDeepImplicitSurface2021}{7}{]} which can represent non-watertight surfaces. Winding number fields {[}\protect\hyperlink{ref-chiGarmentNetsCategorylevelPose2021}{8}{]} which can represent self-intersecting geometry. Closest surface point fields {[}\protect\hyperlink{ref-venkateshDeepImplicitSurface2021}{7}{]} which map to the nearest point on the surface (\(\mathbb R^3\to\partial\mathcal O\)). Deep medial fields {[}\protect\hyperlink{ref-rebainDeepMedialFields2021}{9}{]} which represent the local feature size of the nearest surface. Category-level shape descriptor fields, used in robotic manipulation {[}\protect\hyperlink{ref-simeonovNeuralDescriptorFields2021}{10}{]}.

Neural fields prove effective at learning complex mappings one would consider ill-posed to optimize {[}\protect\hyperlink{ref-frankleLotteryTicketHypothesis2019}{11},\protect\hyperlink{ref-xieNeuralFieldsVisual2021}{12}{]}. Atzmon et al. {[}\protect\hyperlink{ref-atzmonSALSignAgnostic2020}{13}{]} show this by learning the SDF without inside/outside supervision. This ability is thanks in part to over-parameterization, and in part to being analytically differentiable w.r.t. input coordinates {[}\protect\hyperlink{ref-baydinAutomaticDifferentiationMachine2018}{14}{]}. Using double back-propagation one may fit neural fields to satisfy Partial Differential Equations (PDEs), or supervise the field gradient with sensor data, enabling data-driven discovery of PDEs {[}\protect\hyperlink{ref-raissiPhysicsinformedNeuralNetworks2019}{15}--\protect\hyperlink{ref-lindellAutoIntAutomaticIntegration2021}{18}{]}. For SDFs such a PDE is the Eikonal equation, which constrains the field gradient \(\nabla_{\mathbf x}d_{\partial\mathcal O}^\pm(\mathbf x)\) to be of unit length {[}\protect\hyperlink{ref-groppImplicitGeometricRegularization2020}{16}{]}. Its direction equals the normal vector near the surface boundary. Network differentiation is also useful during inference, as one may compute differential geometric quantities such as surface normals and curvature {[}\protect\hyperlink{ref-yangGeometryProcessingNeural2021}{19},\protect\hyperlink{ref-novelloExploringDifferentialGeometry2022}{20}{]}.

Neural fields excel in high-dimensional problems, since their size grows with target signal complexity instead of resolution. Mildenhall et al. {[}\protect\hyperlink{ref-mildenhallNeRFRepresentingScenes2020a}{21}{]} show this with NeRFs: a mapping with 5 Degrees of Freedom (DoF). NeRFs represent both the density and anisotropic (view-dependent) radiance field of 3D scenes. By ray marching these fields they achieve realistic novel-view synthesis from registered 2D image supervision. NeRF advancements and applications are plentiful {[}\protect\hyperlink{ref-xieNeuralFieldsVisual2021}{12},\protect\hyperlink{ref-tewariStateArtNeural2020}{22},\protect\hyperlink{ref-tewariAdvancesNeuralRendering2021}{23}{]}, including adaptation to low-light High Dynamic Range scenarios {[}\protect\hyperlink{ref-mildenhallNeRFDarkHigh2022}{24},\protect\hyperlink{ref-mildenhallLocalLightField2019}{25}{]}, modeling complex materials {[}\protect\hyperlink{ref-baatzNeRFTexNeuralReflectance2021}{26}{]} and registration {[}\protect\hyperlink{ref-goliNerf2nerfPairwiseRegistration2022}{27}{]}. Guo et al. {[}\protect\hyperlink{ref-guoObjectCentricNeuralScene2020}{28}{]} extend NeRF to consider the incident light direction, making for a 7 DoF mapping. Video NeRFs {[}\protect\hyperlink{ref-parkHyperNeRFHigherDimensionalRepresentation2021}{29}--\protect\hyperlink{ref-tschernezkiNeuralDiffSegmenting3D2021}{32}{]} go further by adding a temporal dimension, an axis along which both density and radiance may undergo extensive changes, showcasing impressive results.

For further details on neural fields, we encourage the reader to view the excellent overview of Xie et al. {[}\protect\hyperlink{ref-xieNeuralFieldsVisual2021}{12}{]}.

\hypertarget{speeding-up-rendering.}{%
\paragraph*{Speeding up Rendering.}\label{speeding-up-rendering.}}
\addcontentsline{toc}{paragraph}{Speeding up Rendering.}

A \emph{forward map} relates (e.g.~3D volume) neural fields to domains where sensor data is available (e.g.~2D maps). In \emph{volume rendering} the forward map may integrate the color contribution along rays cast through a 3D field {[}\protect\hyperlink{ref-maxOpticalModelsDirect1995}{33}{]}, while in \emph{surface rendering} it may locate the first ray-surface intersection or nearest distance {[}\protect\hyperlink{ref-hartSphereTracingGeometric1996}{34},\protect\hyperlink{ref-knodtNeuralRayTracingLearning2021}{35}{]}. One such forward map is ray-marching, which samples points equidistantly along the ray to numerically approximate the ray integral. This requires in the order of tens or hundreds of field evaluations, proving expensive with neural fields.

Several works address this problem which we divide into three categories:

The first category is making field evaluations cheaper. \emph{Local} methods achieve this by subdividing the field into simple patches or chunks represented by smaller separate networks {[}\protect\hyperlink{ref-chibaneImplicitFunctionsFeature2020}{36}--\protect\hyperlink{ref-rebainDeRFDecomposedRadiance2021}{42}{]}, while \emph{hybrid} methods decode a grid of conditioning vectors with a simple decoder network {[}\protect\hyperlink{ref-lindellBACONBandlimitedCoordinate2022}{43}--\protect\hyperlink{ref-mullerInstantNeuralGraphics2022}{48}{]}. Local and hybrid methods in effect bypass the difficulty of globally learning shapes with a single network. \emph{Tabulation} methods forgo the neural network in favor of discrete data structures and interpolation {[}\protect\hyperlink{ref-yuPlenoxelsRadianceFields2021}{49}--\protect\hyperlink{ref-karnewarReLUFieldsLittle2022}{51}{]}. Works in this category often achieve higher fidelity but are often unable to learn global shape priors. Some global methods \emph{bake} their fields offline before rendering, where one essentially extracts a tabulation {[}\protect\hyperlink{ref-hedmanBakingNeuralRadiance2021}{52},\protect\hyperlink{ref-reiserMERFMemoryEfficientRadiance2023}{53}{]}.

The second category seeks to algorithmically reduce the number of field evaluations, avoiding sampling empty or obscured regions. Surface rendering methods often opt to learn a distance field {[}\protect\hyperlink{ref-parkDeepSDFLearningContinuous2019}{5},\protect\hyperlink{ref-rebainDeepMedialFields2021}{9},\protect\hyperlink{ref-yarivVolumeRenderingNeural2021}{54},\protect\hyperlink{ref-oechsleUNISURFUnifyingNeural2021}{55}{]} such as the SDF which permits sphere-tracing {[}\protect\hyperlink{ref-hartSphereTracingGeometric1996}{34}{]} (visualized in Fig.~\ref{fig:sphere-tracing}). Volume rendering methods may perform a coarse pre-evaluation to produce an index {[}\protect\hyperlink{ref-reiserMERFMemoryEfficientRadiance2023}{53},\protect\hyperlink{ref-liRTNeRFRealTimeOnDevice2022}{56},\protect\hyperlink{ref-linEfficientNeuralRadiance2022}{57}{]}, or construct a Monte-Carlo estimate of the ray integral {[}\protect\hyperlink{ref-morozovDifferentiableRenderingReparameterized2023}{58}{]}.

The third category is our focus: directly predicting the ray integral, discussed in the next subsection.

\hypertarget{neural-ray-fields.}{%
\paragraph*{Neural Ray Fields.}\label{neural-ray-fields.}}
\addcontentsline{toc}{paragraph}{Neural Ray Fields.}

To represent a ray field one must parametrize the domain of rays. Rays can be represented in a plethora of ways, the simplest being a tuple of two 3D points through which the ray passes. Front-facing neural light fields {[}\protect\hyperlink{ref-fengSIGNETEfficientNeural2021}{59},\protect\hyperlink{ref-attalLearningNeuralLight2022}{60}{]} prove with such a representation able to map rays to observed colors in highly realistic scenes. They consider rays cast between points on the near and far plane, which can not represent \(360^\circ\) ray fields. The challenge is how to uniquely encode rays without symmetries.

A 3D line \(\ell(t) = \mathbf o + t\hat{\mathbf q}\) parametrized by some origin \(\mathbf o\) and direction \(\hat{\mathbf q}\), has 4 DoF: compared to the 6 DoFs of 3D rigid bodies, lines lose two being invariant to translations along the line direction and rotations about the line axis. Rays technically gain a DoF over lines, featuring a starting point, but both we and the prior works discussed below discard this DoF and consider rays and lines equivalently. This effectively places the observer infinitely far away.

It is impossible to represent the space of rays/lines in a 4D vector space that is uniform and without singular directions, discontinuities or special cases {[}\protect\hyperlink{ref-rentelnManifoldsTensorsForms2013}{61}--\protect\hyperlink{ref-sitzmann2021lfns}{63}{]}. Naively using the uniform 6D vector \((\mathbf o, \hat{\mathbf q})\) however produces a highly symmetric space that hinders learning.

Lindell et al. {[}\protect\hyperlink{ref-lindellAutoIntAutomaticIntegration2021}{18}{]} learn segments of the volume rendering equation integral {[}\protect\hyperlink{ref-maxOpticalModelsDirect1995}{33}{]} by splitting the ray into \(n\) sections. They sample \(k\) points along each section and feed them along with ray direction \(\hat{\mathbf q}\) into their network. While this ray parametrization enables the use of positional encoding {[}\protect\hyperlink{ref-mildenhallNeRFRepresentingScenes2020a}{21},\protect\hyperlink{ref-tancikFourierFeaturesLet2020a}{64}{]}, it is sensitive to the ray sampling positions and to the number of segments chosen. Mukund et al. {[}\protect\hyperlink{ref-mukundAttentionAllThat2023}{65}{]} also sample \(k\) points along the ray, but associate each point with colors from multiple source views, then interpolate between them with a transformer model.

Neff et al. {[}\protect\hyperlink{ref-neffDONeRFRealTimeRendering2021}{66}{]} accelerate rendering NeRFs by training an accompanying \emph{oracle} network which predicts, given a ray, the salient segments along that ray to be further ray-marched. Yenamandra et al. {[}\protect\hyperlink{ref-yenamandraFIReFastInverse2022}{67}{]} accelerate sphere-tracing neural SDFs by training an accompanying network to predict, given a ray, some initial starting point. They both represent \(\ell\) as the 6D vector \((\mathbf o, \hat{\mathbf q})\), and remove two DoF by normalizing \(\hat{\mathbf q}\) to be of unit size, and limit \(\mathbf o\) to lie on the sphere (with fixed radius \(r\)) circumscribed around the reconstruction volume. This in effect restricts the 6D vector to a 4D subspace, or manifold, embedded in 6D: \((r^{-1}\mathbf o, \hat{\mathbf q})\in S^2\smash\times S^2\subset\mathbb R^6\) where \(S^2\) is the unit 2-sphere. This representation has a finite reconstruction volume determined by \(r\).

Sitzmann et al. {[}\protect\hyperlink{ref-sitzmann2021lfns}{63}{]} forgo the Cartesian radiance field and learn \(360^\circ\) neural light fields (LFN) directly. They represent rays using 6D Plücker coordinates {[}\protect\hyperlink{ref-jiaPluckerCoordinatesLines2020}{68}{]}. Normalized Plücker coordinates encode the ray \(\ell\) as \((\hat{\mathbf q}, \mathbf m)\), where \(\mathbf m=\mathbf o\times \hat{\mathbf q}\) is the \emph{moment} vector of the ray origin \(\mathbf o\) about the coordinate system origin. Plücker coordinates are thus restricted to \(S^2\smash\times T^2_{\mathbf o}\subset\mathbb R^6\), where \(T_{\mathbf d}^n=\{\mathbf x\in\mathbb R^{n+1}:\mathbf x\cdot \mathbf d=0\}\) is the tangent space orthogonal to \(\mathbf d\ne\mathbf 0\), containing the coordinate system origin.

Feng et al. {[}\protect\hyperlink{ref-fengPRIFPrimaryRaybased2022}{69}{]} target with PRIF surface rendering instead of light fields. They trade the moment \(\mathbf m\) for the more geometrically grounded perpendicular foot \(\mathbf o_\perp=\hat{\mathbf q}\times\mathbf m\), i.e.~the orthogonal projection of the coordinate origin onto the ray \(\ell\). \((\hat{\mathbf q}, \mathbf o_\perp)\in S^2\smash\times T^2_{\mathbf m}\). We visualize both \(\mathbf m\) and \(\mathbf o_\perp\) in Fig.~\ref{fig:ray-canonicalization}.

Both {[}\protect\hyperlink{ref-yenamandraFIReFastInverse2022}{67}{]} and {[}\protect\hyperlink{ref-fengPRIFPrimaryRaybased2022}{69}{]} compute the ray-surface intersection point by predicting the (signed) displacement along the ray from their normalized ray origin. This does not represent \emph{whether} the ray intersects or not, which both works solve with a separate network output classifying hit/miss rays. While it does allow representing non-watertight geometry, it is an independent quantity that is not geometrically grounded, which generalizes poorly to unseen views.

Ray fields struggle to represent surface discontinuities, common near boundaries and overlapping geometry, due to neural networks being Lipschitz continuous on their inputs. The works of {[}\protect\hyperlink{ref-neffDONeRFRealTimeRendering2021}{66},\protect\hyperlink{ref-yenamandraFIReFastInverse2022}{67},\protect\hyperlink{ref-fengPRIFPrimaryRaybased2022}{69}{]} all produce interpolation artifacts near surface discontinuities, and is addressed in two ways: {[}\protect\hyperlink{ref-neffDONeRFRealTimeRendering2021}{66},\protect\hyperlink{ref-yenamandraFIReFastInverse2022}{67}{]} reduce the impact of discontinuities by relaxing the task to aid sampling a Cartesian network, and {[}\protect\hyperlink{ref-fengPRIFPrimaryRaybased2022}{69}{]} opt to filter outlier predictions with high gradients. Neff et al. {[}\protect\hyperlink{ref-neffDONeRFRealTimeRendering2021}{66}{]} note how multiple depth predictions do not improve their results.

An open problem for neural ray fields is \emph{multi-view consistency}. Neural fields feature an inductive bias inherited from the structure of the input domain, and Cartesian fields find success thanks to this and to being multi-view consistent by construction. Ray fields do not share these qualities. Their extra DoF may cause a predicted point to change appearance across views. {[}\protect\hyperlink{ref-sitzmann2021lfns}{63}{]} address this with meta-learning {[}\protect\hyperlink{ref-finnModelagnosticMetalearningFast2017}{70}--\protect\hyperlink{ref-sitzmannMetaSDFMetaLearningSigned2020}{72}{]}, learning a latent space of light fields that are multi-view consistent. Sticking the latent manifold, they achieve few-shot single-view reconstruction through latent vector optimization.

Instead, we propose to address multi-view consistency by modeling reconstructions in a dual domain which jointly determines hit/miss classification, where multiple predictions are geometrically grounded (addressing surface discontinuities), and whose quantities are stable w.r.t. changes in incident viewing direction. This domain is the \emph{medial axis}.

\hypertarget{the-medial-axis.}{%
\paragraph*{The Medial Axis.}\label{the-medial-axis.}}
\addcontentsline{toc}{paragraph}{The Medial Axis.}

The Medial Axis Transform \(\operatorname{MAT}(\mathcal O)\) is a complete descriptor of shape \(\mathcal O\subset\mathbb R ^3\). The MAT is a set of 3D points and radii which together form \emph{medial atoms} (spheres) that are \emph{maximally inscribed} in \(\mathcal O\). The MAT is invertible, since reconstructing \(\mathcal O\) from \(\operatorname{MAT}(\mathcal O)\) amounts to taking the union of the medial atoms.

The MAT has various downstream uses, including 3D shape retrieval {[}\protect\hyperlink{ref-kimGraphRepresentationMedial2001}{73},\protect\hyperlink{ref-he3DShapeDescriptor2015}{74}{]}, segmentation {[}\protect\hyperlink{ref-linSEGMAT3DShape2022}{75}{]}, and manipulation {[}\protect\hyperlink{ref-duMedialAxisExtraction2004}{76}{]}. MAT-inspired sphere representations have further applications in constructing simplified static {[}\protect\hyperlink{ref-thierySphereMeshesShapeApproximation2013}{77}{]} and dynamic {[}\protect\hyperlink{ref-thieryAnimatedMeshApproximation2016}{78}{]} shapes, closest point computations {[}\protect\hyperlink{ref-tkachSpheremeshesRealtimeHand2016}{79}{]}, and volumetric physics simulation {[}\protect\hyperlink{ref-anglesVIPERVolumeInvariant2019}{80}{]}.

Classically, Bouix et al. {[}\protect\hyperlink{ref-bouixDivergenceBasedMedialSurfaces2000}{81}{]} compute the MAT from voxel models, Du et al. {[}\protect\hyperlink{ref-duMedialAxisExtraction2004}{76}{]} and Tam et al. {[}\protect\hyperlink{ref-tamShapeSimplificationBased2003}{82}{]} compute the MAT from surface meshes, and Rebain et al. {[}\protect\hyperlink{ref-rebainLSMATLeastSquares2019}{83}{]} iteratively approximate the MAT from oriented point clouds by phrasing the inscription and maximality constraint as optimization energies. The MAT is unstable under noise {[}\protect\hyperlink{ref-rebainLSMATLeastSquares2019}{83},\protect\hyperlink{ref-attaliStabilityComputationMedial2009}{84}{]}, but Tam et al. {[}\protect\hyperlink{ref-tamShapeSimplificationBased2003}{82}{]} show how one may prune medial axis branches while maintaining the salient features of the shape.

In the neural literature, Yang et al. {[}\protect\hyperlink{ref-yangP2MATNETLearningMedial2020}{85}{]} predict a set of medial atoms given a sparse surface point cloud, showing how data-driven approaches fare better on sparse and noisy data thanks to its learned priors. Rebain et al. {[}\protect\hyperlink{ref-rebainDeepMedialFields2021}{9}{]} learn a relaxed MAT as a neural field. They model a \(\mathbb R^3\to\mathbb R\) \emph{medial field} mapping spatial coordinates to the radius of the medial atom tangential the nearest surface.

There are many ways to define and apply the MAT, and we encourage the reader to view excellent overview by Tagliasacchi et al. {[}\protect\hyperlink{ref-tagliasacchi3DSkeletonsStateoftheArt2016}{86}{]}. In short there are four MAT definitions: (1) the set of maximally inscribed balls tangent to the surface, (2) the ridges of the (signed) distance \(d_{\partial\mathcal O}^\pm\) (i.e.~the grassfire transform), (3) the Maxwell set, i.e.~the set of points with more than one nearest surface neighbor, (4) all local axes of reflectional symmetry, i.e.~the set of all bi-tangent spheres. In this work we use the first definition.

\hypertarget{sec:method}{%
\section{Method}\label{sec:method}}

We start by establishing a mathematical framework for oriented ray intersection fields in Section~\ref{sec:oriented-ray-fields}. We then define our proposed Medial Atom Ray Field (MARF) in Section~\ref{sec:medial-fields}, followed by a discussion on how MARFs address the challenges in learning ray fields. We define our neural architecture in Section~\ref{sec:neural-network}, and outline training data pre-processing, losses and optimization strategy in Section~\ref{sec:training}.

\hypertarget{sec:oriented-ray-fields}{%
\subsection{Oriented Ray Intersection Fields}\label{sec:oriented-ray-fields}}

Consider a closed 3D shape \(\mathcal O\subset \mathbb R^3\) with regular surface boundary \(\partial\mathcal O\), and the oriented ray as the line \(\ell\in\mathcal R\) with origin \(\mathbf o\) and unit direction \(\hat{\mathbf q}\):

\begin{equation}\protect\hypertarget{eq:ell}{}{ \ell(t) = \mathbf o + t\hat{\mathbf q}
}\label{eq:ell}\end{equation}

\noindent The \emph{oriented ray intersection field} \(f_{\mathcal O}:\mathcal R\to\mathbb R^3\) maps 3D oriented rays to their nearest intersection points on the surface \(\partial\mathcal O\). \(f_{\mathcal O}\) is in essence a single-ray ray caster. Formally \(f_{\mathcal O}\) maps \(\ell\) to the point \(\mathbf p_\ell\in\partial\mathcal O\) along \(\ell(t)\) minimizing \(t\):

\begin{equation}\protect\hypertarget{eq:oriented-ray-field}{}{
  f_{\mathcal O}(\ell) = \mathbf p_\ell = \ell\left(
    \vcenter{\hbox{\math
      \underset{t \ :\ \ell(t) \in \partial\mathcal O}{\arg\min} t
    \endmath}}
  \right)
}\label{eq:oriented-ray-field}\end{equation}

\(f_{\mathcal O}\) is a partial map, since not all rays intersect with the shape. In such cases we may still observe by how much a ray misses, dubbed the \emph{silhouette distance} \(s_\ell\), which exhibits the property \(\nexists\mathbf p_\ell \Leftrightarrow s_\ell > 0\):

\begin{equation}\protect\hypertarget{eq:silhouette-distance}{}{
  s_\ell = \min_{t\in\mathbb R,\ \mathbf x\in \partial\mathcal O} \|\ell(t) - \mathbf x\|
}\label{eq:silhouette-distance}\end{equation}

\hypertarget{differential-geometry-in-rays-fields.}{%
\paragraph*{Differential Geometry in Rays Fields.}\label{differential-geometry-in-rays-fields.}}
\addcontentsline{toc}{paragraph}{Differential Geometry in Rays Fields.}

A \emph{surface normal} \(\hat{\mathbf n}_\ell\) is a unit vector (in the 2-sphere \(S^2\)) whose orientation is orthogonal to the plane tangent at point \(\mathbf p_\ell\) on the surface \(\partial\mathcal O\), and whose direction determines the shape exterior (\(\mathbb R^3\smash\setminus\mathcal O\)). Computing normals is not straight forward given the view-dependent ray parametrization \(\ell(t)=\mathbf o+t\hat{\mathbf q}\). We first compute, for each coordinate axis \(\hat{\mathbf e}_i\), a surface tangent vector as the partial derivative \(\mathbf t_i\smash=\partial\mathbf p_\ell/\partial o_i\), visualized in Fig.~\ref{fig:ray-normal}. The normal \(\hat{\mathbf n}_\ell\), orthogonal to the tangent space, is determined by the cross product of two tangents. But our tangents may, depending on view direction, become zero (since \(\hat{\mathbf q}\|\hat{\mathbf e}_i\Rightarrow\mathbf t_i\smash=\mathbf 0\)) or change the cross-product handedness determining the exterior. We thus compute the cross product of all three tangent pairs, then modulate their sign and contribution using the viewing direction before summation and normalization:

\begin{equation}\protect\hypertarget{eq:oriented-ray-field-normal}{}{ \begin{aligned}
  \hat{\mathbf n}_\ell = \tfrac{\mathbf n_\ell'}{\|\mathbf n_\ell'\|},
  \ \mathbf n_\ell' =\
    & -\hat{q}_1 \left( \mathbf t_2 \times \mathbf t_3 \right)\ =& -\hat{q}_1 \left( \tfrac{\partial \mathbf p_\ell}{\partial o_2} \times \tfrac{\partial \mathbf p_\ell}{\partial o_3} \right) \\
    & -\hat{q}_2 \left( \mathbf t_3 \times \mathbf t_1 \right)   & -\hat{q}_2 \left( \tfrac{\partial \mathbf p_\ell}{\partial o_3} \times \tfrac{\partial \mathbf p_\ell}{\partial o_1} \right) \\
    & -\hat{q}_3 \left( \mathbf t_1 \times \mathbf t_2 \right)   & -\hat{q}_3 \left( \tfrac{\partial \mathbf p_\ell}{\partial o_1} \times \tfrac{\partial \mathbf p_\ell}{\partial o_2} \right) \\
\end{aligned} }\label{eq:oriented-ray-field-normal}\end{equation}

\noindent where \(o_i\) and \(\hat{q}_i\) are the \(i^\text{th}\) scalar components of \(\mathbf o\) and \(\hat{\mathbf q}\) from Eq.~\ref{eq:ell}.

\begin{pandoccrossrefsubfigures}

\subfloat[Moment \(\mathbf m\) and foot \(\mathbf o_\perp\)]{\includegraphics{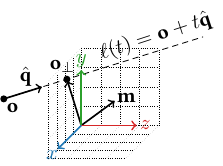}\label{fig:ray-canonicalization}} \subfloat[Normal derived from surface tangents.]{\includegraphics{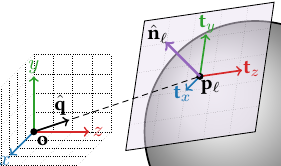}\label{fig:ray-normal}}

\caption[{\textsc{(a)} illustrates the relation between ray origin \(\mathbf o\) and direction \(\hat{\mathbf q}\) to the moment \(\mathbf m\) and perpendicular foot \(\mathbf o_\perp\) which are invariant to changes in the length of \(\mathbf q\) and translation of \(\mathbf o\) along \(\mathbf q\). \textsc{(b)} illustrates how Eq.~\ref{eq:oriented-ray-field-normal} determines the normal vector \(\hat{\mathbf n}_\ell\), orthogonal to the tangent space, by projecting the \(x,y,z\) coordinate unit vectors onto the plane tangent at \(\mathbf p_\ell\) where ray \(\ell\) intersects.}]{\textsc{(a)} illustrates the relation between ray origin \(\mathbf o\) and direction \(\hat{\mathbf q}\) to the moment \(\mathbf m\) and perpendicular foot \(\mathbf o_\perp\) which are invariant to changes in the length of \(\mathbf q\) and translation of \(\mathbf o\) along \(\mathbf q\). \textsc{(b)} illustrates how Eq.~\ref{eq:oriented-ray-field-normal} determines the normal vector \(\hat{\mathbf n}_\ell\), orthogonal to the tangent space, by projecting the \(x,y,z\) coordinate unit vectors onto the plane tangent at \(\mathbf p_\ell\) where ray \(\ell\) intersects.}

\label{fig:feet-and-normals}

\end{pandoccrossrefsubfigures}

\emph{Curvature} describes how a surface deviates from the tangent plane and is intrinsic to the shape. The curvature \(\kappa\) along a single direction is the reciprocal of the radius of an osculating circle, where positive curves osculate inside. Curvature on 3D surfaces (i.e.~2-manifolds) may be expressed as two \emph{principal} curvatures \(\kappa_1, \kappa_2\), respectively the maximum and minimum curvatures. The principal directions of curvature are always perpendicular, except at umbilical points and on flat surfaces where \(\kappa_1\smash=\kappa_2\).

Curvature is contained in the \emph{shape operator} \(\mathcal D\hat{\mathbf n}_\ell\), defined as the total derivative of the unit normal \(\hat{\mathbf n}_\ell\) along the tangent space {[}\protect\hyperlink{ref-igehyTracingRayDifferentials1999}{87}{]}. The ray origin \(\mathbf o\), our input, is not restricted to the tangent space. As such we compute \(\mathcal D\hat{\mathbf n}\) by projecting the total derivative onto the tangent plane:

\begin{equation}\protect\hypertarget{eq:shape-operator}{}{ \mathcal D\hat{\mathbf n}_\ell = \left(\mathbf I - \hat{\mathbf n}_\ell\hat{\mathbf n}_\ell^\top\right)\nabla_{\mathbf o}\hat{\mathbf n}_\ell
}\label{eq:shape-operator}\end{equation}

The principal curvatures and directions equal the maximal and minimal eigenvalues of the shape operator \(\mathcal D\hat{\mathbf n}_\ell\) and associated eigenvectors. In our case the total derivative \(\nabla_{\mathbf o}\hat{\mathbf n}_\ell\) is a \(3\smash\times 3\) matrix with three eigenvectors, but the eigenvector with the smallest absolute eigenvalue is associated with the normal and can be discarded {[}\protect\hyperlink{ref-novelloExploringDifferentialGeometry2022}{20}{]}. The mean curvature is half the trace of the shape operator \(\mathcal D\hat{\mathbf n_\ell}\), and the Gaussian curvature is its determinant {[}\protect\hyperlink{ref-yangGeometryProcessingNeural2021}{19}{]}.

We can compute these differentials analytically for continuous neural representations, unlike for meshes which do not admit a continuous normal field in turn requiring an approximation like the discrete shape operator {[}\protect\hyperlink{ref-cohen-steinerRestrictedDelaunayTriangulations2003}{88}{]}. Computing curvature requires a sufficiently smooth activation function {[}\protect\hyperlink{ref-sitzmannImplicitNeuralRepresentations2020}{17},\protect\hyperlink{ref-novelloExploringDifferentialGeometry2022}{20}{]} since piecewise linear activations have no second derivative.

\hypertarget{sec:medial-fields}{%
\subsection{The Medial Atom Ray Field (MARF)}\label{sec:medial-fields}}

We propose learning \emph{Medial Atom Ray Fields} (MARFs), a dual field that also represents the ray intersection field. A MARF \(\mathcal M_{\mathcal O}\) maps an oriented ray \(\ell\in\mathcal R\) to a \emph{medial atom} (sphere) with center \(\mathbf c_\ell\in\mathbb R^3\) and radius \(r_\ell\in\mathbb R^+\), such that the atom:

\begin{itemize}
\tightlist
\item
  \emph{intersect} \(\ell\) at the same point (\(\mathbf p_\ell\)) where \(\ell\) intersects the surface \(\partial\mathcal O\) of \(\mathcal O\),
\item
  is \emph{tangential} to surface \(\partial\mathcal O\) at \(\mathbf p_\ell\) (i.e.~share \(\hat{\mathbf n}_\ell\) from Eq.~\ref{eq:oriented-ray-field-normal}),
\item
  is fully \emph{inscribed} in shape \(\mathcal O\), and
\item
  is \emph{maximal}.
\end{itemize}

\noindent The atoms of MARF are thus members of the Medial Axis Transform (MAT) {[}\protect\hyperlink{ref-tagliasacchi3DSkeletonsStateoftheArt2016}{86}{]} of \(\mathcal O\). The MARF \(\mathcal M_{\mathcal O}\) relates to the ``ray-caster'' \(f_{\mathcal O}\) and its normal \(\hat{\mathbf n}_\ell\) (Eqs.~\ref{eq:oriented-ray-field}, \ref{eq:oriented-ray-field-normal}) as follows:

\begin{equation}\protect\hypertarget{eq:marf}{}{
  \mathcal M_{\mathcal O}(\ell) = (\mathbf c_\ell, r_\ell)
  \ :\
  \|f_{\mathcal O}(\ell)-\mathbf c_\ell\|=r_\ell,
  \hat{\mathbf n}_\ell=\frac{f_{\mathcal O}(\ell)-\mathbf c_\ell}{\|f_{\mathcal O}(\ell)-\mathbf c_\ell\|}
}\label{eq:marf}\end{equation}

To determine the point \(\mathbf p_\ell\) where the ray \(\ell\) intersects a medial atom \((\mathbf c_\ell, r_\ell)\) we solve the system \(\mathbf p_\ell\smash{=}\ell(t),\ \|\ell(t)-\mathbf c_\ell\|\smash{=}r_\ell\):

\begin{equation}\protect\hypertarget{eq:medial-atom-intersection}{}{ \begin{aligned}
  \mathbf p_\ell &= \mathbf o+\hat{\mathbf q}\left(-(\hat{\mathbf q}\cdot (\mathbf o-\mathbf c_\ell))\pm \sqrt{\delta_\ell } \right)
  \\
  \text{where }\ \delta_\ell    &= (\hat{\mathbf q}\cdot (\mathbf o-\mathbf c_\ell))^2-(\|\mathbf o-\mathbf c_\ell\|^2-r_\ell^2)
\end{aligned} }\label{eq:medial-atom-intersection}\end{equation}

\noindent This phrasing of intersection \(\mathbf p_\ell\) yields up to two real solutions (a near and far hit) when \(\ell\) hits (\(\delta_\ell\smash\ge 0\)), and a complex solution when \(\ell\) misses (\(\delta_\ell\smash<0\)). We can as such use \(\delta_\ell\) to determine ray hit/miss classification.

When ray \(\ell\) misses, the real component of \(\mathbf p_\ell\) equals the orthogonal projection of \(\mathbf c_\ell\) onto \(\ell\), yielding the following relation with the silhouette \(s_\ell\) from Eq.~\ref{eq:silhouette-distance}:

\begin{equation}\protect\hypertarget{eq:medial-silhouette-distance}{}{ s_\ell = \|\operatorname{Re}(\mathbf p_\ell) -\mathbf c_\ell\| - r_\ell
}\label{eq:medial-silhouette-distance}\end{equation}

It is cheaper to compute the surface normal \(\hat{\mathbf n}_\ell\) using the medial atom than to compute the differential in Eq.~\ref{eq:oriented-ray-field-normal}. From here on we use the term ``analytical normal'' to tell Eq.~\ref{eq:oriented-ray-field-normal} apart from this ``medial normal'':

\begin{equation}\protect\hypertarget{eq:marf-normal}{}{ \hat{\mathbf n}_\ell=\frac{\mathbf p_\ell-\mathbf c_\ell}{\|\mathbf p_\ell-\mathbf c_\ell\|}
}\label{eq:marf-normal}\end{equation}

By construction the medial normal will naturally ``roll off'' as the ray approaches the edge of the represented shape.

\hypertarget{learning-the-medial-axis.}{%
\paragraph*{Learning the Medial Axis.}\label{learning-the-medial-axis.}}
\addcontentsline{toc}{paragraph}{Learning the Medial Axis.}

We do not assume the medial axis is available for supervision, meaning the network must discover it during training. Inspired by Rebain et al. {[}\protect\hyperlink{ref-rebainLSMATLeastSquares2019}{83}{]} we phrase the medial axis conditions --- maximality and inscription --- as optimization energies. The \emph{maximality} energy induces a positive pressure on radius \(r_\ell\), increasing the atom size, whereas the \emph{inscription} energy penalizes any medial atom candidate visible from the outside, violating the inscription constraint. We define these losses in Section~\ref{sec:training}. We omit their \emph{pinning} energy, since we pin atoms tangential to the surface hit point \(\mathbf p_\ell\).

\hypertarget{representing-surface-discontinuities.}{%
\paragraph*{Representing Surface Discontinuities.}\label{representing-surface-discontinuities.}}
\addcontentsline{toc}{paragraph}{Representing Surface Discontinuities.}

Neural fields produce interpolation artifacts near sharp edges and discontinuities due to being Lipschitz continuous on their inputs. We address this by predicting multiple medial atom \emph{candidates} (in this work we predict 16), the winner of which we chose using the following metric:

\begin{equation}\protect\hypertarget{eq:candidate-metric}{}{ m_{\ell,i} = \begin{cases}
  {\hat{\mathbf q}\cdot\left(\mathbf p_{\ell,i}-\mathbf o \right)}
    &\text{if }s_{\ell,i}= 0\\
  {\infty}
    &\text{if }s_{\ell,i}>0\ \wedge\ \exists k(s_{\ell,k} = 0)\\
  {s_{\ell,i}}
    &\text{if }\forall k(s_{\ell,k} > 0)\\
\end{cases} }\label{eq:candidate-metric}\end{equation}

\noindent The first case is when the \(i^\text{th}\) atom candidate intersects ray \(\ell\), computing the signed displacement with regard to the ray origin \(\mathbf o\). The second case is when candidate \(i\) misses \(\ell\) but at least one other do hit. The final case is when \emph{all} candidates miss \(\ell\), in which case the metric falls back on the silhouette distance.

\begin{figure}
\hypertarget{fig:supervision-cases}{%
\centering
\includegraphics{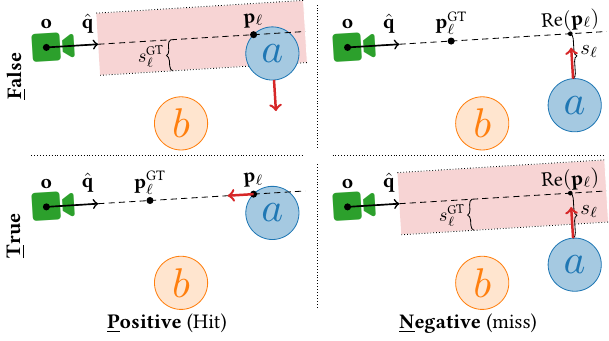}
\caption{Four MARF supervision scenarios. For each ray \(\ell\) with origin \(\mathbf o\) and direction \(\hat{\mathbf q}\) we predict \(n\) atom candidates then pick and supervise the one (\(a\)) that minimizes metric \(m_\ell\) (see Eq.~\ref{eq:candidate-metric}). In short the metric favors the atom closest to the ray, then the atom closest to the camera. The \textbf{TP} case supervises toward a target intersection point \(\mathbf p_\ell^\text{GT}\), applying a pressure along \(\ell\) (visualized as red arrows), while the \textbf{FP}, \textbf{FN} and \textbf{TN} cases all supervise toward a target silhouette distance \(s_\ell^\text{GT}\), shown as a red cylinder lathed about the ray, applying an orthogonal pressure to the atom. Of interest is how atom \(b\) might be a better supervision candidate than \(a\) in \textbf{TP}, \textbf{TN} and \textbf{FN}, demonstrating a shortcoming of metric \(m_\ell\).}\label{fig:supervision-cases}
}
\end{figure}

Under this metric we may supervise \(\mathbf p_{\ell,i}, \mathbf n_{\ell,i}\) and \(s_{\ell,i}\) for candidate \(\arg\min_i m_{\ell,i}\). It is not perfect, as discussed in Fig.~\ref{fig:supervision-cases}, but it ensures the validity of the aforementioned quantities. From here on if we omit the \(i\) subscript, then only the winning atom is concerned.

This metric alone is not enough. Neff et al. {[}\protect\hyperlink{ref-neffDONeRFRealTimeRendering2021}{66}{]} discuss how multiple network outputs do not alone improve their results. This is likely because only the winning output receives supervision while the others drift. We observed atoms either going unused, or ``fighting'' to represent the same geometry. As such we add a ``specialization'' regularization which incentivizes each candidate to target separate regions. We do so by enforcing a spherical prior distribution, centered in the per-candidate centroid, further detailed in Section~\ref{sec:training}. All atom candidates are subject to the medial axis inscription constraint.

\hypertarget{enforcing-multi-view-consistency.}{%
\paragraph*{Enforcing Multi-View Consistency.}\label{enforcing-multi-view-consistency.}}
\addcontentsline{toc}{paragraph}{Enforcing Multi-View Consistency.}

Ray fields are not multi-view consistent by construction. We observe the following trait of multi-view consistency: surface hit points do not move when the viewing angle changes. For any oriented ray intersection field \(f_{\mathcal O}\) (Eq.~\ref{eq:oriented-ray-field}) this means that if we shift the ray origin \(\mathbf o\), i.e.~its pivot point, to the hit \(\mathbf p_\ell\), then its derivative w.r.t. view direction must be zero:

\begin{equation}\protect\hypertarget{eq:mv-property}{}{
  \exists \mathbf p_\ell
    \Rightarrow \left\|\nabla_{\hat{\mathbf q}} f_{\mathcal O}(\mathbf p_\ell, \hat{\mathbf q})\right\|
    = \left\|\nabla_{\hat{\mathbf q}} \mathbf p_\ell\right\|
    = 0
}\label{eq:mv-property}\end{equation}

\noindent This is not a trivial property for signed displacement methods like {[}\protect\hyperlink{ref-yenamandraFIReFastInverse2022}{67},\protect\hyperlink{ref-fengPRIFPrimaryRaybased2022}{69}{]} to learn, as they must learn the inverse of the change in displacement origin. But it extends cleanly to fixing the medial atom in place:

\begin{equation}\protect\hypertarget{eq:mat-mv-property}{}{
  \exists \mathbf p_\ell
    \Rightarrow \left\|\nabla_{\hat{\mathbf q}} \mathcal M_{\mathcal O}(\mathbf p_\ell, \hat{\mathbf q})\right\|
    \le \left\|\nabla_{\hat{\mathbf q}} \mathbf c_\ell\right\| + \left\|\nabla_{\hat{\mathbf q}} r_\ell\right\|
    = 0
}\label{eq:mat-mv-property}\end{equation}

\noindent In Section~\ref{sec:training} we express Eqs.~\ref{eq:mv-property}, \ref{eq:mat-mv-property} as loss functions.

\hypertarget{sec:neural-network}{%
\subsection{Network Architecture}\label{sec:neural-network}}

We model a neural network \(\Phi_\theta\) with learned parameters \(\theta\), optionally conditioned on latent codes, to fit the medial atom ray field \(\mathcal M_{\mathcal O_i}\) from Eq.~\ref{eq:marf} for shapes \(\{\mathcal O_i\smash\subset\mathbb R^3\}_{i=1}^n\). Shown in Fig.~\ref{fig:nn-architecture} we model the network as a Multi-Layer Perceptron (MLP) with skip connections (inspired by {[}\protect\hyperlink{ref-parkDeepSDFLearningContinuous2019}{5},\protect\hyperlink{ref-sitzmann2021lfns}{63}{]}) to the middle and final hidden layer. Formally:

\begin{equation}{ \begin{aligned}
  \Phi_\theta(\mathbf x) &= \mathbf W_k\left(\phi_{k-1}\circ\phi_{k-2}\circ\cdots\circ\phi_0\right)+\mathbf b_k
  \\
  \phi_i(\mathbf x_i) &= \sigma_i\left(\mathbf W_i\mathbf x_i+\mathbf b_i\right)
  \\
  \theta &= \{(\mathbf W_i,\mathbf b_i)\}_{i=0}^k
\end{aligned} }\end{equation}

\noindent where \(\Phi_\theta\) is the composition of \(k\) layers where \(\phi_i:\mathbb R^{m_i}\to\mathbb R^{n_i}\) is the \(i^\text{th}\) network layer, each applying some affine transformation/linear map on intermediate activation \(\mathbf x_i\) followed by an element-wise application of an activation \(\sigma_i\). For all \(\sigma_i\) we use Leaky ReLU and layer normalization {[}\protect\hyperlink{ref-baLayerNormalization2016}{89}{]}, but the middle and final \(\sigma_i\) also concatenates the original input \(\mathbf x_0\), forming skip connections {[}\protect\hyperlink{ref-parkDeepSDFLearningContinuous2019}{5}{]}. During training, \(\sigma_i\) also applies dropout.

We encode the ray \(\ell\) as the following 9D input vector with 4 DoF:

\begin{equation}\protect\hypertarget{eq:ray-input-embedding}{}{ \mathbf x
  = (
      \hat{\mathbf q},
      \mathbf m,
      \mathbf o_\perp
  ),
  \ \text{where }\ \ \ \begin{aligned}
    \hat{\mathbf q} &= \mathbf q / \|\mathbf q\| \\
    \mathbf m       &= \mathbf o\times\hat{\mathbf q}  \\
    \mathbf o_\perp &= \hat{\mathbf q}\times\mathbf m  \\
  \end{aligned}
}\label{eq:ray-input-embedding}\end{equation}

\noindent where \(\mathbf m\) is the moment proposed by {[}\protect\hyperlink{ref-sitzmann2021lfns}{63}{]} and \(\mathbf o_\perp\) is the perpendicular foot used in PRIF {[}\protect\hyperlink{ref-fengPRIFPrimaryRaybased2022}{69}{]}. Either \(\mathbf o_\perp\) or \(\mathbf m\) would have sufficed: they are of equal length and separated by a \(90^\circ\) rotation, virtually equivalent to the linear maps that neural networks learn. Redundant information however improves learning {[}\protect\hyperlink{ref-rahamanSpectralBiasNeural2019}{2}{]}, so we combine them in turn forming an orthogonal basis when \(\ell\) does not pass though the origin. With this in mind we add an extra skip connection to the final layer, since our network predicts points in \(\mathbb R ^3\), improving performance.

The final output is \(4\smash\times n\) features wide, split into \(n\) medial atom centers and radii \(\{(\mathbf c_{\ell, i}, r_{\ell, i})\}_{i=1}^n\), ensuring \(r_\ell\in\mathbb R^+\) by using the absolute predicted value. Inspired by {[}\protect\hyperlink{ref-atzmonSALSignAgnostic2020}{13},\protect\hyperlink{ref-ben-shabatDiGSDivergenceGuided2021}{90}{]} we propose a principled initialization strategy for MARFs. As customary we sample the network parameters \(\theta\) from a uniform distribution according to {[}\protect\hyperlink{ref-heDelvingDeepRectifiers2015}{91}{]}, but we then scale the final layer weights \(\mathbf W_k\) by \(0.05\), in effect reducing the variance of the final predictions. We initialize the final bias \(\mathbf b_k\) to \(n\) random atoms \(0.6\) units away from the origin and with \(0.1\) radius. This initialization is multi-view consistent, with each atom candidate starting in a different region as opposed to them all clustering near the origin.

To condition the network on multiple shapes we use the auto-decoder framework by Park et al. {[}\protect\hyperlink{ref-parkDeepSDFLearningContinuous2019}{5}{]}, where the latent vector \(\mathbf z_i\in\mathbb R^k\) which represents shape \(\mathcal O_i\) is concatenated with the input coordinates before being fed into the network and are optimized alongside the network weights. We concatenate \(\mathbf z_i\) at two sites: the initial input and at the middle skip connection. We do not condition the final skip connection.

\hypertarget{sec:training}{%
\subsection{Training}\label{sec:training}}

We train unconditioned (i.e.~single-object) MARFs to represent the Stanford \emph{Armadillo, Buddha, Bunny, Dragon,} and \emph{Lucy} {[}\protect\hyperlink{ref-stanford3DScanning}{92}{]}, and conditioned (i.e.~multi-object) MARFs to represent the \emph{four-legged} object class in COSEG {[}\protect\hyperlink{ref-wangActiveCoanalysisSet2012}{93}{]}.

\hypertarget{data-pre-processing.}{%
\paragraph*{Data Pre-Processing.}\label{data-pre-processing.}}
\addcontentsline{toc}{paragraph}{Data Pre-Processing.}

We sample ground truth data points from 3D triangle meshes for training. We scale and translate the meshes to fit inside the unit sphere, then render \(200\smash\times 200\) depth and normal maps with the rendering pipeline of {[}\protect\hyperlink{ref-kleinebergAdversarialGenerationContinuous2020}{94}{]} from 50 equidistant virtual camera views. We unproject the depth maps to 3D points, then compute silhouette distances. We accelerate this with a ball tree index {[}\protect\hyperlink{ref-pedregosaScikitlearnMachineLearning2011}{95}{]} on the hit-points which we sphere-trace along the miss-rays with a 25\% step-length. The smallest observed distance approximates the silhouette.

On non-watertight meshes we classify back-face depth pixels as neither hits nor misses, to avoid training on missing data visible as the black holes in Fig.~\ref{fig:batchstep-bunny}. We designate these ``missing'' Ground Truth (GT) pixels as non-hits, i.e.~\(\nexists\mathbf p_\ell^\text{GT}\). We still subject these rays to regularization during training -- maintaining a fixed batch size in the process -- but they otherwise provide no direct supervision. This violates the property \(\nexists\mathbf p_\ell \Leftrightarrow s_\ell > 0\); we thus introduce the notational convenience in Eq.~\ref{eq:hit_miss_gt_gates} to ``gate'' losses where ground truth rays \emph{hit} (\(h_\ell^\text{GT}\)), \emph{miss} (\(m_\ell^\text{GT}\)), or are hitting but the intersection data is \emph{missing} (\(\bar h_\ell^\text{GT}\bar m_\ell^\text{GT}\)). Some losses supervise only true hits, denoted \(h_\ell h_\ell^\text{GT}\).

\begin{equation}\protect\hypertarget{eq:hit_miss_gt_gates}{}{
  h_\ell^\text{GT} = \begin{cases}
    {1}&\text{if }\exists \mathbf p_\ell^\text{GT} \\
    {0}&\text{if }\nexists\mathbf p_\ell^\text{GT} \\
  \end{cases}
  ,\ \ \
  m_\ell^\text{GT} = \begin{cases}
    {1}&\text{if }s_\ell^\text{GT}>0 \\
    {0}&\text{otherwise} \\
  \end{cases}
}\label{eq:hit_miss_gt_gates}\end{equation}

\hypertarget{loss.}{%
\paragraph*{Loss.}\label{loss.}}
\addcontentsline{toc}{paragraph}{Loss.}

We use the following losses to train our network:

\begin{itemize}
\item
  \textbf{Intersection loss} \(\mathcal L_{\mathbf p}\) and \textbf{normal loss} \(\mathcal L_{\mathbf n}\): When ray \(\ell\) hits we supervise the Euclidean distance between the hit \(\mathbf p_\ell\) (from Eq.~\ref{eq:medial-atom-intersection}) and ground truth \(\mathbf p_\ell^\text{GT}\). In addition we supervise the cosine similarity between medial normal \(\hat{\mathbf n}_\ell\) (from Eq.~\ref{eq:marf-normal}) and ground truth \(\hat{\mathbf n}_\ell^\text{GT}\):

  \begin{equation}\protect\hypertarget{eq:loss-intersection}{}{ \mathcal L_{\mathbf p} =
    \frac{1}{|B|}\sum_{\ell\in B}
    h_\ell
    h_\ell^\text{GT}
    \left\|
      \mathbf p_\ell
      -
      \mathbf p_\ell^\text{GT}
    \right\|
  }\label{eq:loss-intersection}\end{equation}

  \begin{equation}\protect\hypertarget{eq:loss-normal}{}{ \mathcal L_{\mathbf n} =
    \frac{1}{|B|}\sum_{\ell\in B}
    h_\ell
    h_\ell^\text{GT}
    \frac{
      \hat{\mathbf n}_\ell
      \cdot
      \hat{\mathbf n}_\ell^\text{GT}
    }{
      \|\hat{\mathbf n}_\ell\|
      \|\hat{\mathbf n}_\ell^\text{GT}\|
    }
  }\label{eq:loss-normal}\end{equation}
\item
  \textbf{Silhouette loss} \(\mathcal L_s\) and \(\mathcal L_{h}\): We supervise the silhouette distance \(s_\ell\) from Eq.~\ref{eq:medial-silhouette-distance} with ground truth \(s_\ell^\text{GT}\):

  \begin{equation}\protect\hypertarget{eq:loss-silhouette}{}{ \mathcal L_s =
    \frac{1}{|B|}\sum_{\ell\in B}
    m_\ell^\text{GT}
    \left(s_\ell-s_\ell^\text{GT}\right)^2
  }\label{eq:loss-silhouette}\end{equation}

  \noindent \(\mathcal L_s\) only supervise misses, since we found it alone to be insufficient to ensure rays hit when they should. We introduce this additional loss gated on hits, whose strength we tune with a separate hyperparameter:

  \begin{equation}\protect\hypertarget{eq:loss-should-hit}{}{ \mathcal L_h =
    \frac{1}{|B|}\sum_{\ell\in B}
    h_\ell^\text{GT}
    s_\ell^2
  }\label{eq:loss-should-hit}\end{equation}
\item
  \textbf{Maximality regularization} \(\mathcal L_r\): To ensure the maximality property of medial atoms we apply a constant positive pressure to the radius of all predicted atom candidates, inspired by Rebain et al. {[}\protect\hyperlink{ref-rebainLSMATLeastSquares2019}{83}{]}:

  \begin{equation}\protect\hypertarget{eq:loss-maximality}{}{ \mathcal L_r =
    \frac{1}{|B|n}
    \sum_{\ell\in B}
    \sum_{i=1}^n
    \left| \left(\operatorname{sg}(r_{\ell,i})+1\right) - r_{\ell,i} \right|
  }\label{eq:loss-maximality}\end{equation}

  \noindent where \(\operatorname{sg}(\cdot)\) returns its input detached from the auto-differentiation graph such that it is considered a constant during back-propagation.
\item
  \textbf{Inscription loss} \(\mathcal L_{ih}\) and \(\mathcal L_{im}\): To enforce the inscription requirement of medial atoms we supervise all predicted atom candidates w.r.t. a second ray. We randomly permute the order of the training batch of rays \(B\) into \(K\), and use \(K\) to compute intersections and silhouettes against all \(n\) atom candidates of \(B\). \(\mathcal L_{ih}\) penalizes atoms that obscure the target intersection of the second ray, while \(\mathcal L_{im}\) penalizes atoms closer than the second ray silhouette permits.

  \begin{equation}\protect\hypertarget{eq:loss-inscription-hit}{}{ \mathcal L_{ih} =
    \sum_{\substack{\ell_a\in B \\ \ell_b = \rho(\ell_a)}}
    \sum_{i=1}^n
    h_{\ell_b}^\text{GT}
    h_{\ell_{b|a},i}
    \frac{
      \max\left(
        0,\
        \hat{\mathbf q}_b
        \smash\cdot
        \left( \mathbf p_{\ell_b}^\text{GT} - \mathbf p_{\ell_{b|a},i} \right)
      \right)
    }{|B|n}
  }\label{eq:loss-inscription-hit}\end{equation}

  \begin{equation}\protect\hypertarget{eq:loss-inscription-miss}{}{ \mathcal L_{im} =
    \sum_{\substack{\ell_a\in B \\ \ell_b = \rho(\ell_a)}}
    \sum_{i=1}^n
    m_{\ell_b}^\text{GT}
    \frac{
      \max\left(
        0,\
        s_{\ell_b}^\text{GT} - s_{\ell_{b|a},i}
      \right)^2
    }{|B|n}
  }\label{eq:loss-inscription-miss}\end{equation}

  \noindent Where \(\rho : B \to K\) is a random bijection (a one-to-one mapping) from \(B\) to \(K\), and \(\mathbf p_{\ell_{b|a},i}\) and \(s_{\ell_{a|b},i}\) denote the intersection or silhouette of the \(i^\text{th}\) atom candidate predicted with \(\ell_a\) as the network input, but with the ray-atom intersection tests computed using \(\ell_b\).
\end{itemize}

\begin{itemize}
\item
  \textbf{Specialization regularization} \(\mathcal L_\sigma\): To avoid atom candidates all clustering on top of each other we introduce \(\mathcal L_\sigma\), which incentivizes each atom candidate \(i\) to cluster its predictions to a smaller volume surrounding a per-candidate centroid \(\bar{\mathbf c}_i\).

  \begin{equation}\protect\hypertarget{eq:loss-specialization}{}{
    \mathcal L_\sigma =
      \frac{1}{n|B|}
      \sum_{i=1}^n
      \sum_{\ell\in B}
        \|\mathbf c_{\ell,i}-\bar{\mathbf c}_i\|^2
    ,\text{ where }\
    \bar{\mathbf c}_i =
      \sum_{\ell\in B}
      \frac{
        \mathbf c_{\ell,i}
      }{|B|}
  }\label{eq:loss-specialization}\end{equation}

  \noindent This in effect amounts to learning an unsupervised part segmentation of \(n\) classes. (Imagine each atom candidate targeting separate limbs of the shape.) This regularization assumes the rays in the training batch cover the whole reconstruction volume.
\item
  \textbf{Multi-view loss} \(\mathcal L_\text{mv}\): We phrase the multi-view consistent property in Eqs.~\ref{eq:mv-property}, \ref{eq:mat-mv-property} as a loss penalizing change in predicted geometry with change in viewing direction. It requires \(\mathbf p_\ell^\text{GT}\) being used as the ray origin, becoming its pivot point. For any oriented ray intersection field, it can be phrased as:

  \begin{equation}\protect\hypertarget{eq:loss-multi-view-prif}{}{ \mathcal L_\text{mv} =
  \frac{1}{|B|}\sum_{\ell\in B}
  h_\ell
  h_\ell^\text{GT}
  \left\| \nabla_{\hat{\mathbf q}} \mathbf p_\ell \right\|^2
  \ :\ \mathbf o=\mathbf p_\ell^\text{GT}
  }\label{eq:loss-multi-view-prif}\end{equation}

  \noindent For MARFs we use this simplified loss:

  \begin{equation}\protect\hypertarget{eq:loss-multi-view}{}{ \mathcal L_\text{mv} =
  \frac{1}{|B|}\sum_{\ell\in B}
  h_\ell
  h_\ell^\text{GT}
  \left(
    \left\| \nabla_{\hat{\mathbf q}} \mathbf c_\ell \right\|^2
    +
    \left\| \nabla_{\hat{\mathbf q}} r_\ell \right\|^2
  \right)
  : \mathbf o=\mathbf p_\ell^\text{GT}
  }\label{eq:loss-multi-view}\end{equation}
\item
  \textbf{Latent code regularization} \(\mathcal L_{\mathbf z}\): As customary when training auto-decoders we enforce a prior over the latent space to ensure the \(n\) embeddings \(\{\mathbf z_i\}_{i=1}^n\) do not stray too far apart. Like Park et al. {[}\protect\hyperlink{ref-parkDeepSDFLearningContinuous2019}{5}{]} we use a spherical prior:

  \begin{equation}\protect\hypertarget{eq:loss-latent-reg}{}{ \mathcal L_{\mathbf z} = \frac{1}{n}\sum_{i=1}^n  \|\mathbf z_i\|^2
  }\label{eq:loss-latent-reg}\end{equation}
\item
  \textbf{Total training loss.} The complete training loss \(\mathcal L\) is given by

  \begin{equation}\protect\hypertarget{eq:loss-total}{}{ \begin{aligned}
    \mathcal L_\text{MARF}
    &=   \lambda_{\mathbf p}\mathcal L_{\mathbf p}
    +    \lambda_{\mathbf n}\mathcal L_{\mathbf n}
    +    \lambda_s          \mathcal L_s
    +    \lambda_h          \mathcal L_h
    +    \lambda_r          \mathcal L_r
    \\&+ \lambda_{ih}       \mathcal L_{ih}
    +    \lambda_{im}       \mathcal L_{im}
    +    \lambda_\sigma     \mathcal L_\sigma
    +    \lambda_\text{mv}  \mathcal L_\text{mv}
    +    \lambda_{\mathbf z}\mathcal L_{\mathbf z}
  \end{aligned} }\label{eq:loss-total}\end{equation}

  \noindent where we tune the \(\lambda\) hyperparameters (see Table~\ref{tbl:loss_hparams}) to balance the loss terms such that none dominate. We schedule some hyperparameters to ease either in or out during training. The training starts with high specialization loss eased out as the solution becomes more stable. We ease in the normal, multi-view, and latent code losses, as they prove counterproductive early in training.
\end{itemize}

{\setlength{\tabcolsep}{2pt}

\hypertarget{tbl:loss_hparams}{}
\tablecaption{\label{tbl:loss_hparams}Hyperparameters for Eq.~\ref{eq:loss-total}; some scheduled using a linear (\(\operatorname{e}_l\)) or sinusoidal (\(\operatorname{e}_s\)) easing function (see Eq.~\ref{eq:schedules}) which ease in from 0 to 1 over `\(\text{duration}\)' epochs, starting at `\(\text{offset}\)' which by default is 0.}

\begin{longtable}[]{@{}cccccccccc@{}}
\toprule
\(\lambda_{\mathbf p}\) & \(\lambda_{\mathbf n}\) & \(\lambda_s\) & \(\lambda_h\) & \(\lambda_r\) & \(\lambda_{ih}\) & \(\lambda_{im}\) & \(\lambda_\sigma\) & \(\lambda_\text{mv}\) & \(\lambda_{\mathbf z}\) \\
\midrule
\endhead
\(2\) & \(\frac{\operatorname{e}_s(85, 15)}{4}\) & \(10\) & \(100\) & \(5\smash\times10^{-4}\) & \(20\) & \(300\) & \(\frac{10-9\operatorname{e}_l(40)}{100}\) & \(\frac{\operatorname{e}_l(50)}{10}\) & \(0.01^2\operatorname{e}_l(30)\) \\
\bottomrule
\end{longtable}

}

\begin{equation}\protect\hypertarget{eq:schedules}{}{ \begin{aligned}
  \\ \operatorname{e}_l(\text{duration}, \text{offset})
    &= \operatorname{clamp}\left(\tfrac{\text{epoch}-\text{offset}}{\text{duration}}, 0, 1\right)
  \\ \operatorname{e}_s(\text{duration}, \text{offset})
    &= -\tfrac{1}{2}\left(\cos\left(\pi \operatorname{e}_l(\text{duration}, \text{offset})\right)-1\right)
\end{aligned} }\label{eq:schedules}\end{equation}

\hypertarget{optimization.}{%
\paragraph*{Optimization.}\label{optimization.}}
\addcontentsline{toc}{paragraph}{Optimization.}

We optimize the network in a stochastic gradient descent scheme, iteratively minimizing the loss in Eq.~\ref{eq:loss-total} by tuning the network weights \(\theta\) through back-propagation. We use the Adam optimizer {[}\protect\hyperlink{ref-kingmaAdamMethodStochastic2017}{96}{]} in PyTorch {[}\protect\hyperlink{ref-paszkePyTorchImperativeStyle2019}{97}{]}, with default momentum and \(5\smash\times 10^{-6}\) weight decay, layer normalization {[}\protect\hyperlink{ref-baLayerNormalization2016}{89}{]} and 1\% dropout. We warm up to a learning rate of \(5\smash\times 10^{-4}\) over 100 steps, held for the first 30 epochs, then decay to \(1\smash\times 10^{-4}\) over the next 170 epochs in a cosine annealing scheme. We train for 200 epochs total, clipping loss gradients exceeding a norm of 1.

\begin{widefig}

\begin{pandoccrossrefsubfigures}

\subfloat[Medial Lambertian]{\includegraphics[width=0.14\textwidth,height=\textheight]{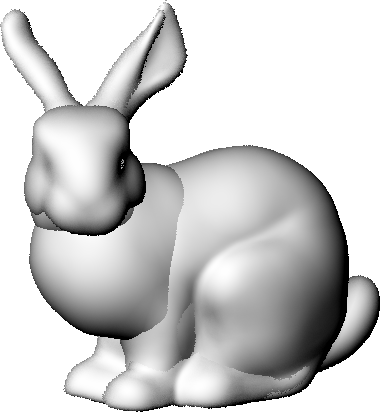}\label{fig:bunny-shadings-lambertian}} \subfloat[Candidate color]{\includegraphics[width=0.14\textwidth,height=\textheight]{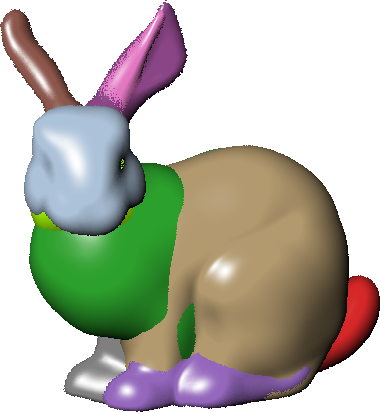}\label{fig:bunny-shadings-candidate-colors}} \subfloat[Medial axis]{\includegraphics[width=0.14\textwidth,height=\textheight]{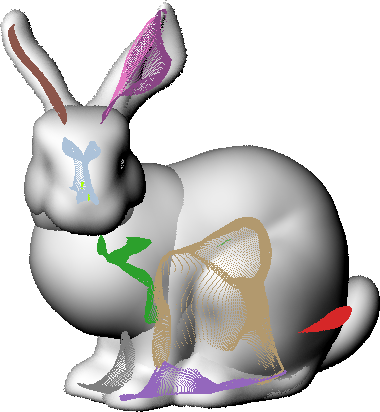}\label{fig:bunny-shadings-medial-axis}} \subfloat[Medial radius]{\includegraphics[width=0.14\textwidth,height=\textheight]{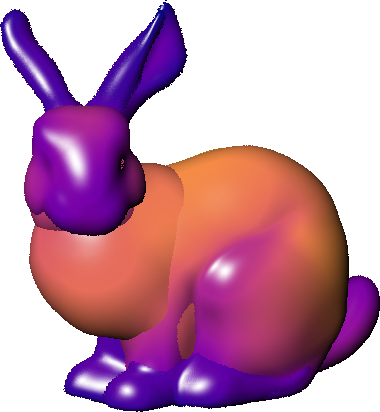}\label{fig:bunny-shadings-medial-radii}} \subfloat[Medial normal]{\includegraphics[width=0.14\textwidth,height=\textheight]{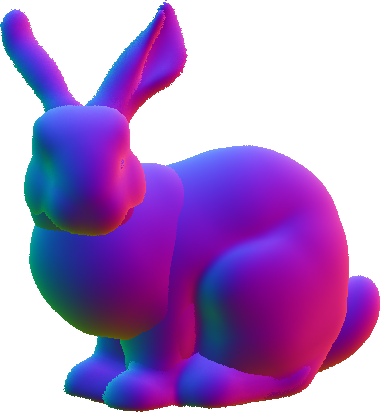}\label{fig:bunny-shadings-medial-normal}} \subfloat[Analytical normal]{\includegraphics[width=0.14\textwidth,height=\textheight]{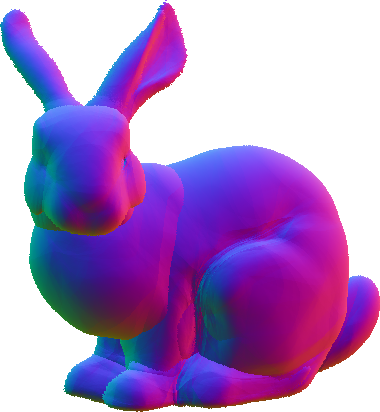}\label{fig:bunny-shadings-analytical-normal}} \subfloat[Medial curvature]{\includegraphics[width=0.14\textwidth,height=\textheight]{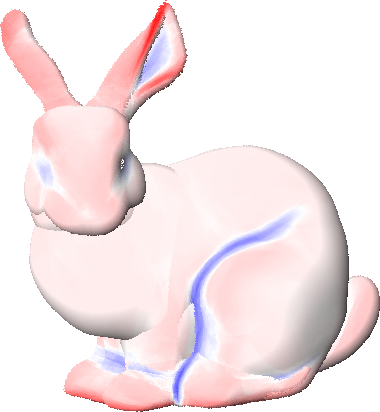}\label{fig:bunny-shadings-medial-mean-curvature}}

\caption[{MARF renderings of the Stanford bunny, visualizing the different network outputs. \textsc{(a)} is a Lambertian shading using the normals shown in \textsc{(e)}. \textsc{(b)} tints the surface with a unique color associated which atom candidate was chosen by metric \(m_{\ell,i}\) in Eq.~\ref{eq:candidate-metric}, indicating a learned unsupervised part segmentation. \textsc{(c)} illustrates the medial axis, also known as the topological skeleton, by superimposing the predicted medial atom centers associated with each hitting camera ray, onto \textsc{(a)}. \textsc{(d)} maps the predicted radius of the intersected medial atom onto a color scale, in effect visualizing \emph{local thickness} which is useful when approximating translucency. \textsc{(e)} visualizes the normals derived from the intersected medial atoms as RGB (see Eq.~\ref{eq:marf-normal}), while \textsc{(f)} visualizes the normals derived by differentiating the whole network (see Eq.~\ref{eq:oriented-ray-field-normal}). \textsc{(g)} visualizes the mean curvature with positive values in red and negative in blue. This curvature is contained in the shape operator (Eq.~\ref{eq:shape-operator}) which is computed by analytical differentiation of the medial normals shown in \textsc{(e)}. \textsc{(a-e)} perform a single network evaluation per pixel, as they are shaded solely using the predicted medial quantities, while \textsc{(f-g)} perform a backward pass to compute the true analytical gradients of the network outputs.}]{MARF renderings of the Stanford bunny, visualizing the different network outputs. \textsc{(a)} is a Lambertian shading using the normals shown in \textsc{(e)}. \textsc{(b)} tints the surface with a unique color associated which atom candidate was chosen by metric \(m_{\ell,i}\) in Eq.~\ref{eq:candidate-metric}, indicating a learned unsupervised part segmentation. \textsc{(c)} illustrates the medial axis, also known as the topological skeleton, by superimposing the predicted medial atom centers associated with each hitting camera ray, onto \textsc{(a)}. \textsc{(d)} maps the predicted radius of the intersected medial atom onto a color scale, in effect visualizing \emph{local thickness} which is useful when approximating translucency. \textsc{(e)} visualizes the normals derived from the intersected medial atoms as RGB (see Eq.~\ref{eq:marf-normal}), while \textsc{(f)} visualizes the normals derived by differentiating the whole network (see Eq.~\ref{eq:oriented-ray-field-normal}). \textsc{(g)} visualizes the mean curvature with positive values in red and negative in blue. This curvature is contained in the shape operator (Eq.~\ref{eq:shape-operator}) which is computed by analytical differentiation of the medial normals shown in \textsc{(e)}. \textsc{(a-e)} perform a single network evaluation per pixel, as they are shaded solely using the predicted medial quantities, while \textsc{(f-g)} perform a backward pass to compute the true analytical gradients of the network outputs.}

\label{fig:bunny-shadings}

\end{pandoccrossrefsubfigures}

\end{widefig}

\begin{pandoccrossrefsubfigures}

\subfloat[Typical \(200\smash\times 200\) training image]{\includegraphics[width=0.13\textwidth,height=\textheight]{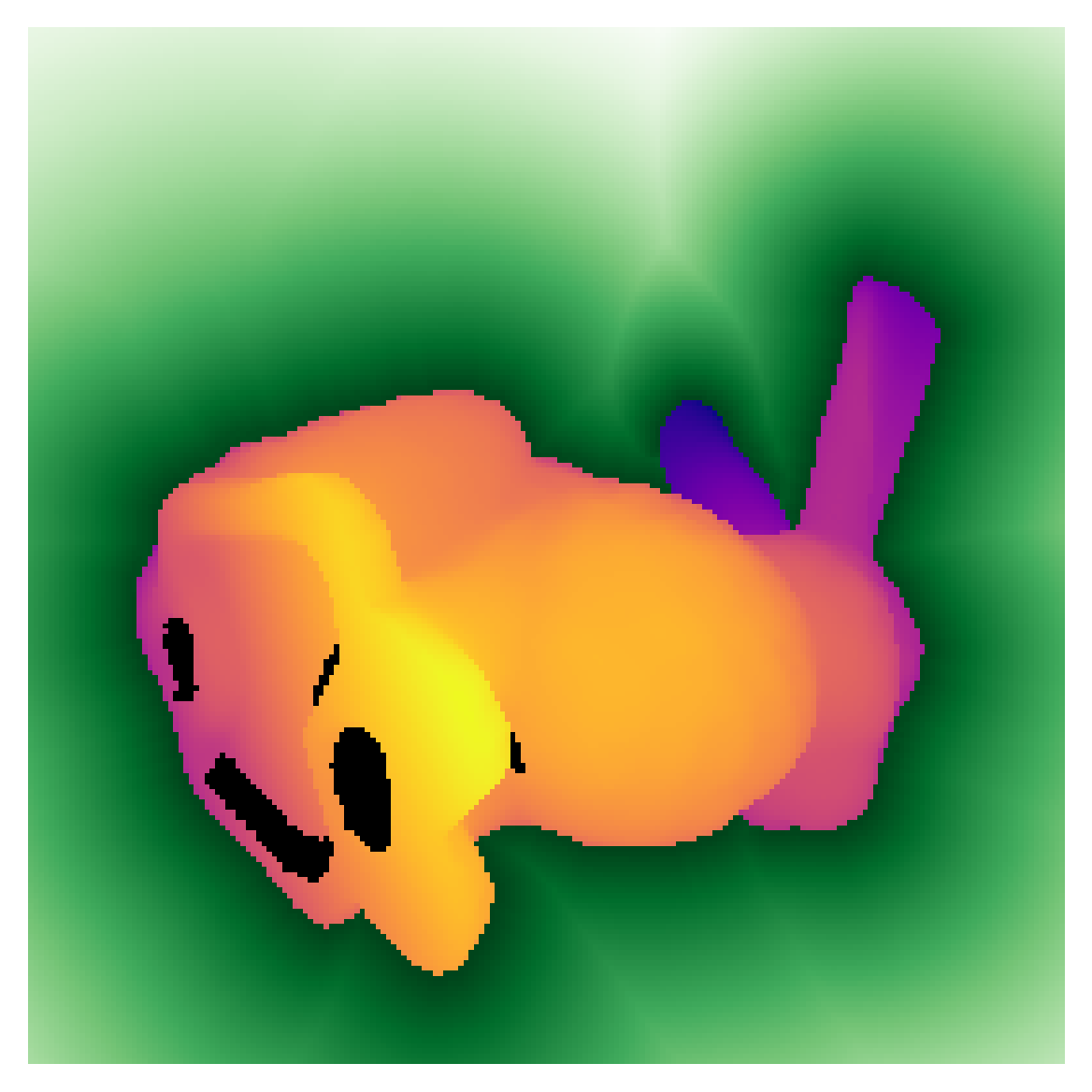}\label{fig:batchstep-bunny}} \subfloat[Simplified \(8\smash\times 8\) image]{\includegraphics[width=0.13\textwidth,height=\textheight]{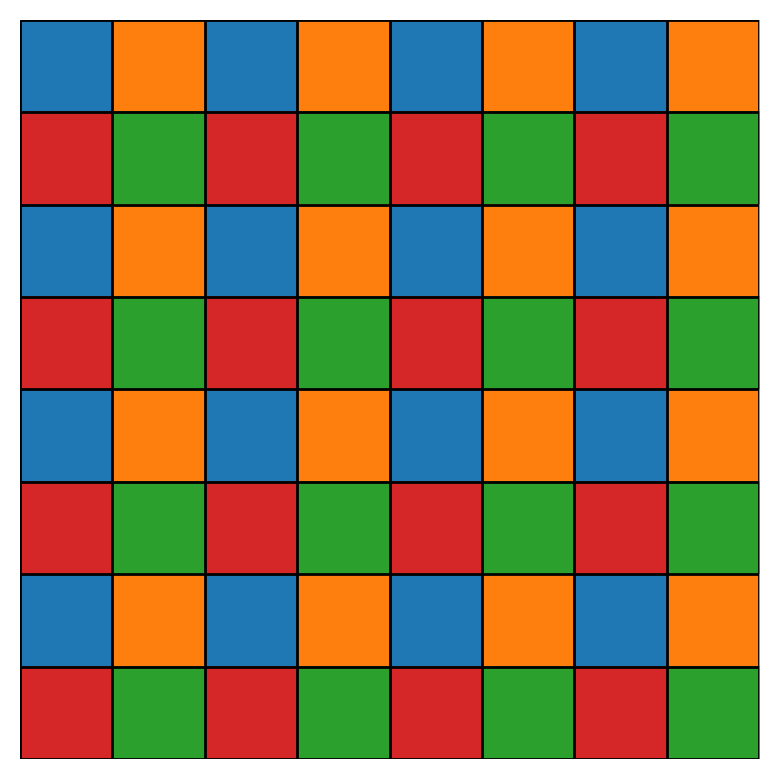}} \subfloat[Four \(4\smash\times 4\) strided sub-images of \textsc{(b)}]{\includegraphics[width=0.1263\textwidth,height=\textheight]{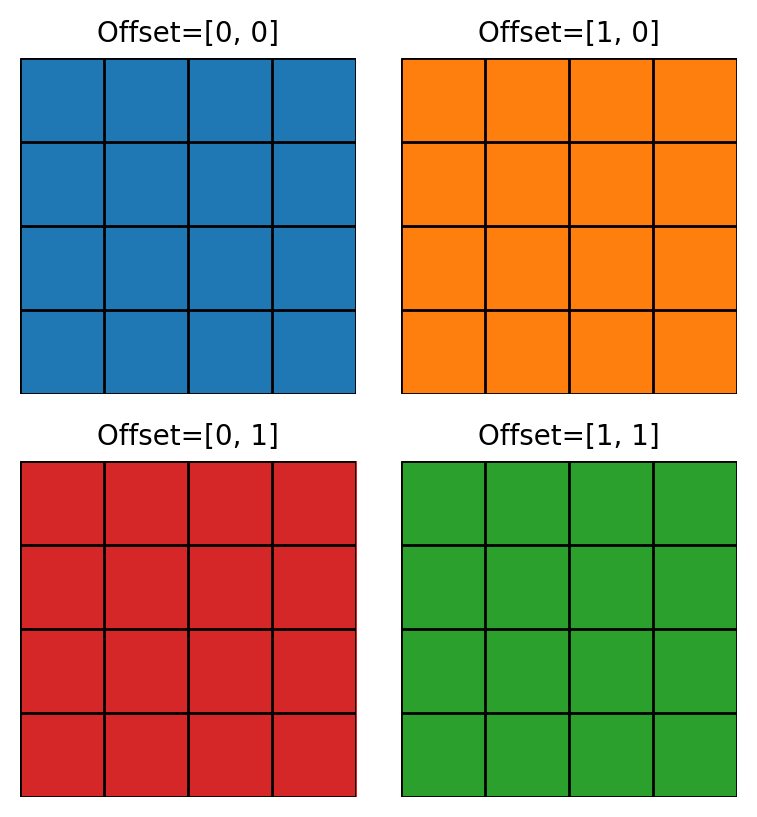}}

\caption[{Splitting single-view depth and silhouette training images \textsc{(a)} into multiple into smaller images during training. Here we show how a simple \(8\times 8\) training image \textsc{(b)}, a representative proxy of \textsc{(a)}, is split into \(2^2\) smaller parts when using a stride size of 2.}]{Splitting single-view depth and silhouette training images \textsc{(a)} into multiple into smaller images during training. Here we show how a simple \(8\times 8\) training image \textsc{(b)}, a representative proxy of \textsc{(a)}, is split into \(2^2\) smaller parts when using a stride size of 2.}

\label{fig:batchstep}

\end{pandoccrossrefsubfigures}

\hypertarget{sec:experiments}{%
\section{Experiments}\label{sec:experiments}}

We detail in Section~\ref{sec:exp-setup} our experimental setup. In Section~\ref{sec:exp-single-shape} evaluation single-shape MARF results, followed by extensive ablation studies in Section~\ref{sec:ablation}. In Section~\ref{sec:exp-rendering} we demonstrate two applications of MARFs in visualization. Finally in Section~\ref{sec:exp-multiple-shape} we present a multi-shape MARF, applicable to inverse rendering applications benefiting from learned shape priors.

\hypertarget{sec:exp-setup}{%
\subsection{Experimental Setup}\label{sec:exp-setup}}

\hypertarget{training.}{%
\paragraph*{Training.}\label{training.}}
\addcontentsline{toc}{paragraph}{Training.}

We model all networks with 8 hidden layers, 512 neurons wide, using PyTorch 1.13 {[}\protect\hyperlink{ref-paszkePyTorchImperativeStyle2019}{97}{]} and train with PyTorch Lightning {[}\protect\hyperlink{ref-williamPyTorchLightning2019}{98}{]} on Python 3.10. To compute analytical derivatives we use the \emph{torch.autograd.grad} function. We reserve 30\% of the 50 virtual camera views for validation while tuning hyperparameters, and train for 200 epochs. While the loss in Eq.~\ref{eq:loss-total} seemingly assumes a single training image, we construct batches of multiple views and multiple objects to make each batch more diverse. We compute their loss independently and average the results. To fit more views and objects in memory per batch while maintaining a sparse set of training views, we split (as illustrated in Fig.~\ref{fig:batchstep}) the \(200\smash\times 200\) training images into \(4^2\) coarser \(50\smash\times 50\) sub-images by using a stride of 4. We randomize the order of sub-images across objects and views into batches of 8 in each epoch. We train with CUDA 11.7 using mixed 16bit float precision and medium matrix multiplication precision. The single-shape MARFs took about 44 minutes to train on an Nvidia A100 GPU, while the 20-shape MARFs took about 7 hours using two A100s, provided by Själander et al. {[}\protect\hyperlink{ref-sjalanderEPICEnergyEfficientHighPerformance2020}{99}{]}.

\hypertarget{evaluation.}{%
\paragraph*{Evaluation.}\label{evaluation.}}
\addcontentsline{toc}{paragraph}{Evaluation.}

To render a MARF we evaluate it on the rays associated with each canvas pixel, discard rays that miss, optionally compute analytical network gradient depending on what we visualize, then compute shading. To visualize the medial axis we superimpose the medial atom centers associated with each hitting ray. For quantitative evaluation we sample ground-truth point clouds from each object mesh by casting rays between 4000 viewpoints spaced equidistantly on the enclosing unit-sphere with PyEmbree {[}\protect\hyperlink{ref-scopatzPyembreePythonWrapper2022}{100}{]}. In effect this means we evaluate using 4000 camera views while training using only 35. We extract point clouds from the MARF and baseline using the same set of rays, and compute the Precision, Recall and Intersection over Union (IoU) of rays that hit. We then sample 30,000 hit points and compute the Chamfer (CD) and cosine similarity (COS) distance with {[}\protect\hyperlink{ref-raviAccelerating3DDeep2020}{101}{]}. Unlike Feng et al. {[}\protect\hyperlink{ref-fengPRIFPrimaryRaybased2022}{69}{]} we do not fit a surface to the hit points to compute the CD. The metrics are defined in the Appendix.

MARF renders \(256\smash\times 256\) resolution images at 18.7 frames/s on an Nvidia GTX 3070 8GB Laptop edition when using medial normals (Eq.~\ref{eq:marf-normal}). With analytical normals (Eq.~\ref{eq:oriented-ray-field-normal}) calls we see MARF renders at 4.8 frames/s, while PRIF renders at 5.2 frames/s.

\hypertarget{baseline.}{%
\paragraph*{Baseline.}\label{baseline.}}
\addcontentsline{toc}{paragraph}{Baseline.}

We train our reproduction of PRIF by Feng et al. {[}\protect\hyperlink{ref-fengPRIFPrimaryRaybased2022}{69}{]} using the same training data, input encoding, learning rate, dropout, normalization and network dimensions as for MARFs. In short, PRIFs predict the signed displacement from the perpendicular foot \(\mathbf o_\perp\) (Eq.~\ref{eq:ray-input-embedding}) along the ray direction \(\hat{\mathbf q}\) -- essentially the \(t\) in Eq.~\ref{eq:ell} -- as well as \emph{whether} the ray intersects with a second network output supervised with binary cross-entropy loss. We train PRIF with its original loss function denoted \(\mathcal L_\text{PRIF}\). As an experiment we also train PRIF with \(2\smash\times\) our normal loss \(\mathcal L_{\mathbf n}\) added (see Eq.~\ref{eq:loss-normal}), where we compute normals through network differentiation using Eq.~\ref{eq:oriented-ray-field-normal}. We also train PRIF with our multi-view loss \(\mathcal L_\text{mv}\) (see Eq.~\ref{eq:loss-multi-view-prif}). Both additions are scaled according to Table~\ref{tbl:loss_hparams}.

\begin{pandoccrossrefsubfigures}

\subfloat[Ground truth]{\includegraphics[width=0.099\textwidth,height=\textheight]{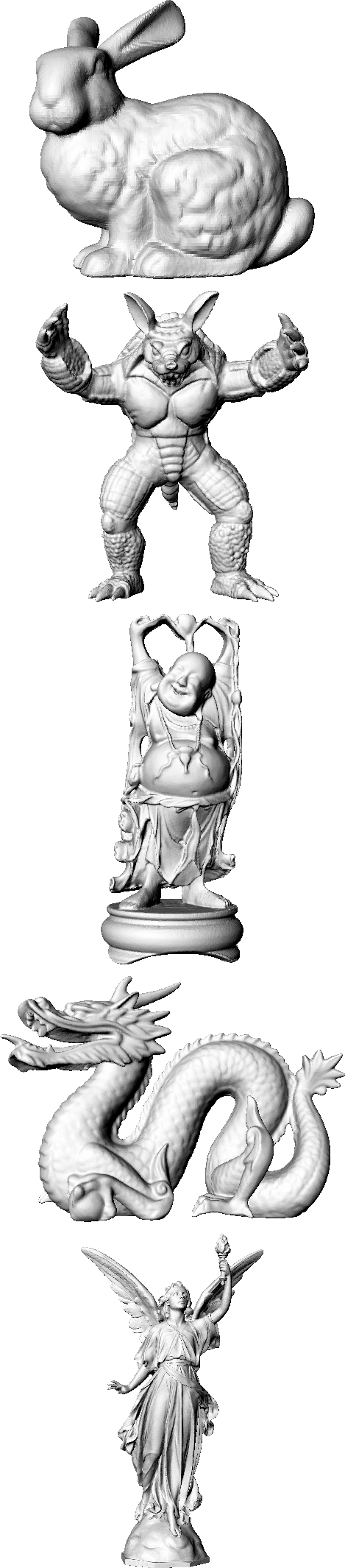}\label{fig:all-single-shapes-gt}} \subfloat[\(\nabla\)PRIF]{\includegraphics[width=0.0962\textwidth,height=\textheight]{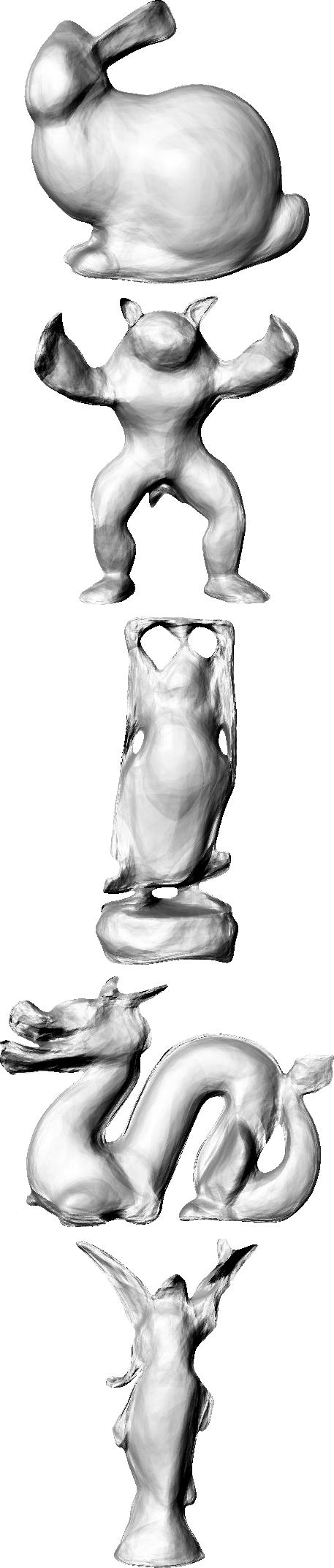}\label{fig:all-single-shapes-prif}} \subfloat[\(\nabla\)MARF]{\includegraphics[width=0.1\textwidth,height=\textheight]{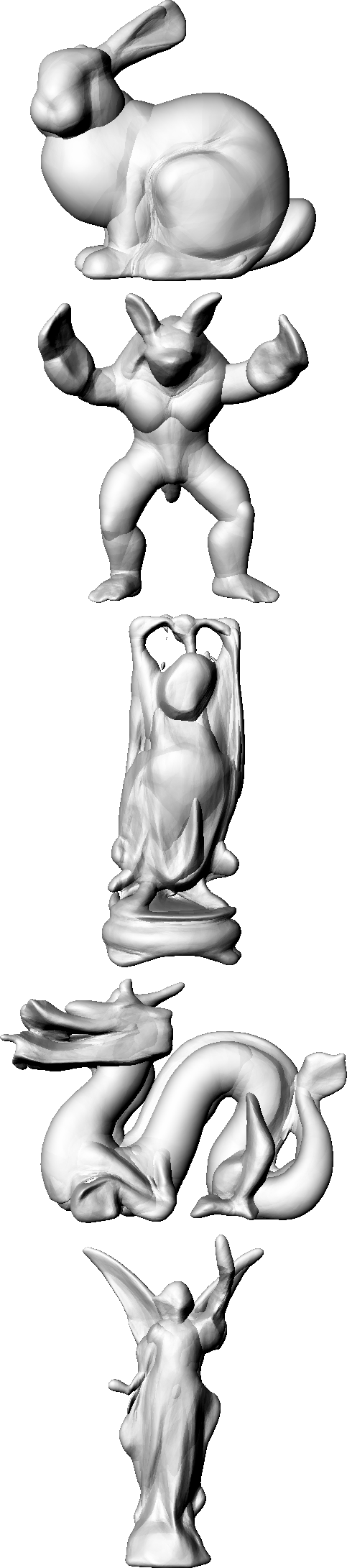}\label{fig:all-single-shapes-marf-nabla}} \subfloat[\(\mathcal M\) MARF]{\includegraphics[width=0.1\textwidth,height=\textheight]{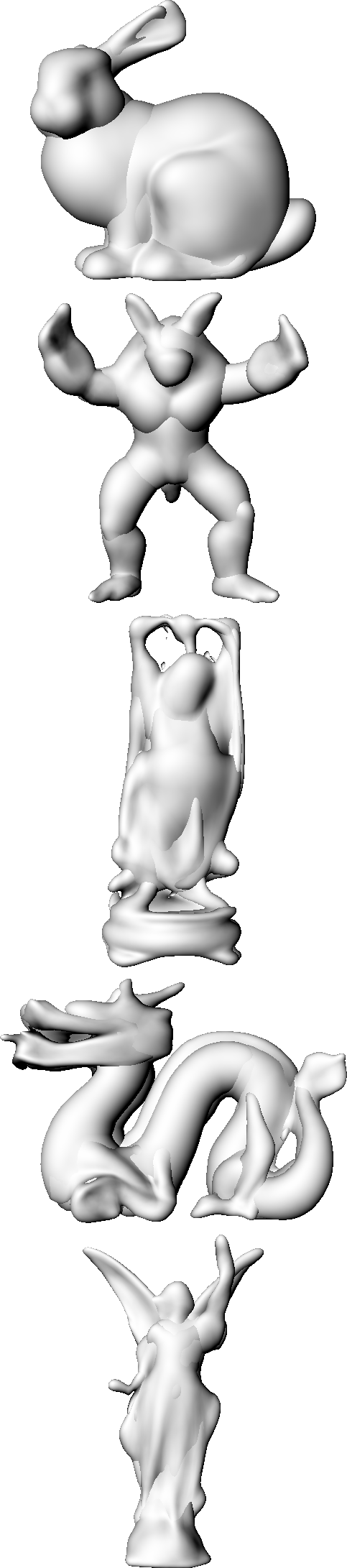}\label{fig:all-single-shapes-marf-medial}} \subfloat[\(\mathcal M\) MARF+axis]{\includegraphics[width=0.1\textwidth,height=\textheight]{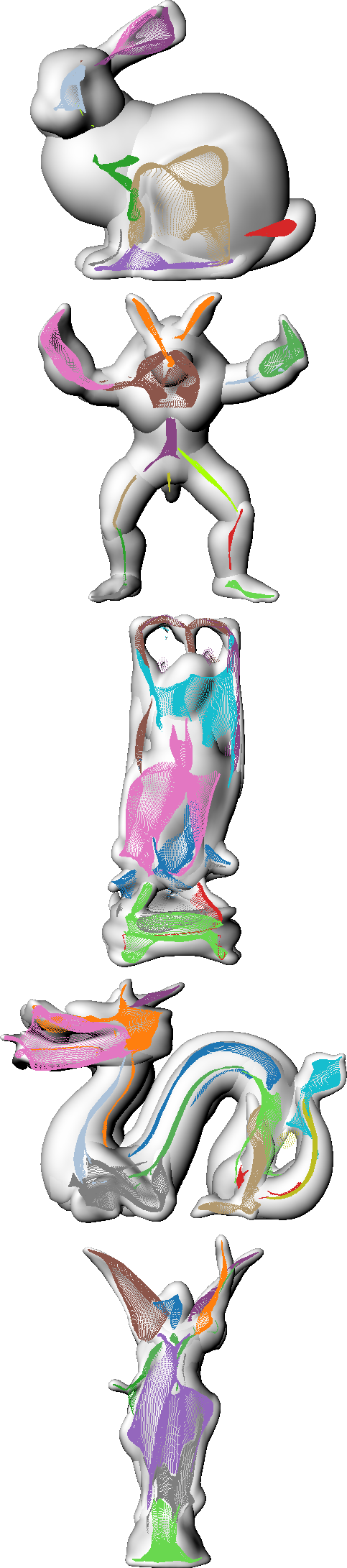}\label{fig:all-single-shapes-marf-axis}}

\caption[{Single-shape MARF and PRIF renderings from a view not present in the training set, with no outlier filtering. \(\nabla\) denotes shading with analytical normals (Eq.~\ref{eq:oriented-ray-field-normal}), and \(\mathcal M\) denotes shading with medial normals (Eq.~\ref{eq:marf-normal}).}]{Single-shape MARF and PRIF renderings from a view not present in the training set, with no outlier filtering. \(\nabla\) denotes shading with analytical normals (Eq.~\ref{eq:oriented-ray-field-normal}), and \(\mathcal M\) denotes shading with medial normals (Eq.~\ref{eq:marf-normal}).}

\label{fig:all-single-shapes}

\end{pandoccrossrefsubfigures}

\hypertarget{sec:exp-single-shape}{%
\subsection{Learning a Single Shape}\label{sec:exp-single-shape}}

Here we examine MARFs trained from scratch to represent a single shape.

\hypertarget{qualitative-results.}{%
\paragraph*{Qualitative Results.}\label{qualitative-results.}}
\addcontentsline{toc}{paragraph}{Qualitative Results.}

We visualize in Fig.~\ref{fig:all-single-shapes} MARFs and PRIFs trained to represent five Stanford 3D Scanning Repository {[}\protect\hyperlink{ref-stanford3DScanning}{92}{]} objects. We visualize the medial quantities represented by MARFs in more detail in Fig.~\ref{fig:bunny-shadings} on the bunny. The reconstructions are convincing and stay consistent across views.

MARFs perform well in areas with positive curvature (shaded red in Fig.~\ref{fig:bunny-shadings-medial-mean-curvature}), where atoms stay relatively still w.r.t. a moving ray. In negatively curved areas however the MARFs must learn to ``swing'' atoms about the curve on the interior, consuming learning capacity. As such MARFs with sufficient number of atom candidates tend to specialize separate atoms to represent each side of sharp negative curves, evident on the body of the bunny in Fig.~\ref{fig:bunny-ears-overhang-good}.

The MARFs allocated atom candidates where needed. The number of discontinuities possible to represent however is upper bounded by number of atom candidates available. 16 candidates proved insufficient for the dragon in Fig.~\ref{fig:all-single-shapes-marf-axis}, which used a single atom candidate to represent both the upper and lower part of its open mouth. On the other end we find atoms going unused. On the bunny in Fig.~\ref{fig:bunny-shadings-candidate-colors} we only see 9 out of 16 total atom candidates. The other atoms are hidden inside the main body. Fully occluded atom candidates receive no supervision, and if occluded early during training they may not get used at all. We believe this is why the Buddha in Fig.~\ref{fig:all-single-shapes-marf-axis} fit a single atom to both hands. Maximality regularization is the only pressure counteracting occlusion, but it tends to slide atoms along medial branches when no intersection loss pins it in place {[}\protect\hyperlink{ref-rebainLSMATLeastSquares2019}{83}{]}.

The PRIF baseline reconstructs the training set admirably but struggles with unseen views. MARFs perform better but some pop-ins can be found. In Fig.~\ref{fig:all-single-shapes-marf-nabla} we see a MARF fail to reconstruct the left ear of the bunny. While visible from most camera angles like Fig.~\ref{fig:bunny-shadings-lambertian}, it is not visible from this one.

MARFs discover sound medial axes on organic shapes like the bunny where the true MAT is simple but struggles on more intricate geometry like the angel Lucy and the Buddha. We shade in Fig.~\ref{fig:all-single-shapes-marf-nabla} using analytically computed normals, which reveal when compared to Fig.~\ref{fig:all-single-shapes-marf-medial} that MARF in such cases ``cheat'' by varying the atom radii instead of properly moving the atoms along the medial axis. The lowered medial axis accuracy results in less multi-view consistency, with some warping visible when moving the camera.

The renders in Fig.~\ref{fig:all-single-shapes-marf-nabla} are more accurate than Fig.~\ref{fig:all-single-shapes-marf-medial}, indicating that MARFs prioritize accurate surface intersections over accurate medial normals. We believe this is because the cosine similarity in the normal loss \(\mathcal L_{\mathbf n}\) (Eq.~\ref{eq:loss-normal}) is in effect a squared distance while the intersection loss \(\mathcal L_{\mathbf p}\) (Eq.~\ref{eq:loss-intersection}) is not. The normal loss proposed in {[}\protect\hyperlink{ref-groppImplicitGeometricRegularization2020}{16}{]} proved unstable however.

\begin{table}[t!] \setlength{\tabcolsep}{2.6pt}

\tablecaption{Single-shape results on five Stanford {[}\protect\hyperlink{ref-stanford3DScanning}{92}{]} objects shown in Fig.~\ref{fig:all-single-shapes}. We present with best in \textbf{bold} mean CD \((\smash\times 10^4)\) and COS scores of reconstruction quality, and IoU scoring ray hit accuracy. We compute IoU on rays cast between 4000 equidistant points, then sample 30,000 hit points to compute CD and COS. We score COS with analytical surface normals \((\nabla)\) using Eq.~\ref{eq:oriented-ray-field-normal}, For MARF we also score COS with medial normals \((\mathcal M)\) using Eq.~\ref{eq:marf-normal}. We present MARFs trained with and without multi-view loss \(\mathcal L_\text{mv}\) (Eq.~\ref{eq:loss-multi-view}), and PRIFs {[}\protect\hyperlink{ref-fengPRIFPrimaryRaybased2022}{69}{]} with its original loss \(\mathcal L_\text{PRIF}\) scored it with and without outlier point filtering. We also train PRIFs with \(\mathcal L_{\mathbf n}\) (Eq.~\ref{eq:loss-normal}) and with \(\mathcal L_\text{mv}\) (Eq.~\ref{eq:loss-multi-view-prif}), scored with filtering. The IoU score considers filtered rays as misses. MARFs are not filtered.}\label{tbl:test2}

\begin{longtable}[]{@{}clcccccccc@{}}
\toprule
\multicolumn{2}{c}{\multirow{2}{*}{Metrics \& Objects}} & \multicolumn{2}{c}{PRIF} & PRIF & PRIF & \multicolumn{2}{c}{\(\text{MARF}\smash+\mathcal L_\text{mv}\)} & \multicolumn{2}{c}{\(\text{MARF}\smash-\mathcal L_\text{mv}\)}\\ \cmidrule(lr){3-4} \cmidrule(lr){5-5} \cmidrule(lr){6-6} \cmidrule(lr){7-8} \cmidrule(lr){9-10}
& & \sout{{Filter}} & {\(\mathcal L_\text{PRIF}\)} & {\(+2\mathcal L_{\mathbf n}\)} & {\(+\mathcal L_\text{mv}\)} & {\(\nabla\)} & {\(\mathcal M\)} & {\(\nabla\)} & {\(\mathcal M\)} \\
\midrule
\endhead
\multirow{5}{*}{CD\(\smash\downarrow\)} & {Armadillo} & {24.578} & {22.705} & {21.653} & {18.015} & \multicolumn{2}{c}{{2.745}} & \multicolumn{2}{c}{\textbf{{2.560}}} \\
& {Buddha} & {12.538} & {12.534} & {13.991} & {9.547} & \multicolumn{2}{c}{{2.996}} & \multicolumn{2}{c}{\textbf{{2.948}}} \\
& {Bunny} & {16.171} & {15.274} & {13.746} & {12.653} & \multicolumn{2}{c}{\textbf{{1.816}}} & \multicolumn{2}{c}{{2.450}} \\
& {Dragon} & {16.484} & {16.028} & {15.180} & {13.713} & \multicolumn{2}{c}{\textbf{{3.187}}} & \multicolumn{2}{c}{{4.046}} \\
& {Lucy} & {11.615} & {10.039} & {9.841} & {7.969} & \multicolumn{2}{c}{\textbf{{2.064}}} & \multicolumn{2}{c}{{2.203}} \\ \midrule
\multirow{5}{*}{COS\(\smash\uparrow\)} & {Armadillo} & {0.597} & {0.603} & {0.632} & {0.644} & \textbf{{0.815}} & {0.788} & {0.793} & {0.762} \\
& {Buddha} & {0.508} & {0.517} & {0.503} & {0.554} & \textbf{{0.715}} & {0.665} & {0.706} & {0.677} \\
& {Bunny} & {0.753} & {0.757} & {0.763} & {0.780} & \textbf{{0.937}} & {0.924} & {0.907} & {0.881} \\
& {Dragon} & {0.558} & {0.562} & {0.582} & {0.583} & \textbf{{0.802}} & {0.768} & {0.743} & {0.698} \\
& {Lucy} & {0.439} & {0.440} & {0.462} & {0.443} & \textbf{{0.630}} & {0.586} & {0.624} & {0.580}\\ \midrule
\multirow{5}{*}{IoU\(\smash\uparrow\)} & {Armadillo} & {84.0\%} & {80.8\%} & {81.5\%} & {81.8\%} & \multicolumn{2}{c}{\textbf{{92.5\%}}} & \multicolumn{2}{c}{{91.3\%}} \\
& {Buddha} & {91.8\%} & {88.2\%} & {88.9\%} & {90.3\%} & \multicolumn{2}{c}{\textbf{{93.1\%}}} & \multicolumn{2}{c}{{91.5\%}} \\
& {Bunny} & {93.2\%} & {90.4\%} & {90.9\%} & {91.3\%} & \multicolumn{2}{c}{\textbf{{95.7\%}}} & \multicolumn{2}{c}{{94.9\%}} \\
& {Dragon} & {88.6\%} & {83.8\%} & {84.7\%} & {86.1\%} & \multicolumn{2}{c}{\textbf{{92.0\%}}} & \multicolumn{2}{c}{{90.5\%}} \\
& {Lucy} & {86.6\%} & {84.0\%} & {85.0\%} & {85.2\%} & \multicolumn{2}{c}{\textbf{{89.8\%}}} & \multicolumn{2}{c}{{88.3\%}} \\
\bottomrule
\end{longtable}

\end{table}

\begin{pandoccrossrefsubfigures}

\subfloat[Trained with \(\mathcal L_{\mathbf n}\)]{\includegraphics[width=0.14\textwidth,height=\textheight]{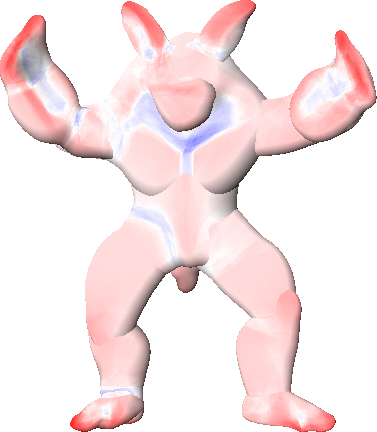}} \subfloat[Trained without \(\mathcal L_{\mathbf n}\)]{\includegraphics[width=0.14\textwidth,height=\textheight]{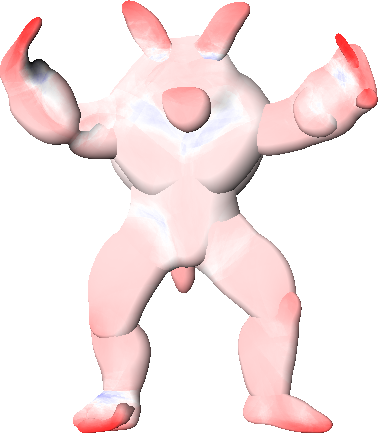}\label{fig:ablation-qualitative-no-normals}} \subfloat[Overfitting to approximated training silhouettes]{\includegraphics[width=0.2\textwidth,height=\textheight]{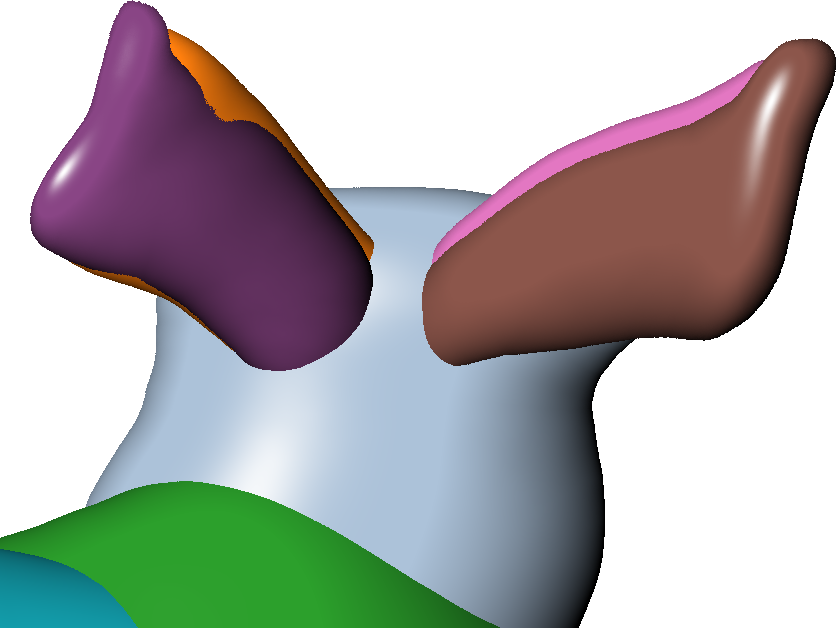}\label{fig:bad-silhouette}}

\caption[{\textsc{(a-b)} compares two MARFs (shaded with mean curvature) trained with and without surface normal supervision. The latter fails to represent negative curvature (blue) in its medial normals (Eq.~\ref{eq:marf-normal}). \textsc{(c)}, shaded with unique colors per atom candidate, illustrates what happens when using too strong silhouette supervision. The MARF overfits small atom candidates near edges to reconstruct the inaccurate ground truth silhouettes, which we approximated using sphere-tracing.}]{\textsc{(a-b)} compares two MARFs (shaded with mean curvature) trained with and without surface normal supervision. The latter fails to represent negative curvature (blue) in its medial normals (Eq.~\ref{eq:marf-normal}). \textsc{(c)}, shaded with unique colors per atom candidate, illustrates what happens when using too strong silhouette supervision. The MARF overfits small atom candidates near edges to reconstruct the inaccurate ground truth silhouettes, which we approximated using sphere-tracing.}

\label{fig:ablation-qualitative}

\end{pandoccrossrefsubfigures}

\begin{pandoccrossrefsubfigures}

\subfloat[Good specialization]{\includegraphics[width=0.22\textwidth,height=\textheight]{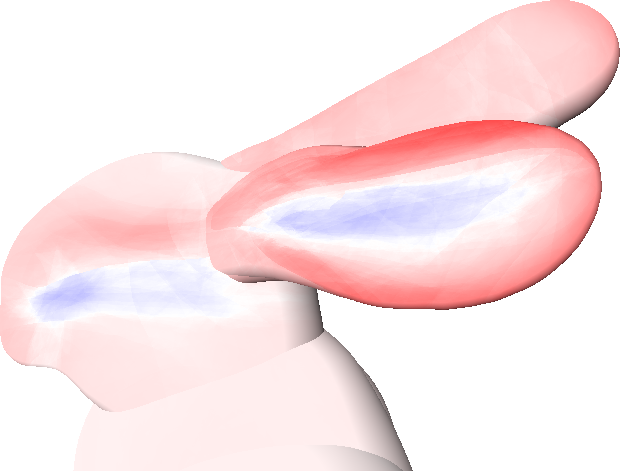}\label{fig:bunny-ears-overhang-good}} \subfloat[Poor specialization]{\includegraphics[width=0.21\textwidth,height=\textheight]{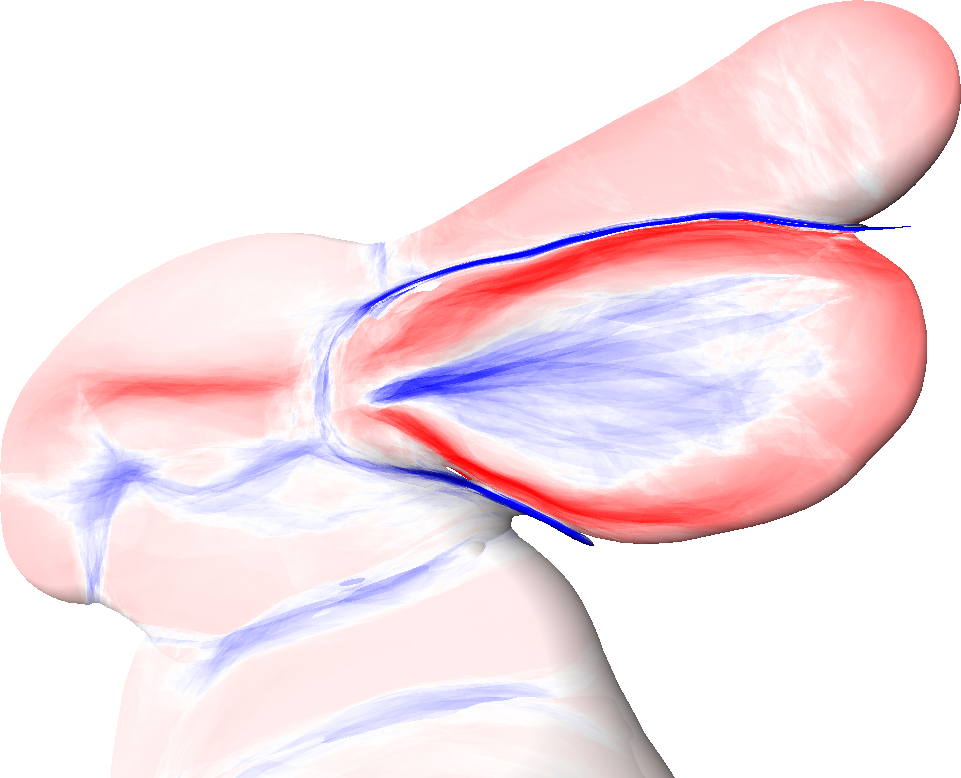}\label{fig:bunny-ears-overhang-bad}}

\caption[{Two Stanford bunny MARFs shaded with mean curvature using medial normals. In \textsc{(a)} we compare the MARF from Fig.~\ref{fig:bunny-shadings} against a MARF trained without our initialization scheme (Section~\ref{sec:neural-network}) in \textsc{(b)}, which failed to specialize its atom candidates to deal with discontinuities. \textsc{(b)} uses a single atom candidate to represent the head and both ears. While negatively curved areas (blue) on the body are better represented, it also produces artifacts where atoms ``jump'' across discontinuities under a Lipschitz bound.}]{Two Stanford bunny MARFs shaded with mean curvature using medial normals. In \textsc{(a)} we compare the MARF from Fig.~\ref{fig:bunny-shadings} against a MARF trained without our initialization scheme (Section~\ref{sec:neural-network}) in \textsc{(b)}, which failed to specialize its atom candidates to deal with discontinuities. \textsc{(b)} uses a single atom candidate to represent the head and both ears. While negatively curved areas (blue) on the body are better represented, it also produces artifacts where atoms ``jump'' across discontinuities under a Lipschitz bound.}

\label{fig:bunny-ears-overhang}

\end{pandoccrossrefsubfigures}

\hypertarget{quantitative-results.}{%
\paragraph*{Quantitative Results.}\label{quantitative-results.}}
\addcontentsline{toc}{paragraph}{Quantitative Results.}

We score MARFs in Table~\ref{tbl:test2} with metrics for both reconstruction quality and ray hit/miss accuracy. There we find that MARF outperform three PRIF {[}\protect\hyperlink{ref-fengPRIFPrimaryRaybased2022}{69}{]} variations trained under the same conditions.

In general, MARFs perform better with multi-view loss \(\mathcal L_\text{mv}\) (Eq.~\ref{eq:loss-multi-view}) than without, justifying the doubled training time thanks to double back-propagation. The Armadillo is the exception, which as shown in Fig.~\ref{fig:all-single-shapes-marf-axis} used a single atom candidate to represent both ears. The multi-view loss spikes as the atom quickly ``jumps'' from one ear to the other. We visualize such a discontinuity in Fig.~\ref{fig:bunny-ears-overhang-bad}.

The benefits of our normal loss \(\mathcal L_{\mathbf n}\) (Eq.~\ref{eq:loss-normal}) and multi-view loss \(\mathcal L_\text{mv}\) transfers over when applied to PRIF. The exception is the Buddha, which sees a decrease in reconstruction quality.

\hypertarget{sec:ablation}{%
\subsection{Ablations}\label{sec:ablation}}

We conduct extensive ablation studies in Table~\ref{tbl:ablation} on the terms of our loss function in Eq.~\ref{eq:loss-total}, as well as on three of our architecture choices: the input encoding scheme, our principled initialization scheme, and number of atom candidates predicted per ray. For each experiment we train five MARFs, one for each of the five Stanford objects explored in Section~\ref{sec:exp-single-shape}.

Our ray input encoding scheme outperforms both LFN {[}\protect\hyperlink{ref-sitzmann2021lfns}{63}{]} and PRIF {[}\protect\hyperlink{ref-fengPRIFPrimaryRaybased2022}{69}{]}. While it does not raise the spectral bias nor provide any additional information it does raise the Lipschitz bound, demonstrating how much a positional encoding {[}\protect\hyperlink{ref-mildenhallNeRFRepresentingScenes2020a}{21},\protect\hyperlink{ref-tancikFourierFeaturesLet2020a}{64}{]} scheme would benefit neural ray fields.

We see accuracy increase and decrease with the number of atom candidates predicted, diminishing in return as it increases. Past 16 candidates we observe a decline in hit precision, where atoms become prune to occlude each other.

Without our initialization scheme or specialization regularization, we find the training getting stuck in a local minima where the atom candidates fail to target separate limbs, illustrated in Fig.~\ref{fig:bunny-ears-overhang}, causing interpolation artifacts near discontinuities.

Without intersection loss we find the atoms still intersect the ray thanks to the silhouette loss, but nothing pins the atoms tangential to the ray-surface intersection point. Inscription loss constrains the atoms to stay on the interior, but no counteracting force ``pulls'' them back out towards the surface. Maximality regularization instead slides the atoms down medial branches where larger medial radii are supported, in effect eroding the represented shape.

Without silhouette loss we observe a large drop in ray hit/miss accuracy. While the intersection and normal losses only supervise true hits, they are still able to gradually ``roll'' atoms to where they are needed thanks to their non-zero size. If we further increase the silhouette loss we find IoU and CD improve, but the normal accuracy decrease. We suspect our approximate silhouette ground truths are too inaccurate, causing overfitting where some atom candidates are specialized to extend the outline, as evident in Fig.~\ref{fig:bad-silhouette}.

Without normal loss we see ray hit precision and surface reconstruction improve, at the cost of medial normal accuracy and hit recall. In Fig.~\ref{fig:ablation-qualitative-no-normals} we show a MARF without normal supervision which fails to represent negative curvatures. In theory the combination of intersection and inscription loss should suffice making atoms ``swing'' about negative curvature, indicating that our inscription testing may be too coarse.

We enforce inscription on each ray using just one other ray in the batch chosen at random. Without inscription loss we find the ray hit recall improving as expected, but at the cost of a lowered ray hit precision. Atom candidates that miss the ray (i.e.~the majority) become free to violate the inscription constraint, becoming prone to getting stuck in local minima as a result. Further increasing inscription loss overpowers the other losses, in effect eroding the represented shape.

Without maximality regularization we observe a non-significant decline on all metrics. In our object-centric setup, its function overlaps with the silhouette loss which stretches atoms to fill in the shape contour from all camera views. When we further increase the effect of maximality regularization we find reconstruction quality improving, but also more false ray hits. This indicates that the atoms grow beyond the confines of the surface boundary. While the amount of inscription loss seems ideal, we believe its resolution may be insufficient.

Without multi-view loss the accuracy drops across all metrics. The surface visibly wobbles as we move the camera, indicating that the MARF is overfitting to the sparse set of training views. Adding too much multi-view loss also causes a drop in accuracy, effectively pruning branches of the medial axis, which is a common strategy to simplify shapes {[}\protect\hyperlink{ref-tamShapeSimplificationBased2003}{82}{]}.

\begin{table}[t!] {\setlength{\tabcolsep}{2.2pt}

\tablecaption{Ablation studies. We present reconstruction quality CD and COS scores, and ray hit IoU, Precision. and Recall scores. Each row is the average score of five single-shape MARFs, one for each object explored in Section~\ref{sec:exp-single-shape}. The first row is our proposed configuration, while the following rows make a single modification each. We mark scores 0.5\% worse than MARF \textbf{\textcolor[RGB]{204,0,0}{red}} and scores 0.5\% better \textbf{\textcolor[RGB]{0,136,0}{green}}. \(\nabla\) denotes analytical normals (Eq.~\ref{eq:oriented-ray-field-normal}) and \(\mathcal M\) denotes medial normals (Eq.~\ref{eq:marf-normal}).}\label{tbl:ablation}

\begin{longtable}[]{@{}llcccccc@{}}
\toprule
\multicolumn{2}{l}{\multirow{2}{*}{Configuration}} & \multirow{2}{*}{IoU\(\smash\uparrow\)} & \multirow{2}{*}{P\(\smash\uparrow\)} & \multirow{2}{*}{R\(\smash\uparrow\)} & CD\(\smash\downarrow\) & \multicolumn{2}{c}{COS\(\smash\uparrow\)}\\ \cmidrule(lr){7-8}
& & & & & {\(\smash\times 10^4\)} & {\(\nabla\)} & {\(\mathcal M\)} \\
\midrule
\endhead
MARF & Table~\ref{tbl:loss_hparams} & {92.6\%} & {95.1\%} & {97.2\%} & {2.56} & {0.780} & {0.746} \\
\midrule LFN {[}\protect\hyperlink{ref-sitzmann2021lfns}{63}{]} encoding & \(\mathbf x\smash=(\hat{\mathbf q},\mathbf m)\) & {92.2\%} & \textbf{\textcolor[RGB]{204,0,0}{94.5\%}} & {97.4\%} & \textbf{\textcolor[RGB]{204,0,0}{2.81}} & \textbf{\textcolor[RGB]{204,0,0}{0.761}} & \textbf{\textcolor[RGB]{204,0,0}{0.724}} \\
PRIF {[}\protect\hyperlink{ref-fengPRIFPrimaryRaybased2022}{69}{]} encoding & \(\mathbf x\smash=(\hat{\mathbf q},\mathbf o_\perp)\) & \textbf{\textcolor[RGB]{204,0,0}{92.1\%}} & {94.7\%} & {97.1\%} & \textbf{\textcolor[RGB]{204,0,0}{2.70}} & \textbf{\textcolor[RGB]{204,0,0}{0.771}} & \textbf{\textcolor[RGB]{204,0,0}{0.737}} \\
No init scheme. & Section~\ref{sec:neural-network} & \textbf{\textcolor[RGB]{204,0,0}{91.8\%}} & {94.8\%} & \textbf{\textcolor[RGB]{204,0,0}{96.7\%}} & \textbf{\textcolor[RGB]{204,0,0}{2.79}} & \textbf{\textcolor[RGB]{204,0,0}{0.763}} & \textbf{\textcolor[RGB]{204,0,0}{0.724}} \\
\midrule 1 atom candidate & & \textbf{\textcolor[RGB]{204,0,0}{87.4\%}} & {95.2\%} & \textbf{\textcolor[RGB]{204,0,0}{91.5\%}} & \textbf{\textcolor[RGB]{204,0,0}{4.54}} & \textbf{\textcolor[RGB]{204,0,0}{0.720}} & \textbf{\textcolor[RGB]{204,0,0}{0.679}} \\
4 atom candidates & & \textbf{\textcolor[RGB]{204,0,0}{90.7\%}} & {95.0\%} & \textbf{\textcolor[RGB]{204,0,0}{95.2\%}} & \textbf{\textcolor[RGB]{204,0,0}{3.36}} & \textbf{\textcolor[RGB]{204,0,0}{0.761}} & \textbf{\textcolor[RGB]{204,0,0}{0.722}} \\
8 atom candidates & & \textbf{\textcolor[RGB]{204,0,0}{91.9\%}} & {95.1\%} & \textbf{\textcolor[RGB]{204,0,0}{96.4\%}} & \textbf{\textcolor[RGB]{204,0,0}{2.67}} & \textbf{\textcolor[RGB]{204,0,0}{0.770}} & \textbf{\textcolor[RGB]{204,0,0}{0.736}} \\
32 atom candidates & & {92.7\%} & {94.7\%} & {97.7\%} & \textbf{\textcolor[RGB]{204,0,0}{2.60}} & {0.778} & {0.749} \\
64 atom candidates & & {92.7\%} & \textbf{\textcolor[RGB]{204,0,0}{94.6\%}} & \textbf{\textcolor[RGB]{0,136,0}{97.8\%}} & \textbf{\textcolor[RGB]{0,136,0}{2.39}} & {0.778} & {0.747} \\
\midrule No intersection loss & \(0\lambda_{\mathbf p}\) & \textbf{\textcolor[RGB]{204,0,0}{91.3\%}} & \textbf{\textcolor[RGB]{204,0,0}{93.9\%}} & {97.0\%} & \textbf{\textcolor[RGB]{204,0,0}{12.29}} & \textbf{\textcolor[RGB]{204,0,0}{0.546}} & \textbf{\textcolor[RGB]{204,0,0}{0.528}} \\
No silhouette loss & \(0\lambda_{s}\ 0\lambda_{h}\) & \textbf{\textcolor[RGB]{204,0,0}{87.6\%}} & \textbf{\textcolor[RGB]{204,0,0}{90.5\%}} & \textbf{\textcolor[RGB]{204,0,0}{96.5\%}} & \textbf{\textcolor[RGB]{204,0,0}{3.69}} & \textbf{\textcolor[RGB]{204,0,0}{0.744}} & \textbf{\textcolor[RGB]{204,0,0}{0.709}} \\
More silhouette loss & \(5\lambda_{s}\ 5\lambda_{h}\) & \textbf{\textcolor[RGB]{0,136,0}{93.4\%}} & \textbf{\textcolor[RGB]{0,136,0}{96.2\%}} & {97.0\%} & \textbf{\textcolor[RGB]{0,136,0}{2.45}} & \textbf{\textcolor[RGB]{204,0,0}{0.772}} & \textbf{\textcolor[RGB]{204,0,0}{0.738}} \\
No normal loss & \(0\lambda_{\mathbf n}\) & {92.7\%} & \textbf{\textcolor[RGB]{0,136,0}{96.2\%}} & \textbf{\textcolor[RGB]{204,0,0}{96.3\%}} & \textbf{\textcolor[RGB]{0,136,0}{2.15}} & {0.782} & \textbf{\textcolor[RGB]{204,0,0}{0.725}} \\
No inscription loss & \(0\lambda_{ih}\ 0\lambda_{im}\) & {92.3\%} & \textbf{\textcolor[RGB]{204,0,0}{93.7\%}} & \textbf{\textcolor[RGB]{0,136,0}{98.4\%}} & \textbf{\textcolor[RGB]{204,0,0}{2.67}} & {0.778} & {0.747} \\
More inscription loss & \(5\lambda_{ih}\ 5\lambda_{im}\) & {92.3\%} & {95.5\%} & \textbf{\textcolor[RGB]{204,0,0}{96.5\%}} & \textbf{\textcolor[RGB]{204,0,0}{2.62}} & {0.776} & \textbf{\textcolor[RGB]{204,0,0}{0.742}} \\
No maximality reg. & \(0\lambda_{r}\) & {92.4\%} & {95.0\%} & {97.1\%} & {2.56} & {0.776} & {0.744} \\
More maximality reg. & \(100\lambda_{r}\) & {92.5\%} & \textbf{\textcolor[RGB]{204,0,0}{94.5\%}} & \textbf{\textcolor[RGB]{0,136,0}{97.8\%}} & \textbf{\textcolor[RGB]{0,136,0}{2.54}} & {0.781} & {0.748} \\
No specialization reg. & \(0\lambda_{\sigma}\) & {92.5\%} & {95.0\%} & {97.3\%} & \textbf{\textcolor[RGB]{204,0,0}{2.58}} & \textbf{\textcolor[RGB]{204,0,0}{0.776}} & {0.743} \\
No multi-view loss & \(0\lambda_{\text{mv}}\) & \textbf{\textcolor[RGB]{204,0,0}{91.3\%}} & \textbf{\textcolor[RGB]{204,0,0}{94.6\%}} & \textbf{\textcolor[RGB]{204,0,0}{96.3\%}} & \textbf{\textcolor[RGB]{204,0,0}{2.84}} & \textbf{\textcolor[RGB]{204,0,0}{0.755}} & \textbf{\textcolor[RGB]{204,0,0}{0.720}} \\
More multi-view loss & \(2\lambda_{\text{mv}}\) & {92.3\%} & {94.8\%} & {97.2\%} & \textbf{\textcolor[RGB]{204,0,0}{2.66}} & \textbf{\textcolor[RGB]{204,0,0}{0.769}} & \textbf{\textcolor[RGB]{204,0,0}{0.734}} \\
\bottomrule
\end{longtable}

} \end{table}

\hypertarget{sec:exp-rendering}{%
\subsection{Applications in Visualization}\label{sec:exp-rendering}}

In this section we showcase two real-time applications in visualization made possible due to the medial quantities predicted by MARFs.

\hypertarget{translucency.}{%
\paragraph*{Translucency.}\label{translucency.}}
\addcontentsline{toc}{paragraph}{Translucency.}

Light traveling inside translucent objects attenuates and scatters rapidly. How this phenomenon appears on the surface is commonly approximated using some measure of local thickness, for which the medial radius predicted by MARFs, also known as the local feature size, is an excellent candidate. We showcase in Fig.~\ref{fig:sss} approximate translucency, using the shading model of Barré-Brisebois et al. {[}\protect\hyperlink{ref-barre-briseboisApproximatingTranslucencyFast2011}{102}{]}. It contributes the following shading coefficient at each point \(\mathbf p_\ell\) on surface \(\partial\mathcal O\):

\begin{equation}\protect\hypertarget{eq:sss}{}{
  k_\text{translucency} =
    \frac{
      1
    }{
      r_\ell + \epsilon
    }
    \cdot
    \operatorname{max}\left(
      \hat{\mathbf q}
      \cdot
      \left(
        s\hat{\mathbf n}_\ell
        -
        \hat{\mathbf l}
      \right),
      0
    \right)^p
}\label{eq:sss}\end{equation}

\noindent where \(r_\ell\) is the thickness (medial radius) at \(\mathbf p_\ell\), \(\epsilon=0.05\) avoids division by zero, \(\hat{\mathbf q}\) is the unit ray direction (Eq.~\ref{eq:ell}), \(s=0.08\) is a distortion determining the amount of subsurface scattering, \(\mathbf n_\ell\) is the medial normal (as per Eq.~\ref{eq:marf-normal}), \(\hat{\mathbf l}\) is the incident light unit vector, and \(p=16\) is a sharpness coefficient.

\begin{pandoccrossrefsubfigures}

\subfloat[Medial radius thickness]{\includegraphics[width=0.24\textwidth,height=\textheight]{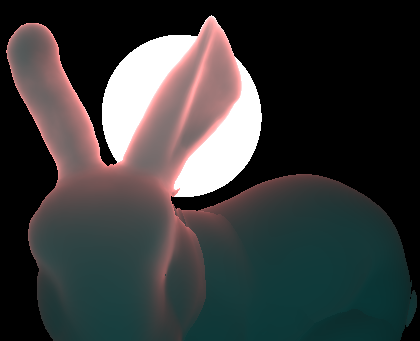}\label{fig:sss-medial}} \subfloat[Uniform thickness]{\includegraphics[width=0.24\textwidth,height=\textheight]{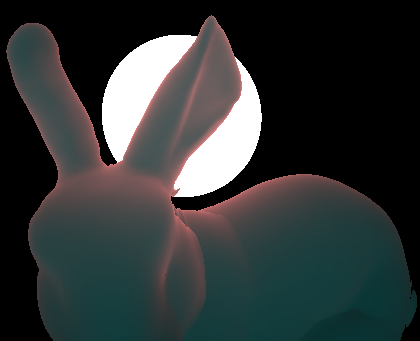}\label{fig:sss-uniform}}

\caption[{Approximate translucency and subsurface scattering. In \textsc{(a)} we use the medial radius (shown in Fig.~\ref{fig:bunny-shadings-medial-radii}) as a measure of local thickness, while \textsc{(b)} assumes uniform thickness, here set to the mean medial radius for a fair comparison. We render each pixel independently using only a single network evaluation and no differentiation.}]{Approximate translucency and subsurface scattering. In \textsc{(a)} we use the medial radius (shown in Fig.~\ref{fig:bunny-shadings-medial-radii}) as a measure of local thickness, while \textsc{(b)} assumes uniform thickness, here set to the mean medial radius for a fair comparison. We render each pixel independently using only a single network evaluation and no differentiation.}

\label{fig:sss}

\end{pandoccrossrefsubfigures}

\hypertarget{anisotrophy.}{%
\paragraph*{Anisotrophy.}\label{anisotrophy.}}
\addcontentsline{toc}{paragraph}{Anisotrophy.}

To show how we can compute the full shape operator \(\mathcal D\hat{\mathbf n}_\ell\) (from Eq.~\ref{eq:shape-operator}) of a MARF using only a \emph{single} network differentiation, we shade in Fig.~\ref{fig:anisotropic-bunny} a MARF with the anisotropic specular reflectance model of Ward {[}\protect\hyperlink{ref-ward1992measuring}{103}{]}. Anisotropic materials feature view-dependent properties, in our case reflectance, whose distribution Ward determines using two perpendicular surface tangents. For these a good fit are the principal directions of curvature \(\hat{\mathbf v}_1\) and \(\hat{\mathbf v}_2\), the eigenvectors of \(\mathcal D\hat{\mathbf n}_\ell\). The Ward model contributes the following specular coefficient at each point \(\mathbf p_\ell\) on surface \(\partial O\):

\begin{equation}\protect\hypertarget{eq:ward-anisotropic-spec}{}{
  k_\text{specular} =
  \frac{
      \exp\left(\vcenter{\hbox{\math
        -2\dfrac{
          \left( \frac{\hat{\mathbf h}\cdot\hat{\mathbf v}_1}{a_1} \right)^2
          +
          \left( \frac{\hat{\mathbf h}\cdot\hat{\mathbf v}_2}{a_2} \right)^2
        }{
            1+\hat{\mathbf n}_\ell \cdot \hat{\mathbf h}
        }
      \endmath}}\right)
  }{
    4\pi
    a_1a_2
    \sqrt{
      ( \hat{\mathbf n}_\ell\cdot\hat{\mathbf l})
      ( \hat{\mathbf n}_\ell\cdot\hat{\mathbf q})
    }
  }
}\label{eq:ward-anisotropic-spec}\end{equation}

\noindent where \(\hat{\mathbf n}_\ell\) is the medial normal from Eq.~\ref{eq:marf-normal}, \(\hat{\mathbf l}\) is the incident light unit vector, \(\hat{\mathbf h}=(\hat{\mathbf l}+\hat{\mathbf q})/\|\hat{\mathbf l}+\hat{\mathbf q}\|\), and \(a_1\) and \(a_2\) are the standard deviations of anisotropy along principal directions of curvature \(\hat{\mathbf v}_1\) and \(\hat{\mathbf v}_2\).

\begin{pandoccrossrefsubfigures}

\subfloat[\(a_1\smash=0.05,\ \ a_2\smash=0.3\)]{\includegraphics[width=0.2\textwidth,height=\textheight]{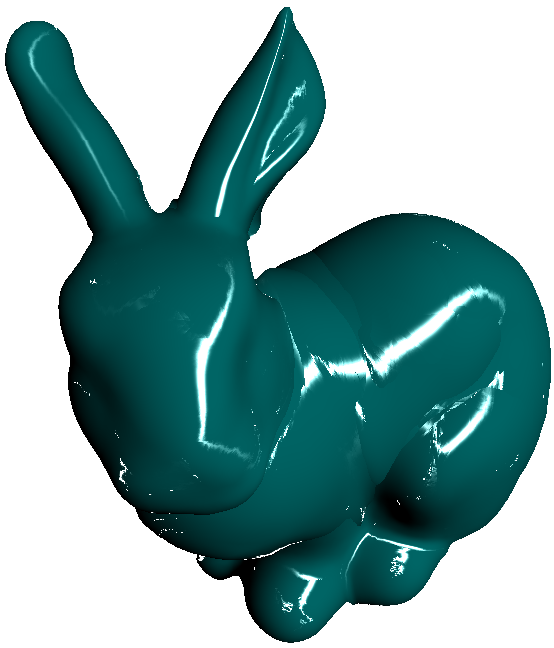}\label{fig:anisotropic-bunny-max}} \subfloat[\(a_1\smash=0.3,\ \ a_2\smash=0.05\)]{\includegraphics[width=0.2\textwidth,height=\textheight]{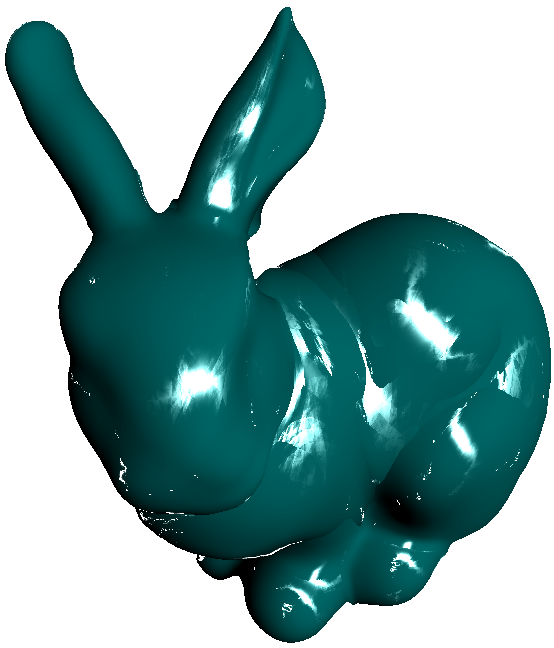}\label{fig:anisotropic-bunny-min}}

\caption[{A Stanford bunny MARF shaded with Ward anisotropic specular reflectance {[}\protect\hyperlink{ref-ward1992measuring}{103}{]} determined by principal directions of curvature. We render each pixel independently with a single forward and backward pass. \(a_1\) and \(a_2\) (see Eq.~\ref{eq:ward-anisotropic-spec}) determine the anisotropic deviation along each principal direction.}]{A Stanford bunny MARF shaded with Ward anisotropic specular reflectance {[}\protect\hyperlink{ref-ward1992measuring}{103}{]} determined by principal directions of curvature. We render each pixel independently with a single forward and backward pass. \(a_1\) and \(a_2\) (see Eq.~\ref{eq:ward-anisotropic-spec}) determine the anisotropic deviation along each principal direction.}

\label{fig:anisotropic-bunny}

\end{pandoccrossrefsubfigures}

\hypertarget{sec:exp-multiple-shape}{%
\subsection{Learning Multiple Shapes}\label{sec:exp-multiple-shape}}

Here we examine a MARF trained to represent multiple shapes. We visualize in Fig.~\ref{fig:coseg-four-legged-interpolations} MARF reconstructions of the \emph{four-legged} COSEG {[}\protect\hyperlink{ref-wangActiveCoanalysisSet2012}{93}{]} object class, reconstructed from learned auto-decoding latent vectors in \(\mathbb R^{16}\). We also show, to demonstrate how smooth the latent space is, in-between interpolations in latent space. This MARF has a CD score of \(2.424\smash\times 10^{-4}\), an analytical COS score of 0.868, and a medial COS score of 0.843, and a 90.9\% IoU score.

MARFs proves able to represent a space of multiple species with different articulations with a consistent part segmentation. The latent space appears smooth despite a sparse training set, with meaningful interpolations. On some in-betweens the predicted atoms fail to intersect the ray, visible on the legs leg of the giraffe-dromedary interpolation and on the dogs. This should improve with more training shapes.

\begin{widefig}

\begin{figure}
\hypertarget{fig:coseg-four-legged-interpolations}{%
\centering
\includegraphics{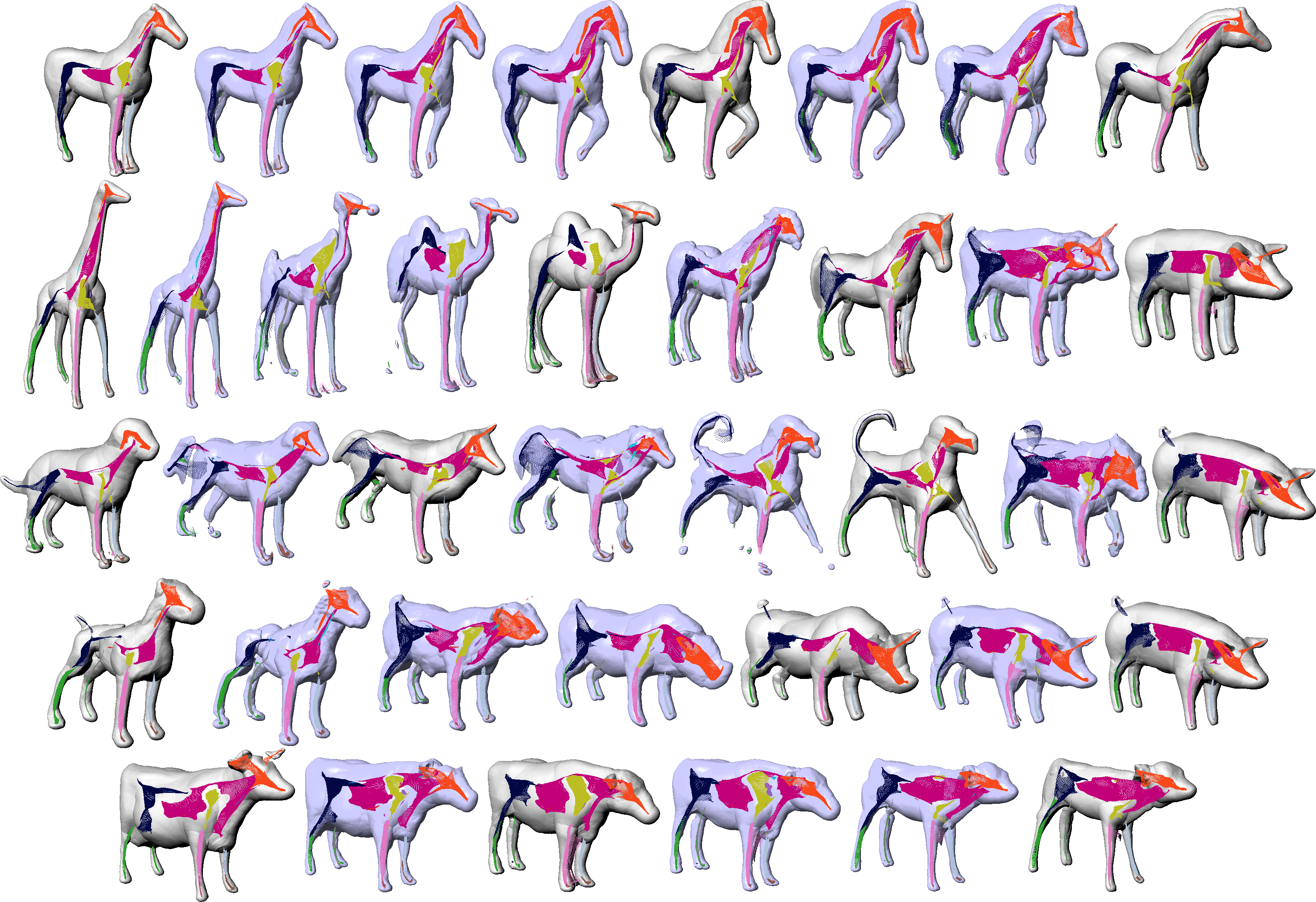}
\caption{Row-wise MARF interpolations in latent space. We illustrate the medial axis by superimposing the medial atom centers associated with each hitting camera ray on top of a Lambertian shading using analytical normals (Eq.~\ref{eq:oriented-ray-field-normal}). We trained this MARF on the COSEG {[}\protect\hyperlink{ref-wangActiveCoanalysisSet2012}{93}{]} ``four-legged'' object class, featuring a total of 20 shapes. Gray renders use known auto-decoder latent vectors, while the blue-tinted renders are in-betweens interpolations. Despite the sparse dataset, we find the MARF propose smooth and meaningful in-betweens.}\label{fig:coseg-four-legged-interpolations}
}
\end{figure}

\end{widefig}

\hypertarget{sec:conclusion}{%
\section{Conclusion}\label{sec:conclusion}}

The novel 3D object representation MARF is a neural ray-to-surface mapping that outperforms prior work, achieving accurate surface rendering with a single network evaluation per camera ray. The geometrically grounded medial representation of MARFs offers more insight while benefiting reconstruction quality, multi-view consistency, and representing discontinuities. We demonstrated how its medial quantities can be used in visualization and inform part-based segmentation. While learning ray-fields remains a difficult problem, we find our results exciting, warranting further study.

\hypertarget{limitations.}{%
\paragraph*{Limitations.}\label{limitations.}}
\addcontentsline{toc}{paragraph}{Limitations.}

Like prior neural ray fields, MARFs assume the camera ray is cast from infinitely far away. This makes rendering views where the camera is placed in-between occluders, such as overhangs, impossible. While this does not affect inter-object scatter rays if adapted to a global-illumination ray-tracing setup, it will affect intra-object bounces whose contribution to illumination must be learned/baked.

\hypertarget{future-work.}{%
\paragraph*{Future work.}\label{future-work.}}
\addcontentsline{toc}{paragraph}{Future work.}

There are many challenges to address concerning both MARFs and neural ray fields in general. Ray fields lack an analog to both positional encoding and local conditioning common in Cartesian neural fields, which drastically improve their fidelity. Our proposed multi-view loss requires 3D supervision, in turn requiring two forward passes if adapted to 2D data. For MARFs in particular, we look to explore less naive candidate selection strategies that select the atom candidate best suited to receive supervision, which is not necessarily the one closest to the ray. We would further like to explore alternatives to fully opaque atoms such that no atom are fully occluded from supervision. Finally, work is needed to reduce the number of MARF loss terms, reducing the effort required to balance their contribution.

\hypertarget{biblography}{%
\section*{Biblography}\label{biblography}}
\addcontentsline{toc}{section}{Biblography}

\small

\ifx\printbibliography\undefined

\hypertarget{refs}{}
\begin{CSLReferences}{0}{0}
\leavevmode\vadjust pre{\hypertarget{ref-bartlettSpectrallynormalizedMarginBounds2017}{}}%
\CSLLeftMargin{{[}1{]} }
\CSLRightInline{Bartlett PL, Foster DJ, Telgarsky M. Spectrally-normalized margin bounds for neural networks. Proceedings of the 31st {International Conference} on {Neural Information Processing Systems}, {Red Hook, NY, USA}: {Curran Associates Inc.}; 2017, p. 6241--50.}

\leavevmode\vadjust pre{\hypertarget{ref-rahamanSpectralBiasNeural2019}{}}%
\CSLLeftMargin{{[}2{]} }
\CSLRightInline{Rahaman N, Baratin A, Arpit D, Draxler F, Lin M, Hamprecht F, et al. \href{https://proceedings.mlr.press/v97/rahaman19a.html}{On the spectral bias of neural networks}. In: Chaudhuri K, Salakhutdinov R, editors. Proceedings of the 36th international conference on machine learning, vol. 97, {PMLR}; 2019, p. 5301--10.}

\leavevmode\vadjust pre{\hypertarget{ref-meschederOccupancyNetworksLearning2019}{}}%
\CSLLeftMargin{{[}3{]} }
\CSLRightInline{Mescheder L, Oechsle M, Niemeyer M, Nowozin S, Geiger A. Occupancy {Networks}: {Learning 3D Reconstruction} in {Function Space}. 2019 {IEEE}/{CVF Conference} on {Computer Vision} and {Pattern Recognition} ({CVPR}), {Long Beach, CA, USA}: {IEEE}; 2019, p. 4455--65. \url{https://doi.org/10.1109/CVPR.2019.00459}.}

\leavevmode\vadjust pre{\hypertarget{ref-chenLearningImplicitFields2019a}{}}%
\CSLLeftMargin{{[}4{]} }
\CSLRightInline{Chen Z, Zhang H. \href{http://arxiv.org/abs/1812.02822}{Learning {Implicit Fields} for {Generative Shape Modeling}} 2019.}

\leavevmode\vadjust pre{\hypertarget{ref-parkDeepSDFLearningContinuous2019}{}}%
\CSLLeftMargin{{[}5{]} }
\CSLRightInline{Park JJ, Florence P, Straub J, Newcombe R, Lovegrove S. {DeepSDF}: {Learning Continuous Signed Distance Functions} for {Shape Representation}. 2019 {IEEE}/{CVF Conference} on {Computer Vision} and {Pattern Recognition} ({CVPR}), {Long Beach, CA, USA}: {IEEE}; 2019, p. 165--74. \url{https://doi.org/10.1109/CVPR.2019.00025}.}

\leavevmode\vadjust pre{\hypertarget{ref-chibaneNeuralUnsignedDistance2020}{}}%
\CSLLeftMargin{{[}6{]} }
\CSLRightInline{Chibane J, Mir A, Pons-Moll G. Neural unsigned distance fields for implicit function learning. Advances in neural information processing systems ({NeurIPS}), 2020.}

\leavevmode\vadjust pre{\hypertarget{ref-venkateshDeepImplicitSurface2021}{}}%
\CSLLeftMargin{{[}7{]} }
\CSLRightInline{Venkatesh R, Karmali T, Sharma S, Ghosh A, Babu RV, Jeni LA, et al. Deep implicit surface point prediction networks. Proceedings of the {IEEE}/{CVF International Conference} on {Computer Vision}, 2021, p. 12653--62.}

\leavevmode\vadjust pre{\hypertarget{ref-chiGarmentNetsCategorylevelPose2021}{}}%
\CSLLeftMargin{{[}8{]} }
\CSLRightInline{Chi C, Song S. {GarmentNets}: {Category-level} pose estimation for garments via canonical space shape completion. The {IEEE} international conference on computer vision ({ICCV}), 2021, p. 10.}

\leavevmode\vadjust pre{\hypertarget{ref-rebainDeepMedialFields2021}{}}%
\CSLLeftMargin{{[}9{]} }
\CSLRightInline{Rebain D, Li K, Sitzmann V, Yazdani S, Yi KM, Tagliasacchi A. \href{http://arxiv.org/abs/2106.03804}{Deep {Medial Fields}} 2021.}

\leavevmode\vadjust pre{\hypertarget{ref-simeonovNeuralDescriptorFields2021}{}}%
\CSLLeftMargin{{[}10{]} }
\CSLRightInline{Simeonov A, Du Y, Tagliasacchi A, Tenenbaum JB, Rodriguez A, Agrawal P, et al. \href{http://arxiv.org/abs/2112.05124}{Neural {Descriptor Fields}: {SE}(3)-{Equivariant Object Representations} for {Manipulation}} 2021.}

\leavevmode\vadjust pre{\hypertarget{ref-frankleLotteryTicketHypothesis2019}{}}%
\CSLLeftMargin{{[}11{]} }
\CSLRightInline{Frankle J, Carbin M. \href{https://openreview.net/forum?id=rJl-b3RcF7}{The {Lottery Ticket Hypothesis}: {Finding Sparse}, {Trainable Neural Networks}}. International {Conference} on {Learning Representations}, 2019.}

\leavevmode\vadjust pre{\hypertarget{ref-xieNeuralFieldsVisual2021}{}}%
\CSLLeftMargin{{[}12{]} }
\CSLRightInline{Xie Y, Takikawa T, Saito S, Litany O, Yan S, Khan N, et al. \href{http://arxiv.org/abs/2111.11426}{Neural {Fields} in {Visual Computing} and {Beyond}} 2021.}

\leavevmode\vadjust pre{\hypertarget{ref-atzmonSALSignAgnostic2020}{}}%
\CSLLeftMargin{{[}13{]} }
\CSLRightInline{Atzmon M, Lipman Y. \href{http://arxiv.org/abs/1911.10414}{{SAL}: {Sign Agnostic Learning} of {Shapes} from {Raw Data}} 2020.}

\leavevmode\vadjust pre{\hypertarget{ref-baydinAutomaticDifferentiationMachine2018}{}}%
\CSLLeftMargin{{[}14{]} }
\CSLRightInline{Baydin AG, Pearlmutter BA, Radul AA, Siskind JM. Automatic {Diﬀerentiation} in {Machine Learning}: A {Survey}. Journal of Marchine Learning Research 2018;18:1--43.}

\leavevmode\vadjust pre{\hypertarget{ref-raissiPhysicsinformedNeuralNetworks2019}{}}%
\CSLLeftMargin{{[}15{]} }
\CSLRightInline{Raissi M, Perdikaris P, Karniadakis GE. Physics-informed neural networks: {A} deep learning framework for solving forward and inverse problems involving nonlinear partial differential equations. Journal of Computational Physics 2019;378:686--707. \url{https://doi.org/10.1016/j.jcp.2018.10.045}.}

\leavevmode\vadjust pre{\hypertarget{ref-groppImplicitGeometricRegularization2020}{}}%
\CSLLeftMargin{{[}16{]} }
\CSLRightInline{Gropp A, Yariv L, Haim N, Atzmon M, Lipman Y. \href{http://arxiv.org/abs/2002.10099}{Implicit {Geometric Regularization} for {Learning Shapes}} 2020.}

\leavevmode\vadjust pre{\hypertarget{ref-sitzmannImplicitNeuralRepresentations2020}{}}%
\CSLLeftMargin{{[}17{]} }
\CSLRightInline{Sitzmann V, Martel JNP, Bergman AW, Lindell DB, Wetzstein G. Implicit {Neural Representations} with {Periodic Activation Functions}. Proc. {NeurIPS}, 2020.}

\leavevmode\vadjust pre{\hypertarget{ref-lindellAutoIntAutomaticIntegration2021}{}}%
\CSLLeftMargin{{[}18{]} }
\CSLRightInline{Lindell DB, Martel JNP, Wetzstein G. {AutoInt}: {Automatic Integration} for {Fast Neural Volume Rendering}. Proceedings of the conference on computer vision and pattern recognition ({CVPR}), {Nashville, TN, USA}: {IEEE}; 2021, p. 14551--60. \url{https://doi.org/10.1109/CVPR46437.2021.01432}.}

\leavevmode\vadjust pre{\hypertarget{ref-yangGeometryProcessingNeural2021}{}}%
\CSLLeftMargin{{[}19{]} }
\CSLRightInline{Yang G, Belongie S, Hariharan B, Koltun V. \href{https://proceedings.neurips.cc/paper/2021/hash/bd686fd640be98efaae0091fa301e613-Abstract.html}{Geometry {Processing} with {Neural Fields}}. Advances in {Neural Information Processing Systems}, vol. 34, {Curran Associates, Inc.}; 2021, p. 22483--97.}

\leavevmode\vadjust pre{\hypertarget{ref-novelloExploringDifferentialGeometry2022}{}}%
\CSLLeftMargin{{[}20{]} }
\CSLRightInline{Novello T, Schardong G, Schirmer L, da Silva V, Lopes H, Velho L. Exploring differential geometry in neural implicits. Computers \& Graphics 2022;108:49--60. \url{https://doi.org/10.1016/j.cag.2022.09.003}.}

\leavevmode\vadjust pre{\hypertarget{ref-mildenhallNeRFRepresentingScenes2020a}{}}%
\CSLLeftMargin{{[}21{]} }
\CSLRightInline{Mildenhall B, Srinivasan PP, Tancik M, Barron JT, Ramamoorthi R, Ng R. \href{http://arxiv.org/abs/2003.08934}{{NeRF}: {Representing Scenes} as {Neural Radiance Fields} for {View Synthesis}} 2020.}

\leavevmode\vadjust pre{\hypertarget{ref-tewariStateArtNeural2020}{}}%
\CSLLeftMargin{{[}22{]} }
\CSLRightInline{Tewari A, Fried O, Thies J, Sitzmann V, Lombardi S, Sunkavalli K, et al. \href{http://arxiv.org/abs/2004.03805}{State of the {Art} on {Neural Rendering}} 2020.}

\leavevmode\vadjust pre{\hypertarget{ref-tewariAdvancesNeuralRendering2021}{}}%
\CSLLeftMargin{{[}23{]} }
\CSLRightInline{Tewari A, Thies J, Mildenhall B, Srinivasan P, Tretschk E, Wang Y, et al. \href{http://arxiv.org/abs/2111.05849}{Advances in {Neural Rendering}} 2021.}

\leavevmode\vadjust pre{\hypertarget{ref-mildenhallNeRFDarkHigh2022}{}}%
\CSLLeftMargin{{[}24{]} }
\CSLRightInline{Mildenhall B, Hedman P, Martin-Brualla R, Srinivasan PP, Barron JT. {NeRF} in the {Dark}: {High Dynamic Range View Synthesis} from {Noisy Raw Images}. 2022 {IEEE}/{CVF Conference} on {Computer Vision} and {Pattern Recognition} ({CVPR}), {New Orleans, LA, USA}: {IEEE}; 2022, p. 16169--78. \url{https://doi.org/10.1109/CVPR52688.2022.01571}.}

\leavevmode\vadjust pre{\hypertarget{ref-mildenhallLocalLightField2019}{}}%
\CSLLeftMargin{{[}25{]} }
\CSLRightInline{Mildenhall B, Srinivasan PP, Ortiz-Cayon R, Kalantari NK, Ramamoorthi R, Ng R, et al. Local light field fusion: Practical view synthesis with prescriptive sampling guidelines. ACM Trans Graph 2019;38:1--4. \url{https://doi.org/10.1145/3306346.3322980}.}

\leavevmode\vadjust pre{\hypertarget{ref-baatzNeRFTexNeuralReflectance2021}{}}%
\CSLLeftMargin{{[}26{]} }
\CSLRightInline{Baatz H, Granskog J, Papas M, Rousselle F, Novák J. {NeRF-Tex}: {Neural Reflectance Field Textures}. Eurographics {Symposium} on {Rendering}, {The Eurographics Association}; 2021, p. 13.}

\leavevmode\vadjust pre{\hypertarget{ref-goliNerf2nerfPairwiseRegistration2022}{}}%
\CSLLeftMargin{{[}27{]} }
\CSLRightInline{Goli L, Rebain D, Sabour S, Garg A, Tagliasacchi A. Nerf2nerf: {Pairwise Registration} of {Neural Radiance Fields} 2022. \url{https://doi.org/10.48550/arXiv.2211.01600}.}

\leavevmode\vadjust pre{\hypertarget{ref-guoObjectCentricNeuralScene2020}{}}%
\CSLLeftMargin{{[}28{]} }
\CSLRightInline{Guo M, Fathi A, Wu J, Funkhouser T. \href{http://arxiv.org/abs/2012.08503}{Object-{Centric Neural Scene Rendering}} 2020.}

\leavevmode\vadjust pre{\hypertarget{ref-parkHyperNeRFHigherDimensionalRepresentation2021}{}}%
\CSLLeftMargin{{[}29{]} }
\CSLRightInline{Park K, Sinha U, Hedman P, Barron JT, Bouaziz S, Goldman DB, et al. \href{http://arxiv.org/abs/2106.13228}{{HyperNeRF}: {A Higher-Dimensional Representation} for {Topologically Varying Neural Radiance Fields}} 2021.}

\leavevmode\vadjust pre{\hypertarget{ref-chenAnimatableNeuralRadiance2021}{}}%
\CSLLeftMargin{{[}30{]} }
\CSLRightInline{Chen J, Zhang Y, Kang D, Zhe X, Bao L, Jia X, et al. \href{http://arxiv.org/abs/2106.13629}{Animatable {Neural Radiance Fields} from {Monocular RGB Videos}} 2021.}

\leavevmode\vadjust pre{\hypertarget{ref-pengAnimatableNeuralRadiance2021}{}}%
\CSLLeftMargin{{[}31{]} }
\CSLRightInline{Peng S, Dong J, Wang Q, Zhang S, Shuai Q, Zhou X, et al. Animatable neural radiance fields for modeling dynamic human bodies. {ICCV}, 2021.}

\leavevmode\vadjust pre{\hypertarget{ref-tschernezkiNeuralDiffSegmenting3D2021}{}}%
\CSLLeftMargin{{[}32{]} }
\CSLRightInline{Tschernezki V, Larlus D, Vedaldi A. {NeuralDiff}: {Segmenting 3D} objects that move in egocentric videos. Proceedings of the international conference on {3D} vision ({3DV}), 2021.}

\leavevmode\vadjust pre{\hypertarget{ref-maxOpticalModelsDirect1995}{}}%
\CSLLeftMargin{{[}33{]} }
\CSLRightInline{Max N. Optical models for direct volume rendering. IEEE Transactions on Visualization and Computer Graphics 1995;1:99--108. \url{https://doi.org/10.1109/2945.468400}.}

\leavevmode\vadjust pre{\hypertarget{ref-hartSphereTracingGeometric1996}{}}%
\CSLLeftMargin{{[}34{]} }
\CSLRightInline{Hart JC. Sphere tracing: {A} geometric method for the antialiased ray tracing of implicit surfaces. Visual Computer 1996;12:527--45. \url{https://doi.org/10.1007/s003710050084}.}

\leavevmode\vadjust pre{\hypertarget{ref-knodtNeuralRayTracingLearning2021}{}}%
\CSLLeftMargin{{[}35{]} }
\CSLRightInline{Knodt J, Baek S-H, Heide F. \href{http://arxiv.org/abs/2104.13562}{Neural {Ray-Tracing}: {Learning Surfaces} and {Reflectance} for {Relighting} and {View Synthesis}} 2021.}

\leavevmode\vadjust pre{\hypertarget{ref-chibaneImplicitFunctionsFeature2020}{}}%
\CSLLeftMargin{{[}36{]} }
\CSLRightInline{Chibane J, Alldieck T, Pons-Moll G. Implicit {Functions} in {Feature Space} for {3D Shape Reconstruction} and {Completion}. 2020 {IEEE}/{CVF Conference} on {Computer Vision} and {Pattern Recognition} ({CVPR}), {Seattle, WA, USA}: {IEEE}; 2020, p. 6968--79. \url{https://doi.org/10.1109/CVPR42600.2020.00700}.}

\leavevmode\vadjust pre{\hypertarget{ref-genovaLearningShapeTemplates2019}{}}%
\CSLLeftMargin{{[}37{]} }
\CSLRightInline{Genova K, Cole F, Vlasic D, Sarna A, Freeman WT, Funkhouser T. \href{http://arxiv.org/abs/1904.06447}{Learning {Shape Templates} with {Structured Implicit Functions}} 2019.}

\leavevmode\vadjust pre{\hypertarget{ref-jiangLocalImplicitGrid2020}{}}%
\CSLLeftMargin{{[}38{]} }
\CSLRightInline{Jiang CM, Sud A, Makadia A, Huang J, Nießner M, Funkhouser T. \href{http://arxiv.org/abs/2003.08981}{Local {Implicit Grid Representations} for {3D Scenes}} 2020.}

\leavevmode\vadjust pre{\hypertarget{ref-chabraDeepLocalShapes2020}{}}%
\CSLLeftMargin{{[}39{]} }
\CSLRightInline{Chabra R, Lenssen JE, Ilg E, Schmidt T, Straub J, Lovegrove S, et al. Deep {Local Shapes}: {Learning Local SDF Priors} for {Detailed 3D Reconstruction} 2020.}

\leavevmode\vadjust pre{\hypertarget{ref-tretschkPatchNetsPatchBasedGeneralizable2020}{}}%
\CSLLeftMargin{{[}40{]} }
\CSLRightInline{Tretschk E, Tewari A, Golyanik V, Zollhöfer M, Stoll C, Theobalt C. {PatchNets}: {Patch-Based Generalizable Deep Implicit 3D Shape Representations}. In: Vedaldi A, Bischof H, Brox T, Frahm J-M, editors. Computer {Vision} -- {ECCV} 2020, vol. 12361, {Cham}: {Springer International Publishing}; 2020, p. 293--309. \url{https://doi.org/10.1007/978-3-030-58517-4_18}.}

\leavevmode\vadjust pre{\hypertarget{ref-reiserKiloNeRFSpeedingNeural2021}{}}%
\CSLLeftMargin{{[}41{]} }
\CSLRightInline{Reiser C, Peng S, Liao Y, Geiger A. {KiloNeRF}: {Speeding} up neural radiance fields with thousands of tiny {MLPs}. International conference on computer vision ({ICCV}), 2021.}

\leavevmode\vadjust pre{\hypertarget{ref-rebainDeRFDecomposedRadiance2021}{}}%
\CSLLeftMargin{{[}42{]} }
\CSLRightInline{Rebain D, Jiang W, Yazdani S, Li K, Yi KM, Tagliasacchi A. \href{https://openaccess.thecvf.com/content/CVPR2021/html/Rebain_DeRF_Decomposed_Radiance_Fields_CVPR_2021_paper.html}{{DeRF}: {Decomposed Radiance Fields}}, 2021, p. 14153--61.}

\leavevmode\vadjust pre{\hypertarget{ref-lindellBACONBandlimitedCoordinate2022}{}}%
\CSLLeftMargin{{[}43{]} }
\CSLRightInline{Lindell DB, Van Veen D, Park JJ, Wetzstein G. {BACON}: {Band-limited Coordinate Networks} for {Multiscale Scene Representation}. {CVPR}, 2022.}

\leavevmode\vadjust pre{\hypertarget{ref-martel2021acorn}{}}%
\CSLLeftMargin{{[}44{]} }
\CSLRightInline{Martel JNP, Lindell DB, Lin CZ, Chan ER, Monteiro M, Wetzstein G. {ACORN}: {Adaptive} coordinate networks for neural representation. ACM Trans Graph (SIGGRAPH) 2021.}

\leavevmode\vadjust pre{\hypertarget{ref-takikawaNeuralGeometricLevel2021}{}}%
\CSLLeftMargin{{[}45{]} }
\CSLRightInline{Takikawa T, Litalien J, Yin K, Kreis K, Loop C, Nowrouzezahrai D, et al. \href{https://openaccess.thecvf.com/content/CVPR2021/html/Takikawa_Neural_Geometric_Level_of_Detail_Real-Time_Rendering_With_Implicit_3D_CVPR_2021_paper.html}{Neural {Geometric Level} of {Detail}: {Real-Time Rendering With Implicit 3D Shapes}}, 2021, p. 11358--67.}

\leavevmode\vadjust pre{\hypertarget{ref-pengConvolutionalOccupancyNetworks2020}{}}%
\CSLLeftMargin{{[}46{]} }
\CSLRightInline{Peng S, Niemeyer M, Mescheder L, Pollefeys M, Geiger A. Convolutional occupancy networks. European conference on computer vision ({ECCV}), 2020.}

\leavevmode\vadjust pre{\hypertarget{ref-xuDISNDeepImplicit2019}{}}%
\CSLLeftMargin{{[}47{]} }
\CSLRightInline{Xu Q, Wang W, Ceylan D, Mech R, Neumann U. \href{http://papers.nips.cc/paper/8340-disn-deep-implicit-surface-network-for-high-quality-single-view-3d-reconstruction.pdf}{{DISN}: {Deep Implicit Surface Network} for {High-quality Single-view 3D Reconstruction}}. In: Wallach H, Larochelle H, Beygelzimer A, Alché-Buc F d', Fox E, Garnett R, editors. Advances in {Neural Information Processing Systems} 32, {Curran Associates, Inc.}; 2019, p. 492--502.}

\leavevmode\vadjust pre{\hypertarget{ref-mullerInstantNeuralGraphics2022}{}}%
\CSLLeftMargin{{[}48{]} }
\CSLRightInline{Müller T, Evans A, Schied C, Keller A. Instant {Neural Graphics Primitives} with a {Multiresolution Hash Encoding} 2022:13.}

\leavevmode\vadjust pre{\hypertarget{ref-yuPlenoxelsRadianceFields2021}{}}%
\CSLLeftMargin{{[}49{]} }
\CSLRightInline{Fridovich-Keil S, Yu A, Tancik M, Chen Q, Recht B, Kanazawa A. Plenoxels: {Radiance Fields} without {Neural Networks}. 2022 {IEEE}/{CVF Conference} on {Computer Vision} and {Pattern Recognition} ({CVPR}), {New Orleans, LA, USA}: {IEEE}; 2022, p. 5491--500. \url{https://doi.org/10.1109/CVPR52688.2022.00542}.}

\leavevmode\vadjust pre{\hypertarget{ref-yuPlenOctreesRealtimeRendering2021}{}}%
\CSLLeftMargin{{[}50{]} }
\CSLRightInline{Yu A, Li R, Tancik M, Li H, Ng R, Kanazawa A. {PlenOctrees} for real-time rendering of neural radiance fields. {ICCV}, 2021.}

\leavevmode\vadjust pre{\hypertarget{ref-karnewarReLUFieldsLittle2022}{}}%
\CSLLeftMargin{{[}51{]} }
\CSLRightInline{Karnewar A, Ritschel T, Wang O, Mitra N. {ReLU Fields}: {The Little Non-linearity That Could}. Special {Interest Group} on {Computer Graphics} and {Interactive Techniques Conference Proceedings}, {Vancouver BC Canada}: {ACM}; 2022, p. 1--9. \url{https://doi.org/10.1145/3528233.3530707}.}

\leavevmode\vadjust pre{\hypertarget{ref-hedmanBakingNeuralRadiance2021}{}}%
\CSLLeftMargin{{[}52{]} }
\CSLRightInline{Hedman P, Srinivasan PP, Mildenhall B, Barron JT, Debevec P. Baking {Neural Radiance Fields} for {Real-Time View Synthesis}. 2021 {IEEE}/{CVF International Conference} on {Computer Vision} ({ICCV}), {Montreal, QC, Canada}: {IEEE}; 2021, p. 5855--64. \url{https://doi.org/10.1109/ICCV48922.2021.00582}.}

\leavevmode\vadjust pre{\hypertarget{ref-reiserMERFMemoryEfficientRadiance2023}{}}%
\CSLLeftMargin{{[}53{]} }
\CSLRightInline{Reiser C, Szeliski R, Verbin D, Srinivasan PP, Mildenhall B, Geiger A, et al. {MERF}: {Memory-Efficient Radiance Fields} for {Real-time View Synthesis} in {Unbounded Scenes} 2023. \url{http://arxiv.org/abs/2302.12249} (accessed March 24, 2023).}

\leavevmode\vadjust pre{\hypertarget{ref-yarivVolumeRenderingNeural2021}{}}%
\CSLLeftMargin{{[}54{]} }
\CSLRightInline{Yariv L, Gu J, Kasten Y, Lipman Y. \href{http://arxiv.org/abs/2106.12052}{Volume {Rendering} of {Neural Implicit Surfaces}} 2021.}

\leavevmode\vadjust pre{\hypertarget{ref-oechsleUNISURFUnifyingNeural2021}{}}%
\CSLLeftMargin{{[}55{]} }
\CSLRightInline{Oechsle M, Peng S, Geiger A. {UNISURF}: {Unifying} neural implicit surfaces and radiance fields for multi-view reconstruction. International conference on computer vision ({ICCV}), 2021.}

\leavevmode\vadjust pre{\hypertarget{ref-liRTNeRFRealTimeOnDevice2022}{}}%
\CSLLeftMargin{{[}56{]} }
\CSLRightInline{Li C, Li S, Zhao Y, Zhu W, Lin Y. {RT-NeRF}: {Real-Time On-Device Neural Radiance Fields Towards Immersive AR}/{VR Rendering}. Proceedings of the 41st {IEEE}/{ACM International Conference} on {Computer-Aided Design}, {San Diego California}: {ACM}; 2022, p. 1--9. \url{https://doi.org/10.1145/3508352.3549380}.}

\leavevmode\vadjust pre{\hypertarget{ref-linEfficientNeuralRadiance2022}{}}%
\CSLLeftMargin{{[}57{]} }
\CSLRightInline{Lin H, Peng S, Xu Z, Yan Y, Shuai Q, Bao H, et al. Efficient {Neural Radiance Fields} for {Interactive Free-viewpoint Video}. {SIGGRAPH Asia} 2022 {Conference Papers}, {Daegu Republic of Korea}: {ACM}; 2022, p. 1--9. \url{https://doi.org/10.1145/3550469.3555376}.}

\leavevmode\vadjust pre{\hypertarget{ref-morozovDifferentiableRenderingReparameterized2023}{}}%
\CSLLeftMargin{{[}58{]} }
\CSLRightInline{Morozov N, Rakitin D, Desheulin O, Vetrov D, Struminsky K. Differentiable {Rendering} with {Reparameterized Volume Sampling} 2023. \url{https://doi.org/10.48550/arXiv.2302.10970}.}

\leavevmode\vadjust pre{\hypertarget{ref-fengSIGNETEfficientNeural2021}{}}%
\CSLLeftMargin{{[}59{]} }
\CSLRightInline{Feng BY, Varshney A. {SIGNET}: {Efficient Neural Representation} for {Light Fields}. 2021 {IEEE}/{CVF International Conference} on {Computer Vision} ({ICCV}), {Montreal, QC, Canada}: {IEEE}; 2021, p. 14204--13. \url{https://doi.org/10.1109/ICCV48922.2021.01396}.}

\leavevmode\vadjust pre{\hypertarget{ref-attalLearningNeuralLight2022}{}}%
\CSLLeftMargin{{[}60{]} }
\CSLRightInline{Attal B, Huang J-B, Zollhofer M, Kopf J, Kim C. Learning {Neural Light Fields} with {Ray-Space Embedding}. 2022 {IEEE}/{CVF Conference} on {Computer Vision} and {Pattern Recognition} ({CVPR}), {New Orleans, LA, USA}: {IEEE}; 2022, p. 19787--97. \url{https://doi.org/10.1109/CVPR52688.2022.01920}.}

\leavevmode\vadjust pre{\hypertarget{ref-rentelnManifoldsTensorsForms2013}{}}%
\CSLLeftMargin{{[}61{]} }
\CSLRightInline{Renteln P. \href{https://books.google.com?id=uJWGAgAAQBAJ}{Manifolds, {Tensors}, and {Forms}: {An Introduction} for {Mathematicians} and {Physicists}}. {Cambridge University Press}; 2013.}

\leavevmode\vadjust pre{\hypertarget{ref-zhouContinuityRotationRepresentations2019}{}}%
\CSLLeftMargin{{[}62{]} }
\CSLRightInline{Zhou Y, Barnes C, Jingwan L, Jimei Y, Hao L. On the {Continuity} of {Rotation Representations} in {Neural Networks}. The {IEEE Conference} on {Computer Vision} and {Pattern Recognition} ({CVPR}), 2019.}

\leavevmode\vadjust pre{\hypertarget{ref-sitzmann2021lfns}{}}%
\CSLLeftMargin{{[}63{]} }
\CSLRightInline{Sitzmann V, Rezchikov S, Freeman B, Tenenbaum J, Durand F. Light field networks: {Neural} scene representations with single-evaluation rendering. Advances in Neural Information Processing Systems 2021;34:19313--25.}

\leavevmode\vadjust pre{\hypertarget{ref-tancikFourierFeaturesLet2020a}{}}%
\CSLLeftMargin{{[}64{]} }
\CSLRightInline{Tancik M, Srinivasan PP, Mildenhall B, Fridovich-Keil S, Raghavan N, Singhal U, et al. \href{http://arxiv.org/abs/2006.10739}{Fourier {Features Let Networks Learn High Frequency Functions} in {Low Dimensional Domains}} 2020.}

\leavevmode\vadjust pre{\hypertarget{ref-mukundAttentionAllThat2023}{}}%
\CSLLeftMargin{{[}65{]} }
\CSLRightInline{Mukund VT, Wang P, Chen X, Chen T, Venugopalan S, Wang Z. Is {Attention All That NeRF Needs}? 2023. \url{https://doi.org/10.48550/arXiv.2207.13298}.}

\leavevmode\vadjust pre{\hypertarget{ref-neffDONeRFRealTimeRendering2021}{}}%
\CSLLeftMargin{{[}66{]} }
\CSLRightInline{Neff T, Stadlbauer P, Parger M, Kurz A, Mueller JH, Chaitanya CRA, et al. {DONeRF}: {Towards Real-Time Rendering} of {Compact Neural Radiance Fields} using {Depth Oracle Networks}. Computer Graphics Forum 2021;40:45--59. \url{https://doi.org/10.1111/cgf.14340}.}

\leavevmode\vadjust pre{\hypertarget{ref-yenamandraFIReFastInverse2022}{}}%
\CSLLeftMargin{{[}67{]} }
\CSLRightInline{Yenamandra T, Tewari A, Yang N, Bernard F, Theobalt C, Cremers D. {FIRe}: {Fast Inverse Rendering} using {Directional} and {Signed Distance Functions} 2022. \url{https://doi.org/10.48550/arXiv.2203.16284}.}

\leavevmode\vadjust pre{\hypertarget{ref-jiaPluckerCoordinatesLines2020}{}}%
\CSLLeftMargin{{[}68{]} }
\CSLRightInline{Jia Y-B. Plücker {Coordinates} for {Lines} in the {Space} 2020. \href{https://faculty.sites.iastate.edu/jia/files/inline-files/plucker-coordinates.pdf}{https://faculty.sites.iastate.edu/jia/files/inline-files/Plücker-coordinates.pdf}.}

\leavevmode\vadjust pre{\hypertarget{ref-fengPRIFPrimaryRaybased2022}{}}%
\CSLLeftMargin{{[}69{]} }
\CSLRightInline{Feng BY, Zhang Y, Tang D, Du R, Varshney A. {PRIF}: {Primary Ray-Based Implicit Function}. In: Avidan S, Brostow G, Cissé M, Farinella GM, Hassner T, editors. Computer {Vision} -- {ECCV} 2022, vol. 13663, {Cham}: {Springer Nature Switzerland}; 2022, p. 138--55. \url{https://doi.org/10.1007/978-3-031-20062-5_9}.}

\leavevmode\vadjust pre{\hypertarget{ref-finnModelagnosticMetalearningFast2017}{}}%
\CSLLeftMargin{{[}70{]} }
\CSLRightInline{Finn C, Abbeel P, Levine S. Model-agnostic meta-learning for fast adaptation of deep networks. International conference on machine learning, {PMLR}; 2017, p. 1126--35.}

\leavevmode\vadjust pre{\hypertarget{ref-rusuMetaLearningLatentEmbedding2018}{}}%
\CSLLeftMargin{{[}71{]} }
\CSLRightInline{Rusu AA, Rao D, Sygnowski J, Vinyals O, Pascanu R, Osindero S, et al. Meta-{Learning} with {Latent Embedding Optimization}. International {Conference} on {Learning Representations}, 2018, p. 11.}

\leavevmode\vadjust pre{\hypertarget{ref-sitzmannMetaSDFMetaLearningSigned2020}{}}%
\CSLLeftMargin{{[}72{]} }
\CSLRightInline{Sitzmann V, Chan ER, Tucker R, Snavely N, Wetzstein G. {MetaSDF}: {Meta-learning Signed Distance Functions}. Advances in Neural Information Processing Systems 2020;33:10136--47.}

\leavevmode\vadjust pre{\hypertarget{ref-kimGraphRepresentationMedial2001}{}}%
\CSLLeftMargin{{[}73{]} }
\CSLRightInline{Kim DH, Yun ID, Lee SU. Graph representation by medial axis transform for {3D} image retrieval. Three-{Dimensional Image Capture} and {Applications IV}, vol. 4298, {SPIE}; 2001, p. 223--30. \url{https://doi.org/10.1117/12.424910}.}

\leavevmode\vadjust pre{\hypertarget{ref-he3DShapeDescriptor2015}{}}%
\CSLLeftMargin{{[}74{]} }
\CSLRightInline{He S, Choi Y-K, Guo Y, Guo X, Wang W. A {3D} shape descriptor based on spectral analysis of medial axis. Computer Aided Geometric Design 2015;39:50--66. \url{https://doi.org/10.1016/j.cagd.2015.08.004}.}

\leavevmode\vadjust pre{\hypertarget{ref-linSEGMAT3DShape2022}{}}%
\CSLLeftMargin{{[}75{]} }
\CSLRightInline{Lin C, Liu L, Li C, Kobbelt L, Wang B, Xin S, et al. {SEG-MAT}: {3D Shape Segmentation Using Medial Axis Transform}. IEEE Transactions on Visualization and Computer Graphics 2022;28:2430--44. \url{https://doi.org/10.1109/TVCG.2020.3032566}.}

\leavevmode\vadjust pre{\hypertarget{ref-duMedialAxisExtraction2004}{}}%
\CSLLeftMargin{{[}76{]} }
\CSLRightInline{Du H, Qin H. Medial axis extraction and shape manipulation of solid objects using parabolic {PDEs}. Proceedings of the ninth {ACM} symposium on {Solid} modeling and applications, {Goslar, DEU}: {Eurographics Association}; 2004, p. 25--35.}

\leavevmode\vadjust pre{\hypertarget{ref-thierySphereMeshesShapeApproximation2013}{}}%
\CSLLeftMargin{{[}77{]} }
\CSLRightInline{Thiery J-M, Guy É, Boubekeur T. Sphere-{Meshes}: Shape approximation using spherical quadric error metrics. ACM Trans Graph 2013;32:1--2. \url{https://doi.org/10.1145/2508363.2508384}.}

\leavevmode\vadjust pre{\hypertarget{ref-thieryAnimatedMeshApproximation2016}{}}%
\CSLLeftMargin{{[}78{]} }
\CSLRightInline{Thiery J-M, Guy É, Boubekeur T, Eisemann E. Animated {Mesh Approximation With Sphere-Meshes}. ACM Trans Graph 2016;35:1--3. \url{https://doi.org/10.1145/2898350}.}

\leavevmode\vadjust pre{\hypertarget{ref-tkachSpheremeshesRealtimeHand2016}{}}%
\CSLLeftMargin{{[}79{]} }
\CSLRightInline{Tkach A, Pauly M, Tagliasacchi A. Sphere-meshes for real-time hand modeling and tracking. ACM Trans Graph 2016;35:1--1. \url{https://doi.org/10.1145/2980179.2980226}.}

\leavevmode\vadjust pre{\hypertarget{ref-anglesVIPERVolumeInvariant2019}{}}%
\CSLLeftMargin{{[}80{]} }
\CSLRightInline{Angles B, Rebain D, Macklin M, Wyvill B, Barthe L, Lewis J, et al. {VIPER}: {Volume Invariant Position-based Elastic Rods}. Proc ACM Comput Graph Interact Tech 2019;2:1--26. \url{https://doi.org/10.1145/3340260}.}

\leavevmode\vadjust pre{\hypertarget{ref-bouixDivergenceBasedMedialSurfaces2000}{}}%
\CSLLeftMargin{{[}81{]} }
\CSLRightInline{Bouix S, Siddiqi K. Divergence-{Based Medial Surfaces}. Computer {Vision} - {ECCV} 2000, {Berlin, Heidelberg}: {Springer}; 2000, p. 603--18. \url{https://doi.org/10.1007/3-540-45054-8_39}.}

\leavevmode\vadjust pre{\hypertarget{ref-tamShapeSimplificationBased2003}{}}%
\CSLLeftMargin{{[}82{]} }
\CSLRightInline{Tam R, Heidrich W. Shape simplification based on the medial axis transform. {IEEE Visualization}, 2003. {VIS} 2003., 2003, p. 481--8. \url{https://doi.org/10.1109/VISUAL.2003.1250410}.}

\leavevmode\vadjust pre{\hypertarget{ref-rebainLSMATLeastSquares2019}{}}%
\CSLLeftMargin{{[}83{]} }
\CSLRightInline{Rebain D, Angles B, Valentin J, Vining N, Peethambaran J, Izadi S, et al. {LSMAT Least Squares Medial Axis Transform}. Computer Graphics Forum 2019;38:5--18. \url{https://doi.org/10.1111/cgf.13599}.}

\leavevmode\vadjust pre{\hypertarget{ref-attaliStabilityComputationMedial2009}{}}%
\CSLLeftMargin{{[}84{]} }
\CSLRightInline{Attali D, Boissonnat J-D, Edelsbrunner H. Stability and {Computation} of {Medial Axes} - a {State-of-the-Art Report}. In: Möller T, Hamann B, Russell RD, editors. Mathematical {Foundations} of {Scientific Visualization}, {Computer Graphics}, and {Massive Data Exploration}, {Berlin, Heidelberg}: {Springer}; 2009, p. 109--25. \url{https://doi.org/10.1007/b106657_6}.}

\leavevmode\vadjust pre{\hypertarget{ref-yangP2MATNETLearningMedial2020}{}}%
\CSLLeftMargin{{[}85{]} }
\CSLRightInline{Yang B, Yao J, Wang B, Hu J, Pan Y, Pan T, et al. {P2MAT-NET}: {Learning} medial axis transform from sparse point clouds. Computer Aided Geometric Design 2020;80:101874. \url{https://doi.org/10.1016/j.cagd.2020.101874}.}

\leavevmode\vadjust pre{\hypertarget{ref-tagliasacchi3DSkeletonsStateoftheArt2016}{}}%
\CSLLeftMargin{{[}86{]} }
\CSLRightInline{Tagliasacchi A, Delame T, Spagnuolo M, Amenta N, Telea A. {3D Skeletons}: {A State-of-the-Art Report}. Computer Graphics Forum 2016;35:573--97. \url{https://doi.org/10.1111/cgf.12865}.}

\leavevmode\vadjust pre{\hypertarget{ref-igehyTracingRayDifferentials1999}{}}%
\CSLLeftMargin{{[}87{]} }
\CSLRightInline{Igehy H. Tracing ray differentials. Proceedings of the 26th annual conference on {Computer} graphics and interactive techniques - {SIGGRAPH} '99, {Not Known}: {ACM Press}; 1999, p. 179--86. \url{https://doi.org/10.1145/311535.311555}.}

\leavevmode\vadjust pre{\hypertarget{ref-cohen-steinerRestrictedDelaunayTriangulations2003}{}}%
\CSLLeftMargin{{[}88{]} }
\CSLRightInline{Cohen-Steiner D, Morvan J-M. Restricted delaunay triangulations and normal cycle. Proceedings of the nineteenth annual symposium on {Computational} geometry, {New York, NY, USA}: {Association for Computing Machinery}; 2003, p. 312--21. \url{https://doi.org/10.1145/777792.777839}.}

\leavevmode\vadjust pre{\hypertarget{ref-baLayerNormalization2016}{}}%
\CSLLeftMargin{{[}89{]} }
\CSLRightInline{Ba JL, Kiros JR, Hinton GE. \href{http://arxiv.org/abs/1607.06450}{Layer {Normalization}} 2016.}

\leavevmode\vadjust pre{\hypertarget{ref-ben-shabatDiGSDivergenceGuided2021}{}}%
\CSLLeftMargin{{[}90{]} }
\CSLRightInline{Ben-Shabat Y, Koneputugodage CH, Gould S. \href{http://arxiv.org/abs/2106.10811}{{DiGS} : {Divergence} guided shape implicit neural representation for unoriented point clouds} 2021.}

\leavevmode\vadjust pre{\hypertarget{ref-heDelvingDeepRectifiers2015}{}}%
\CSLLeftMargin{{[}91{]} }
\CSLRightInline{He K, Zhang X, Ren S, Sun J. Delving {Deep} into {Rectifiers}: {Surpassing Human-Level Performance} on {ImageNet Classification}. 2015 {IEEE International Conference} on {Computer Vision} ({ICCV}), {Santiago, Chile}: {IEEE}; 2015, p. 1026--34. \url{https://doi.org/10.1109/ICCV.2015.123}.}

\leavevmode\vadjust pre{\hypertarget{ref-stanford3DScanning}{}}%
\CSLLeftMargin{{[}92{]} }
\CSLRightInline{The {Stanford 3D Scanning Repository}, {Stanford Computer Graphic Laboratory Homepage} 2014. \url{https://graphics.stanford.edu/data/3Dscanrep/} (accessed January 27, 2023).}

\leavevmode\vadjust pre{\hypertarget{ref-wangActiveCoanalysisSet2012}{}}%
\CSLLeftMargin{{[}93{]} }
\CSLRightInline{Wang Y, Asafi S, van Kaick O, Zhang H, Cohen-Or D, Chen B. Active co-analysis of a set of shapes. ACM Trans Graph 2012;31:165:1--0. \url{https://doi.org/10.1145/2366145.2366184}.}

\leavevmode\vadjust pre{\hypertarget{ref-kleinebergAdversarialGenerationContinuous2020}{}}%
\CSLLeftMargin{{[}94{]} }
\CSLRightInline{Kleineberg M, Fey M, Weichert F. \href{http://arxiv.org/abs/2002.00349}{Adversarial {Generation} of {Continuous Implicit Shape Representations}} 2020.}

\leavevmode\vadjust pre{\hypertarget{ref-pedregosaScikitlearnMachineLearning2011}{}}%
\CSLLeftMargin{{[}95{]} }
\CSLRightInline{Pedregosa F, Varoquaux G, Gramfort A, Michel V, Thirion B, Grisel O, et al. Scikit-learn: {Machine Learning} in {Python}. Journal of Machine Learning Research 2011;12:2825--30.}

\leavevmode\vadjust pre{\hypertarget{ref-kingmaAdamMethodStochastic2017}{}}%
\CSLLeftMargin{{[}96{]} }
\CSLRightInline{Kingma DP, Ba J. \href{http://arxiv.org/abs/1412.6980}{Adam: {A Method} for {Stochastic Optimization}} 2017.}

\leavevmode\vadjust pre{\hypertarget{ref-paszkePyTorchImperativeStyle2019}{}}%
\CSLLeftMargin{{[}97{]} }
\CSLRightInline{Paszke A, Gross S, Massa F, Lerer A, Bradbury J, Chanan G, et al. \href{http://papers.neurips.cc/paper/9015-pytorch-an-imperative-style-high-performance-deep-learning-library.pdf}{{PyTorch}: {An Imperative Style}, {High-Performance Deep Learning Library}}. In: Wallach H, Larochelle H, Beygelzimer A, Alché-Buc F d', Fox E, Garnett R, editors. Advances in {Neural Information Processing Systems} 32, {Curran Associates, Inc.}; 2019, p. 8024--35.}

\leavevmode\vadjust pre{\hypertarget{ref-williamPyTorchLightning2019}{}}%
\CSLLeftMargin{{[}98{]} }
\CSLRightInline{William F, The PyTorch Lightning team. {PyTorch Lightning}. 2019. \url{https://doi.org/10.5281/zenodo.3828935}.}

\leavevmode\vadjust pre{\hypertarget{ref-sjalanderEPICEnergyEfficientHighPerformance2020}{}}%
\CSLLeftMargin{{[}99{]} }
\CSLRightInline{Själander M, Jahre M, Tufte G, Reissmann N. \href{http://arxiv.org/abs/1912.05848}{{EPIC}: {An Energy-Efficient}, {High-Performance GPGPU Computing Research Infrastructure}} 2020.}

\leavevmode\vadjust pre{\hypertarget{ref-scopatzPyembreePythonWrapper2022}{}}%
\CSLLeftMargin{{[}100{]} }
\CSLRightInline{Scopatz A. \href{https://github.com/adam-grant-hendry/pyembree}{Pyembree: {Python} wrapper for {Intel Embree} 2.17.7}. 2022.}

\leavevmode\vadjust pre{\hypertarget{ref-raviAccelerating3DDeep2020}{}}%
\CSLLeftMargin{{[}101{]} }
\CSLRightInline{Ravi N, Reizenstein J, Novotny D, Gordon T, Lo W-Y, Johnson J, et al. \href{https://arxiv.org/abs/2007.08501}{Accelerating {3D Deep Learning} with {PyTorch3D}} 2020.}

\leavevmode\vadjust pre{\hypertarget{ref-barre-briseboisApproximatingTranslucencyFast2011}{}}%
\CSLLeftMargin{{[}102{]} }
\CSLRightInline{Barré-Brisebois C, Bouchard M. Approximating {Translucency} for a {Fast Cheap} and {Convincing Subsurface Scattering Look}. Game developers conference, vol. 6, 2011.}

\leavevmode\vadjust pre{\hypertarget{ref-ward1992measuring}{}}%
\CSLLeftMargin{{[}103{]} }
\CSLRightInline{Ward GJ. Measuring and modeling anisotropic reflection. Proceedings of the 19th annual conference on {Computer} graphics and interactive techniques, 1992, p. 265--72.}

\end{CSLReferences}

\else \printbibliography[heading=none] \renewcommand{\printbibliography}{} \fi

\normalsize

\appendix \counterwithin{figure}{section}

\hypertarget{appendix}{%
\section{Appendix}\label{appendix}}

\hypertarget{metrics}{%
\subsection{Metrics}\label{metrics}}

\hypertarget{precision-recall-and-intersection-over-union-iou.}{%
\paragraph*{Precision, Recall, and Intersection over Union (IoU).}\label{precision-recall-and-intersection-over-union-iou.}}
\addcontentsline{toc}{paragraph}{Precision, Recall, and Intersection over Union (IoU).}

IoU quantifies the overlap between two binary classifiers, in our case ray hit/miss classification. Precision and recall scores the relevance of the classification. For a batch of rays \(B\) with hits being positive, the Precision, Recall, and IoU is:

\begin{equation}{ \begin{aligned}
  \text{Precision}
    &\ =\ \frac{TP}{TP+FP}
    &= \frac{
      \left|\left\{\ell\in B : s_\ell=0 \wedge s_\ell^\text{GT}=0\right\}\right|
    }{
      \left|\left\{\ell\in B : s_\ell=0\right\}\right|
    }
  \\
  \text{Recall}
    &\ =\ \frac{TP}{TP+FN}
    &= \frac{
      \left|\left\{\ell\in B : s_\ell=0 \wedge s_\ell^\text{GT}=0\right\}\right|
    }{
      \left|\left\{\ell\in B : s_\ell^\text{GT}=0\right\}\right|
    }
  \\
  \text{IoU}
    &\ =\ \frac{TP}{TP+FP+FN}
    &= \frac{
      \left|\left\{\ell\in B : s_\ell=0 \wedge s_\ell^\text{GT}=0\right\}\right|
    }{
      \left|\left\{\ell\in B : s_\ell=0 \vee s_\ell^\text{GT}=0\right\}\right|
    }
\end{aligned} }\end{equation}

\hypertarget{chamfer-distance-cd-and-cosine-similarity-cos.}{%
\paragraph*{Chamfer Distance (CD) and Cosine Similarity (COS).}\label{chamfer-distance-cd-and-cosine-similarity-cos.}}
\addcontentsline{toc}{paragraph}{Chamfer Distance (CD) and Cosine Similarity (COS).}

CD is the ``average-case'' distance between two point clouds \(U\) and \(V\). COS scores orientation using the same matching between \(U\) and \(V\) as CD, and computes the normal vector cosine similarity.

\begin{equation}{ \begin{aligned}
  \operatorname{CD}\
    &= \frac{1}{|U|}\sum_{\mathbf u\in U}\min_{\mathbf v\in V}\left\| \mathbf u -\mathbf v \right\| \\
    &+ \frac{1}{|V|}\sum_{\mathbf v\in V}\min_{\mathbf u\in U}\left\| \mathbf u -\mathbf v \right\| \\
  \operatorname{COS}\
    &= \frac{1}{|U|}\sum_{\mathbf u\in U}\hat{\mathbf n}_{u}\cdot\hat{\mathbf n}_{\arg\min_{\mathbf v\in V}\|\mathbf u-\mathbf v\|} \\
    &+ \frac{1}{|V|}\sum_{\mathbf v\in V}\hat{\mathbf n}_{v}\cdot\hat{\mathbf n}_{\arg\min_{\mathbf u\in U}\|\mathbf u-\mathbf v\|} \\
\end{aligned} }\end{equation}

where \(\hat{\mathbf n}_{\mathbf x}\) is the unit normal vector of oriented point \(\mathbf x\).

\end{document}